[ english version / português abaixo ]

# **New world order, globalism and QAnon communities on Brazilian Telegram:** how conspiracism opens doors to more harmful groups

*Ergon Cugler de Moraes Silva*

Brazilian Institute of Information in
Science and Technology (IBICT)
Brasília, Federal District, Brazil

contato@ergoncugler.com
www.ergoncugler.com

## Abstract

Conspiracy theories involving the New World Order (NWO), Globalism, and QAnon have become central to discussions on Brazilian Telegram, especially during global crises such as the COVID-19 pandemic. Therefore, this study aims to address the research question: **how are Brazilian conspiracy theory communities on new world order, globalism and QAnon topics characterized and articulated on Telegram?** It is worth noting that this study is part of a series of seven studies whose main objective is to understand and characterize Brazilian conspiracy theory communities on Telegram. This series of seven studies is openly and originally available on arXiv at Cornell University, applying a mirrored method across the seven studies, changing only the thematic object of analysis and providing investigation replicability, including with proprietary and authored codes, adding to the culture of free and open-source software. Regarding the main findings of this study, the following were observed: NWO and Globalism have become central catalysts for the dissemination of conspiracy theories; QAnon acts as a "hub narrative" that connects NWO and Globalism; During crises, mentions of NWO have grown exponentially, reflecting distrust in institutions; NWO and Globalism attract followers of other conspiracy theories, such as anti-vaccines, serving as the main gatekeeper of the entire conspiracy theory network; Religious narratives are often used to legitimize NWO, reinforcing ideological cohesion.

## Key findings

- ➔ Especially during and after the COVID-19 pandemic, the New World Order (NWO) and Globalism became central catalysts for the dissemination of conspiracy theories on Brazilian Telegram, with 2,411,003 posts about NWO and 768,176 about Globalism, creating an environment of global distrust that reinforces the cohesion among these narratives;
- ➔ The interconnection between NWO, Globalism, and QAnon forms a dense network of disinformation, where QAnon acts as a "narrative-hub" connecting 531,678 posts, while NWO and Globalism amplify these connections, rapidly exposing members to a vast network;
- ➔ During global crises, such as the COVID-19 pandemic and the 2020 U.S. elections, mentions of NWO and Globalism grew exponentially, with NWO registering a 3,150% increase between 2019 and 2021, reflecting growing distrust in institutions;

- ➔ NWO and Globalism act as gateways for new members already immersed in other conspiracy theories, receiving 8,699 links from General Conspiracies and 6,260 from anti-vaccine communities, functioning as epicenters that attract and retain followers from other narratives;
- ➔ Globalism stands out as an organizing principle of various conspiracy theories, connecting themes such as Anti-Woke, Revisionism, and Climate Change, with 32,640 links received from General Conspiracies, suggesting that all theories are part of a larger plan;
- ➔ QAnon operates as a flexible metanarrative that absorbs and connects other conspiracy theories, receiving 6,308 links from General Conspiracies and 5,033 from NWO, reinforcing the conspiratorial worldview of its followers and making it difficult for contradictory information to enter;
- ➔ NWO consolidates itself as a central interpretive paradigm in the conspiratorial universe, unifying various narratives against a supposed global elite, with 3,488,686 posts, functioning as an aggregator of disinformation and legitimizing a conspiratorial worldview;
- ➔ Religious themes such as “God” and “Bible” are instrumentalized in narratives about NWO and QAnon, suggesting that the fight against NWO and Globalism is a spiritual battle, which strengthens the ideological cohesion of communities and resistance to scientific information;
- ➔ In NWO and Globalism communities, climate change is seen as part of a global conspiracy to control the population, with topics such as “carbon” and “climate” frequently cited, contributing to the rejection of science and polarization;
- ➔ NWO communities perpetuate disinformation about vaccines, associating them with a plan for population control, suggesting that vaccination campaigns are part of a manipulation agenda, which perpetuates vaccine resistance and amplifies disinformation.

## 1. Introduction

After analyzing thousands of Brazilian conspiracy theory communities on Telegram and extracting tens of millions of content pieces from these communities, created and/or shared by millions of users, this study aims to compose a series of seven studies that address the phenomenon of conspiracy theories on Telegram, focusing on Brazil as a case study. Through the identification approaches implemented, it was possible to reach a total of 217 Brazilian conspiracy theory communities on Telegram on new world order, globalism and QAnon topics, summing up 5,545,369 content pieces published between June 2016 (initial publications) and August 2024 (date of this study), with 718,246 users aggregated from within these communities. Thus, this study aims to understand and characterize the communities focused on new world order, globalism and QAnon present in this Brazilian network of conspiracy theories identified on Telegram.

To this end, a mirrored method will be applied across all seven studies, changing only the thematic object of analysis and providing investigation replicability. In this way, we will adopt technical approaches to observe the connections, temporal series, content, and overlaps of themes within the conspiracy theory communities. In addition to this study, the other six are openly and originally available on arXiv at Cornell University. This series paid particular attention to ensuring data integrity and respecting user privacy, as provided by Brazilian legislation (Law No. 13,709/2018 / Brazilian law from 2018).

Therefore, the question arises: **how are Brazilian conspiracy theory communities on new world order, globalism and QAnon topics characterized and articulated on Telegram?**

## 2. Materials and methods

The methodology of this study is organized into three subsections: **2.1. Data extraction**, which describes the process and tools used to collect information from Telegram communities; **2.2. Data processing**, which discusses the criteria and methods applied to classify and anonymize the collected data; and **2.3. Approaches to data analysis**, which details the techniques used to investigate the connections, temporal series, content, and thematic overlaps within conspiracy theory communities.

### 2.1. Data extraction

This project began in February 2023 with the publication of the first version of TelegramScrap (Silva, 2023), a proprietary, free, and open-source tool that utilizes Telegram's Application Programming Interface (API) by Telethon library and organizes data extraction cycles from groups and open channels on Telegram. Over the months, the database was expanded and refined using four approaches to identifying conspiracy theory communities:

**(i) Use of keywords:** at the project's outset, keywords were listed for direct identification in the search engine of Brazilian groups and channels on Telegram, such as "apocalypse", "survivalism", "climate change", "flat earth", "conspiracy theory", "globalism", "new world order", "occultism", "esotericism", "alternative cures", "qAnon" "reptilians", "revisionism", "aliens", among others. This initial approach provided some communities whose titles and/or descriptions of groups and channels explicitly contained terms related to conspiracy theories. However, over time, it was possible to identify many other communities that the listed keywords did not encompass, some of which deliberately used altered characters to make it difficult for those searching for them on the network.

**(ii) Telegram channel recommendation mechanism:** over time, it was identified that Telegram channels (except groups) have a recommendation tab called "similar channels", where Telegram itself suggests ten channels that have some similarity with the channel being observed. Through this recommendation mechanism, it was possible to find more Brazilian conspiracy theory communities, with these being recommended by the platform itself.

**(iii) Snowball approach for invitation identification:** after some initial communities were accumulated for extraction, a proprietary algorithm was developed to identify URLs containing "t.me/", the prefix for any invitation to Telegram groups and channels. Accumulating a frequency of hundreds of thousands of links that met this criterion, the unique addresses were listed, and their repetitions counted. In this way, it was possible to investigate new Brazilian groups and channels mentioned in the messages of those already investigated, expanding the network. This process was repeated periodically to identify new communities aligned with conspiracy theory themes on Telegram.

**(iv) Expansion to tweets published on X mentioning invitations:** to further diversify the sources of Brazilian conspiracy theory communities on Telegram, a proprietary search query was developed to identify conspiracy theory-themed keywords using tweets published on X (formerly Twitter) that, in addition to containing one of the keywords, also included "t.me/", the prefix for any invitation to Telegram groups and channels, "https://x.com/search?q=lang%3Apt%20%22t.me%2F%22%20SEARCH-TERM".

With the implementation of community identification approaches for conspiracy theories developed over months of investigation and method refinement, it was possible to build a project database encompassing a total of 855 Brazilian conspiracy theory communities on Telegram (including other themes not covered in this study). These communities have collectively published 27,227,525 pieces of content from May 2016 (the first publications) to August 2024 (the period of this study), with a combined total of 2,290,621 users across the Brazilian communities. It is important to note that this volume of users includes two elements: first, it is a variable figure, as users can join and leave communities daily, so this value represents what was recorded on the publication extraction date; second, it is possible that the same user is a member of more than one group and, therefore, is counted more than once. In this context, while the volume remains significant, it may be slightly lower when considering the deduplicated number of citizens within these Brazilian conspiracy theory communities.

### 2.2. Data processing

With all the Brazilian conspiracy theory groups and channels on Telegram extracted, a manual classification was conducted considering the title and description of the community. If there was an explicit mention in the title or description of the community related to a specific theme, it was classified into one of the following categories: (i) "Anti-Science"; (ii) "Anti-Woke and Gender"; (iii) "Antivax"; (iv) "Apocalypse and Survivalism"; (v) "Climate Changes"; (vi) "Flat Earth"; (vii) "Globalism"; (viii) "New World Order"; (ix) "Occultism and Esotericism"; (x) "Off Label and Quackery"; (xi) "QAnon"; (xii) "Reptilians and Creatures"; (xiii) "Revisionism and Hate Speech"; (xiv) "UFO and Universe". If there was no explicit mention related to the themes in the title or description of the community, it was classified as (xv) "General Conspiracy". In the following table, we can observe the metrics related to the classification of these conspiracy theory communities in Brazil.

**Table 01.** Conspiracy theory communities in Brazil (metrics up to August 2024)

| Categories | Groups | Users | Contents | Comments | Total |
|---|---|---|---|---|---|
| **Anti-Science** | 22 | 58,138 | 187,585 | 784,331 | 971,916 |
| **Anti-Woke and Gender** | 43 | 154,391 | 276,018 | 1,017,412 | 1,293,430 |
| **Antivax** | 111 | 239,309 | 1,778,587 | 1,965,381 | 3,743,968 |
| **Apocalypse and Survivalism** | 33 | 109,266 | 915,584 | 429,476 | 1,345,060 |
| **Climate Changes** | 14 | 20,114 | 269,203 | 46,819 | 316,022 |
| **Flat Earth** | 33 | 38,563 | 354,200 | 1,025,039 | 1,379,239 |
| **General Conspiracy** | 127 | 498,190 | 2,671,440 | 3,498,492 | 6,169,932 |
| **Globalism** | 41 | 326,596 | 768,176 | 537,087 | 1,305,263 |
| **NWO** | 148 | 329,304 | 2,411,003 | 1,077,683 | 3,488,686 |
| **Occultism and Esotericism** | 39 | 82,872 | 927,708 | 2,098,357 | 3,026,065 |
| **Off Label and Quackery** | 84 | 201,342 | 929,156 | 733,638 | 1,662,794 |
| **QAnon** | 28 | 62,346 | 531,678 | 219,742 | 751,420 |
| **Reptilians and Creatures** | 19 | 82,290 | 96,262 | 62,342 | 158,604 |
| **Revisionism and Hate Speech** | 66 | 34,380 | 204,453 | 142,266 | 346,719 |
| **UFO and Universe** | 47 | 58,912 | 862,358 | 406,049 | 1,268,407 |
| **Total** | **855** | **2,296,013** | **13,183,411** | **14,044,114** | **27,227,525** |

Source: Own elaboration (2024).

With this volume of extracted data, it was possible to segment and present in this study only communities and content related to new world order, globalism and QAnon themes. In parallel, other themes of Brazilian conspiracy theory communities were also addressed with studies aimed at characterizing the extent and dynamics of the network, which are openly and originally available on arXiv at Cornell University.

Additionally, it should be noted that only open communities were extracted, meaning those that are not only publicly identifiable but also do not require any request to access the content, being available to any user with a Telegram account who needs to join the group or channel. Furthermore, in compliance with Brazilian legislation, particularly the General Data Protection Law (LGPD), or Law No. 13,709/2018 (Brazilian law from 2018), which deals with privacy control and the use/treatment of personal data, all extracted data were anonymized for the purposes of analysis and investigation. Therefore, not even the identification of the communities is possible through this study, thus extending the user's privacy by anonymizing their data beyond the community itself to which they submitted by being in a public and open group or channel on Telegram.

## 2.3. Approaches to data analysis

A total of 217 selected communities focused on new world order, globalism and QAnon themes, containing 5,545,369 publications and 718,246 combined users, will be analyzed. Four approaches will be used to investigate the conspiracy theory communities selected for the scope of this study. These metrics are detailed in the following table:

**Table 02.** Selected communities for analysis (metrics up to August 2024)

| Categories | Groups | Users | Contents | Comments | Total |
|---|---|---|---|---|---|
| **NWO** | 148 | 329,304 | 2,411,003 | 1,077,683 | 3,488,686 |
| **QAnon** | 28 | 62,346 | 531,678 | 219,742 | 751,420 |
| **Globalism** | 41 | 326,596 | 768,176 | 537,087 | 1,305,263 |
| **Total** | **217** | **718,246** | **3,710,857** | **1,834,512** | **5,545,369** |

Source: Own elaboration (2024).

**(i) Network:** by developing a proprietary algorithm to identify messages containing the term "t.me/" (inviting users to join other communities), we propose to present volumes and connections observed on how **(a)** communities within the new world order, globalism and QAnon theme circulate invitations for their users to explore more groups and channels within the same theme, reinforcing shared belief systems; and how **(b)** these same communities circulate invitations for their users to explore groups and channels dealing with other conspiracy theories, distinct from their explicit purpose. This approach is valuable for observing whether these communities use themselves as a source of legitimation and reference and/or rely on other conspiracy theory themes, even opening doors for their users to explore other conspiracies. Furthermore, it is worth mentioning the study by Rocha *et al.* (2024), where a network identification approach was also applied in Telegram communities, but by observing similar content based on an ID generated for each unique message and its similar ones;

**(ii) Time series:** the "Pandas" library (McKinney, 2010) is used to organize the investigation data frames, observing **(a)** the volume of publications over the months; and **(b)** the volume of engagement over the months, considering metadata of views, reactions, and comments collected during extraction. In addition to volumetry, the "Plotly" library (Plotly Technologies Inc., 2015) enabled the graphical representation of this variation;

**(iii) Content analysis:** in addition to the general word frequency analysis, time series are applied to the variation of the most frequent words over the semesters—observing from June 2016 (initial publications) to August 2024 (when this study was conducted). With the "Pandas" (McKinney, 2010) and "WordCloud" (Mueller, 2020) libraries, the results are presented both volumetrically and graphically;

**(iv) Thematic agenda overlap:** following the approach proposed by Silva & Sátiro (2024) for identifying thematic agenda overlap in Telegram communities, we used the "BERTopic" model (Grootendorst, 2020). BERTopic is a topic modeling algorithm that facilitates the generation of thematic representations from large amounts of text. First, the algorithm extracts document embeddings using sentence transformer models, such as "all-MiniLM-L6-v2". These embeddings are then reduced in dimensionality using techniques like "UMAP", facilitating the clustering process. Clustering is performed using "HDBSCAN", a density-based technique that identifies clusters of different shapes and sizes, as well as outliers. Subsequently, the documents are tokenized and represented in a bag-of-words structure, which is normalized (L1) to account for size differences between clusters. The topic representation is refined using a modified version of "TF-IDF", called "Class-TF-IDF", which considers the importance of words within each cluster (Grootendorst, 2020). It is important to note that before applying BERTopic, we cleaned the dataset by removing Portuguese "stopwords" using "NLTK" (Loper & Bird, 2002). For topic modeling, we used the "loky" backend to optimize performance during data fitting and transformation.

In summary, the methodology applied ranged from data extraction using the own tool TelegramScrap (Silva, 2023) to the processing and analysis of the collected data, employing various approaches to identify and classify Brazilian conspiracy theory communities on Telegram. Each stage was carefully implemented to ensure data integrity and respect for user privacy, as mandated by Brazilian legislation. The results of this data will be presented below, aiming to reveal the dynamics and characteristics of the studied communities.

## 3. Results

The results are detailed below in the order outlined in the methodology, beginning with the characterization of the network and its sources of legitimation and reference, progressing to the time series, incorporating content analysis, and concluding with the identification of thematic agenda overlap among the conspiracy theory communities.

### 3.1. Network

The analysis of the network of communities related to the New World Order (NWO), Globalism, and QAnon reveals the complexity and interconnectivity of this conspiratorial ecosystem. The first figure, which depicts the internal network of these themes, highlights how these communities are densely intertwined, creating a continuous cycle of mutual reinforcement among their beliefs. The interdependence of these narratives is evidenced by the large nodes that act as epicenters of dissemination, where once followers enter this conspiratorial universe, they are quickly engulfed by a multiplicity of narratives. These interactions make it increasingly difficult for individuals involved to separate reality from conspiracy theory, leading to an environment where ideas about global control, hidden elites, and QAnon are in constant feedback. The second figure explores the communities that serve as gateways to these themes, showing how NWO, Globalism, and QAnon attract new followers by exposing these communities to a vast universe of theories. These communities function as central hubs in the conspiratorial network, facilitating the transition and deepening of followers into other interconnected theories. The significant overlap between these communities highlights that entry into one of them can quickly lead to the adoption of multiple other theories, creating a continuous cycle of reinforcement of these beliefs.

In the third figure, we observe how these themes interconnect with other conspiratorial areas, acting as exit gateways to new narratives. The centrality of NWO communities in the network, for example, shows that they play a crucial role in connecting different theories, functioning as nuclei of dissemination that encourage the exploration of new conspiratorial areas. This indicates that followers of NWO, Globalism, and QAnon not only reinforce their beliefs but are also constantly exposed to new narratives, expanding their engagement within this ecosystem. The flow charts of invitation links, which analyze the centrality of NWO, Globalism, and QAnon, reinforce the idea that these themes are not isolated but deeply intertwined with a vast network of conspiracy theories. NWO, for instance, receives a significant volume of links from General Conspiracies and Anti-Vaccination communities, functioning as a central interpretive paradigm that unifies different narratives against a supposed global elite. Globalism, in turn, acts as an organizing principle of other theories, strengthening the internal cohesion of these communities by suggesting that all conspiracies are part of a larger plan. Meanwhile, QAnon, by connecting with various other theories, functions as a “narrative hub”, synthesizing different beliefs into a coherent whole and reinforcing the original conspiratorial worldview of its followers.

In summary, the networks and flows of links between NWO, Globalism, and QAnon show that these themes operate as central engines within the conspiratorial universe, acting both as gateways and as exit points to other theories, while simultaneously reinforcing the cohesion and expansion of this ecosystem of interconnected beliefs.

[ english version / português abaixo ]

**Figure 01.** Internal network between nwo, globalism and QAnon communities

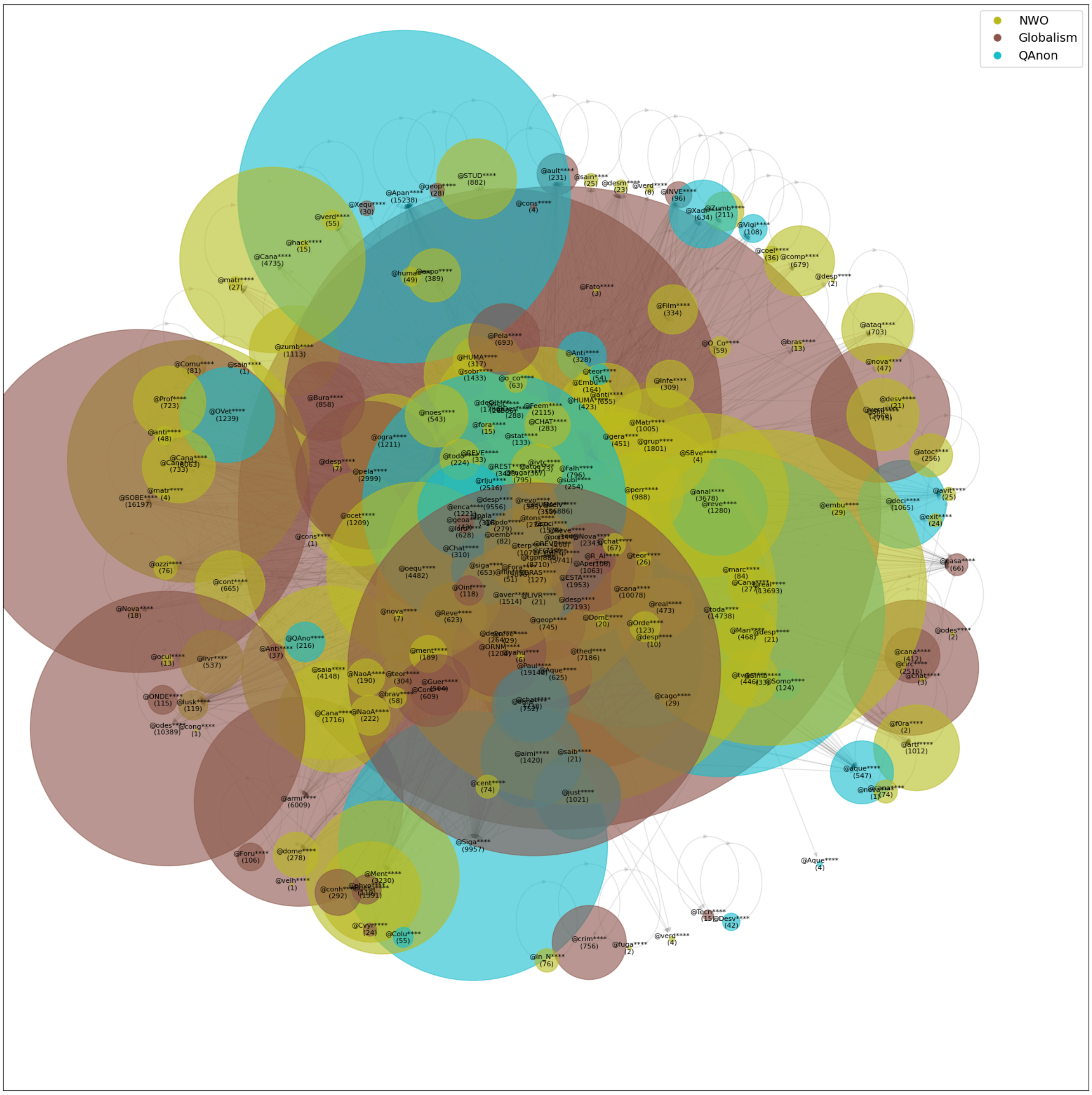

Source: Own elaboration (2024).

This graph reveals a complex and densely connected network among communities discussing topics such as the New World Order, Globalism, and QAnon. The significant overlap between the groups suggests a strong interdependence between these conspiratorial narratives, with communities frequently referencing one another. This network reflects a conspiratorial ecosystem where ideas about global control, hidden elites, and QAnon-related theories are constantly interlinked, creating a continuous cycle of reinforcement of these beliefs. The density and breadth of the network indicate that once followers enter this circle, they are quickly engulfed by a multiplicity of narratives, making it difficult for them to discern reality from conspiracy theories. The large nodes represent the epicenters of these interactions, with a high capacity for content amplification that spreads throughout the network, reinforcing the cohesion among the different narratives.

[ english version / português abaixo ]

**Figure 02.** Network of communities that open doors to the theme (gateway)

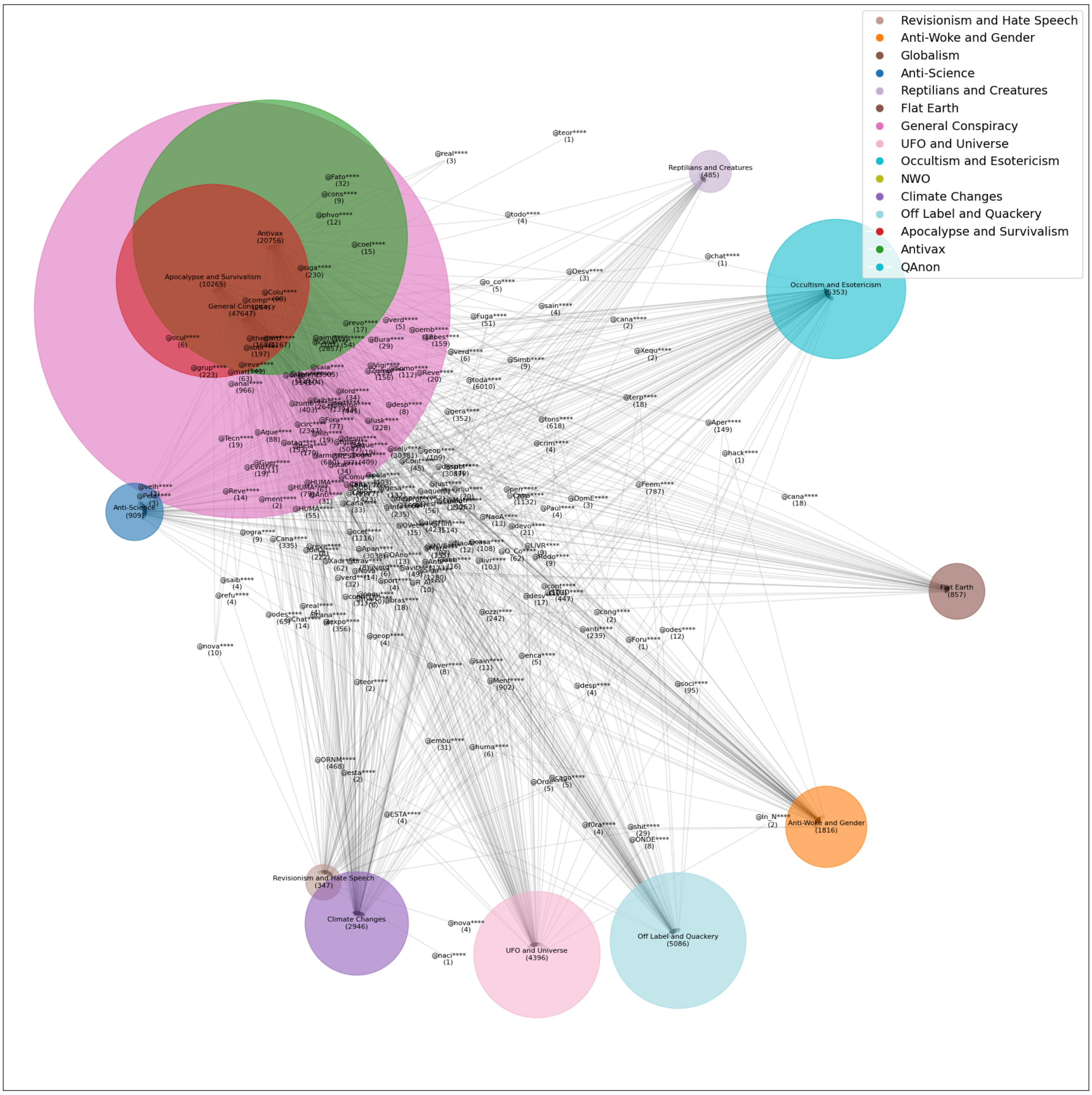

Source: Own elaboration (2024).

The figure shows the network of communities that are gateways to discussions about the New World Order, Globalism, and QAnon. The large spheres representing these communities highlight their importance as central hubs in the conspiratorial network. These themes act as magnets for conspiracy theory followers, who, upon contact with these communities, are exposed to a vast network of other related theories. The significant overlap between these communities suggests that the transition between different theories is common, creating a continuous cycle of reinforcement of these beliefs. Additionally, these communities not only centralize these narratives but also enhance the ability to connect different conspiratorial areas, increasing individual engagement.

[ english version / português abaixo ]

**Figure 03.** Network of communities whose theme opens doors (exit point)

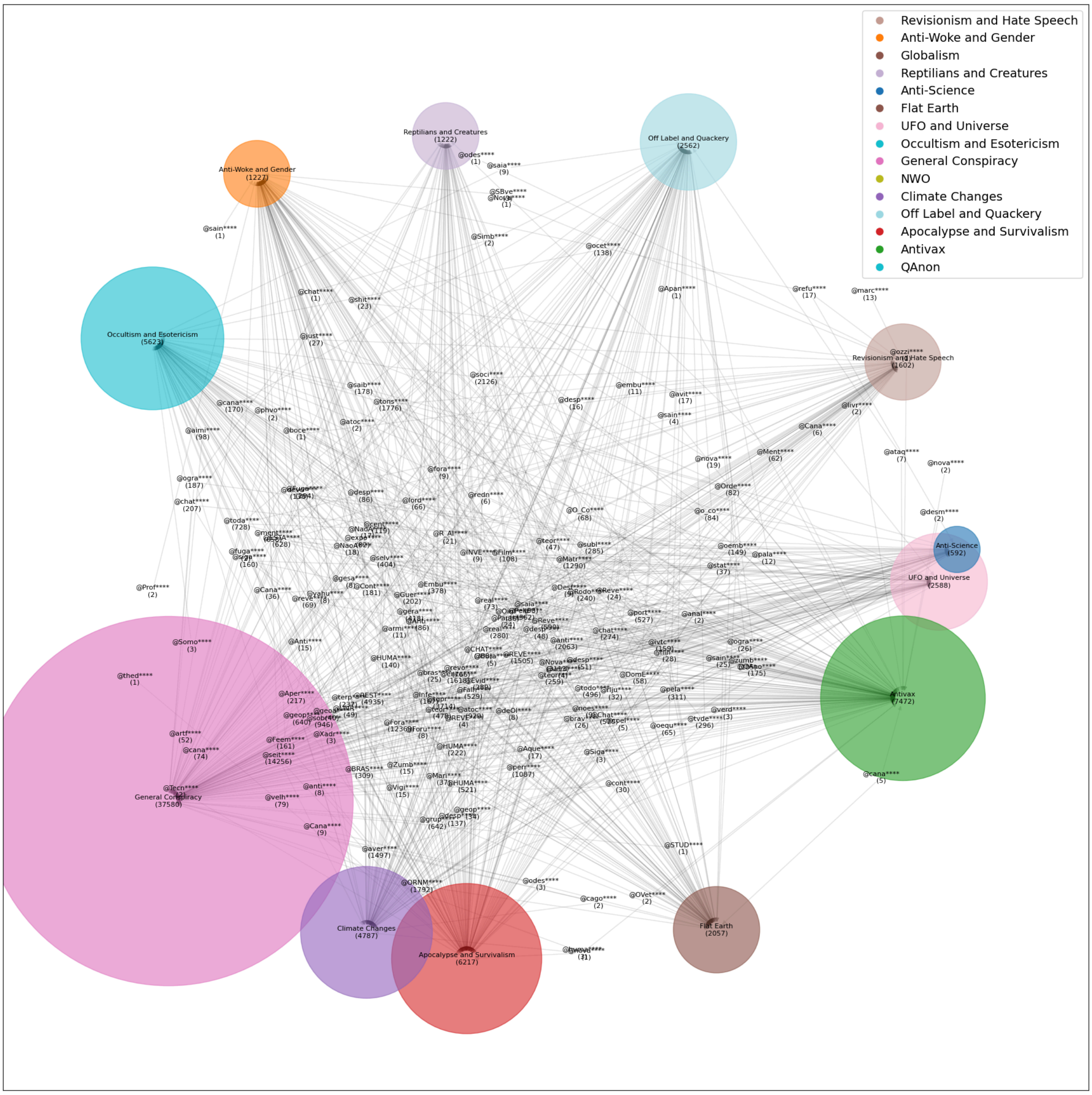

Source: Own elaboration (2024).

This graph highlights the network of communities related to the New World Order (NWO), Globalism, and QAnon, demonstrating how they intertwine with various other conspiratorial themes. The centrality of NWO communities in the network indicates that they play a crucial role in connecting different conspiracy theories. The interactions between NWO and themes such as Globalism, QAnon, and Occultism are particularly noteworthy, suggesting that these communities function as nuclei of disinformation dissemination, which not only reinforce the beliefs of their members but also encourage the exploration of new conspiratorial areas. The dense web of connections indicates that followers of NWO are more likely to engage in a wide range of conspiracy theories, indicating that these communities act as a vital link for the propagation of an amplified and multifaceted conspiratorial worldview.

[ english version / português abaixo ]

**Figure 04.** Flow of invitation links between new world order communities

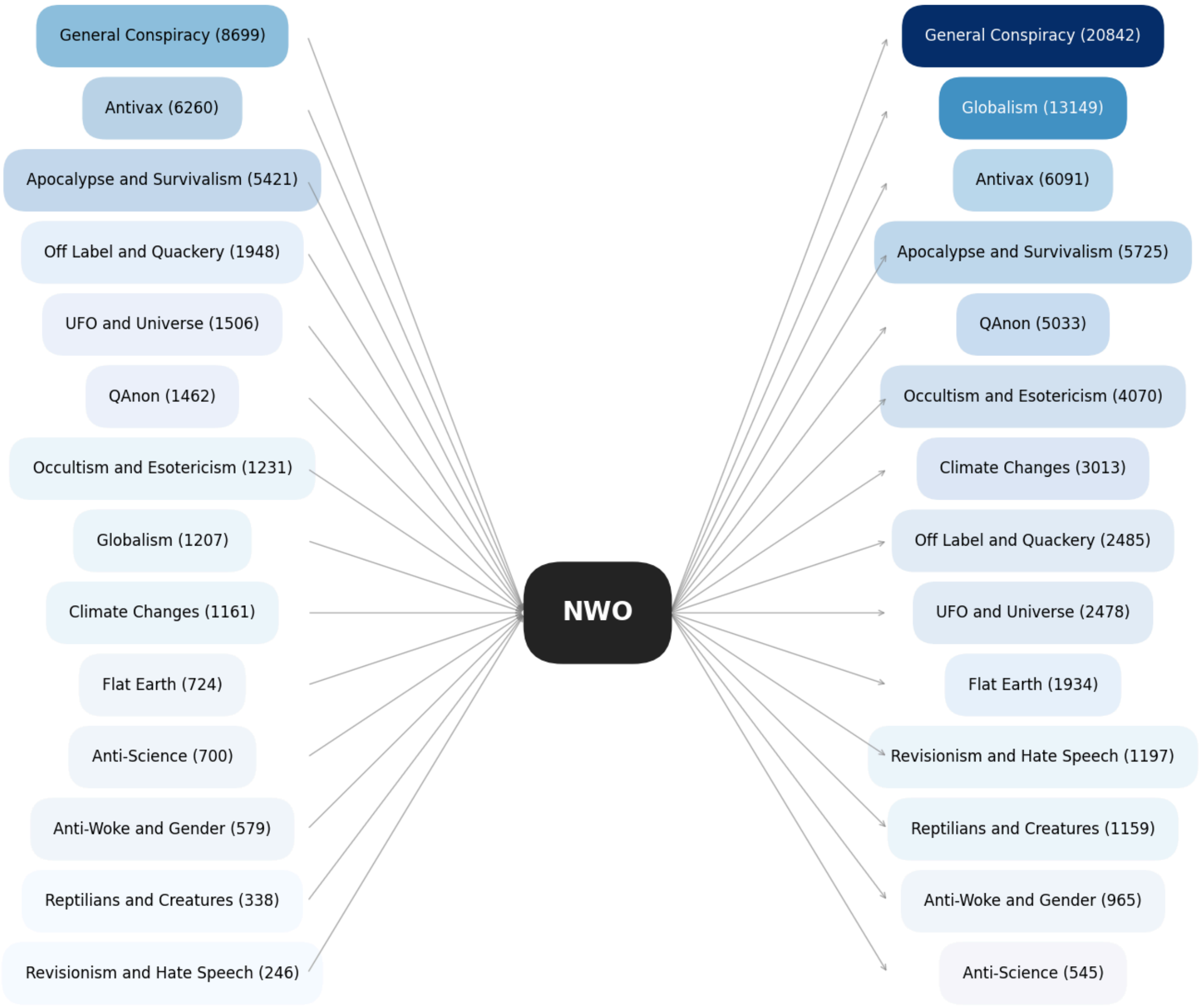

Source: Own elaboration (2024).

The flow chart of invitations on the New World Order (NWO) theme clearly demonstrates its centrality in the conspiratorial universe. With a significant volume of links originating from General Conspiracies (8,699) and Anti-Vaccination (6,260), NWO is configured not only as a central theme but as the main aggregator of narratives that intertwine around distrust against global institutions. NWO functions as an interpretive paradigm, where various narratives find coherence and meaning by positioning themselves against a supposed global elite that manipulates global events. This interconnection is not trivial. It implies that NWO acts as a platform for mutual reinforcement among different conspiracy theories. The theories circulating within this network are mutually reinforcing, creating a cycle where individuals who enter through one of these gateways (such as Anti-Vaccination or Apocalypse and Survivalism) are progressively led to adopt a worldview where everything is interconnected by a common thread of domination and control. This cycle, therefore, is not just an addition of beliefs but a reconfiguration of critical thinking, which begins to interpret all global events as pieces of the same manipulative puzzle.

[ english version / português abaixo ]

**Figure 05.** Flow of invitation links between globalism communities

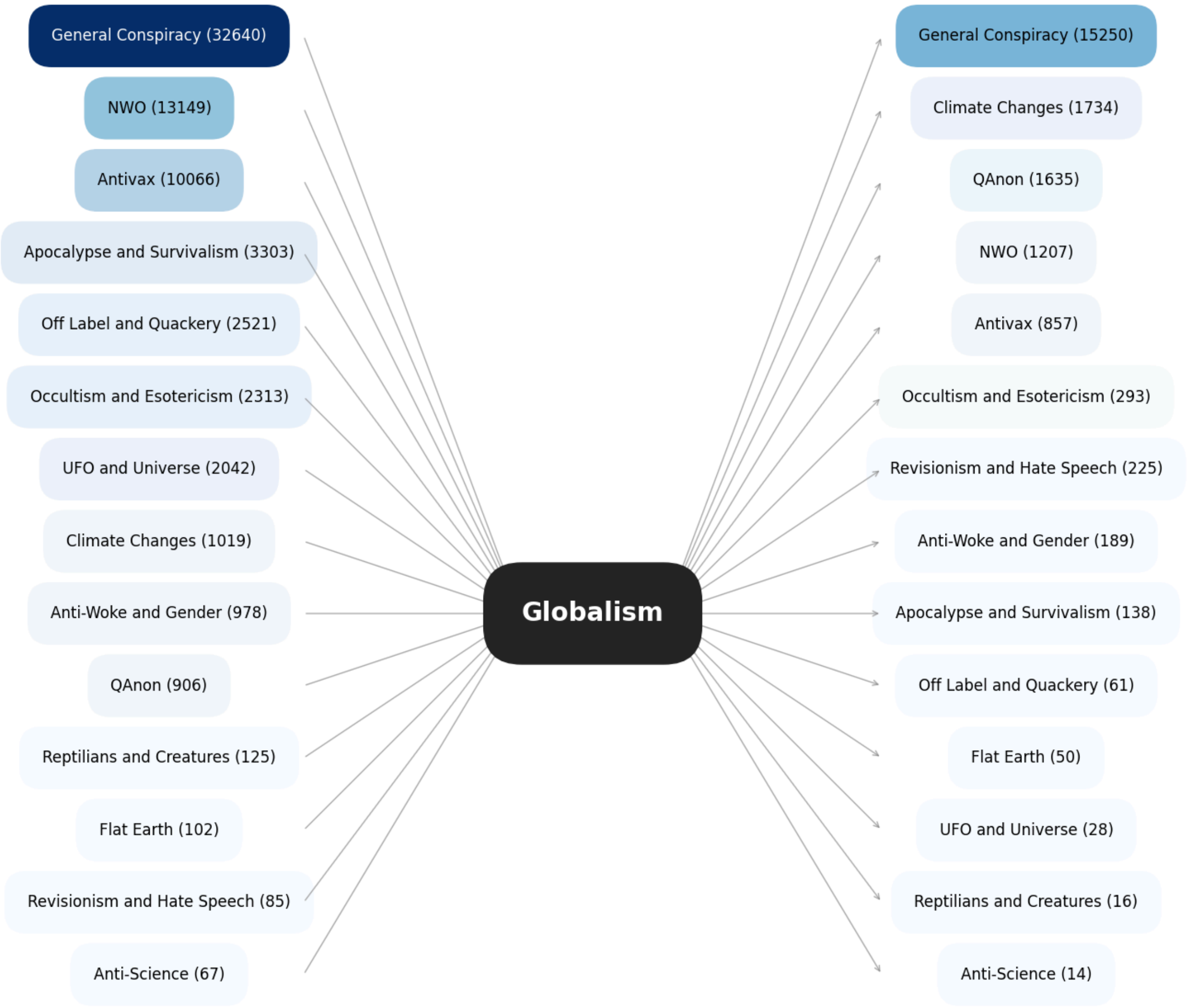

Source: Own elaboration (2024).

The Globalism graph reveals an intrinsic relationship between the theme and General Conspiracies, which sends 32,640 links, highlighting that Globalism is a concept that is strengthened and self-reinforced by various conspiracy theories. This connection indicates that Globalism is not just a narrative but an organizing principle of other theories, a sort of "glue" that unites different discourses of resistance against the established world order. The narrative of Globalism functions as a form of legitimization of other theories by suggesting that all supposed conspiracies are part of a larger plan. As Globalism distributes invitations to General Conspiracies (15,250 links) and Climate Change (1,734 links), it becomes evident that this theory not only attracts but also disseminates a diverse set of beliefs. This dual role suggests that Globalism operates both as an entry point and as a channel for disinformation, building bridges between narratives that might initially seem disconnected. The analysis here points to Globalism as an integrative narrative, which not only facilitates the entry of new adherents into the conspiratorial universe but also reinforces the internal cohesion of these communities by offering a unifying explanation for a myriad of theories.

[ english version / português abaixo ]

**Figure 06.** Flow of invitation links between QAnon communities

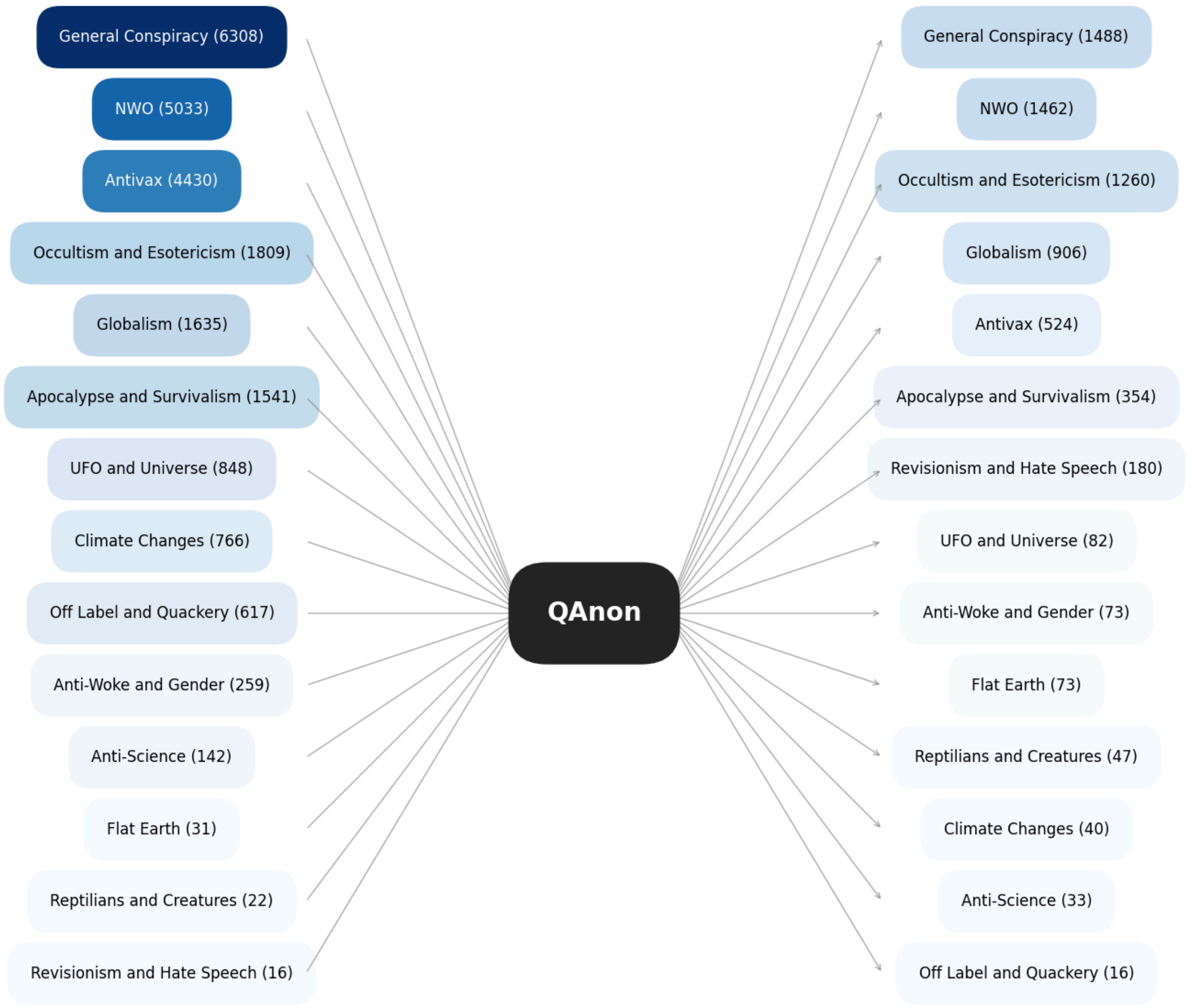

Source: Own elaboration (2024).

QAnon, as shown in the flow chart, is deeply interconnected with a vast network of theories ranging from General Conspiracies (6,308 links) to NWO (5,033 links). This positioning reveals that QAnon is not just one theory among many but a kind of "narrative hub", which connects and synthesizes various other theories into a coherent whole for its followers. QAnon's role in spreading other narratives is evident from the invitations it sends to General Conspiracies (1,488 links) and NWO (1,462 links), indicating that, for QAnon adherents, these theories are not just parallel but complementary. Reflecting on the function of QAnon in the broader context of conspiracy theories, we can see that it acts as a "flexible belief system". Its adherents can integrate new information in a way that always reinforces the original conspiratorial worldview, making it highly resistant to contradictory facts. This suggests that QAnon is more than just a simple theory; it is a metanarrative that offers a model for understanding and organizing other beliefs. Thus, QAnon functions as a fast track to radicalization, absorbing and reconnecting various theories within a comprehensive interpretive framework.

[ english version / português abaixo ]

### 3.2. Time series

Time Series Next, we will explore the time series that highlight the significant growth of conspiracy theories such as the New World Order (NWO) and QAnon, and how these intertwine with narratives about Anti-Woke and Globalism. The following graph illustrates how the COVID-19 pandemic and the 2020 U.S. presidential elections acted as catalysts for the explosion of these mentions. NWO, which already had relevance, expanded its reach, reflecting widespread distrust in global institutions. QAnon followed a similar pattern, solidifying its presence. Over time, we observe a gradual decline, but these theories remain significantly entrenched in public discourse, with NWO maintaining a predominant position until 2024, suggesting that global crises continue to fuel these narratives.

**Figure 07.** Line graph over the period

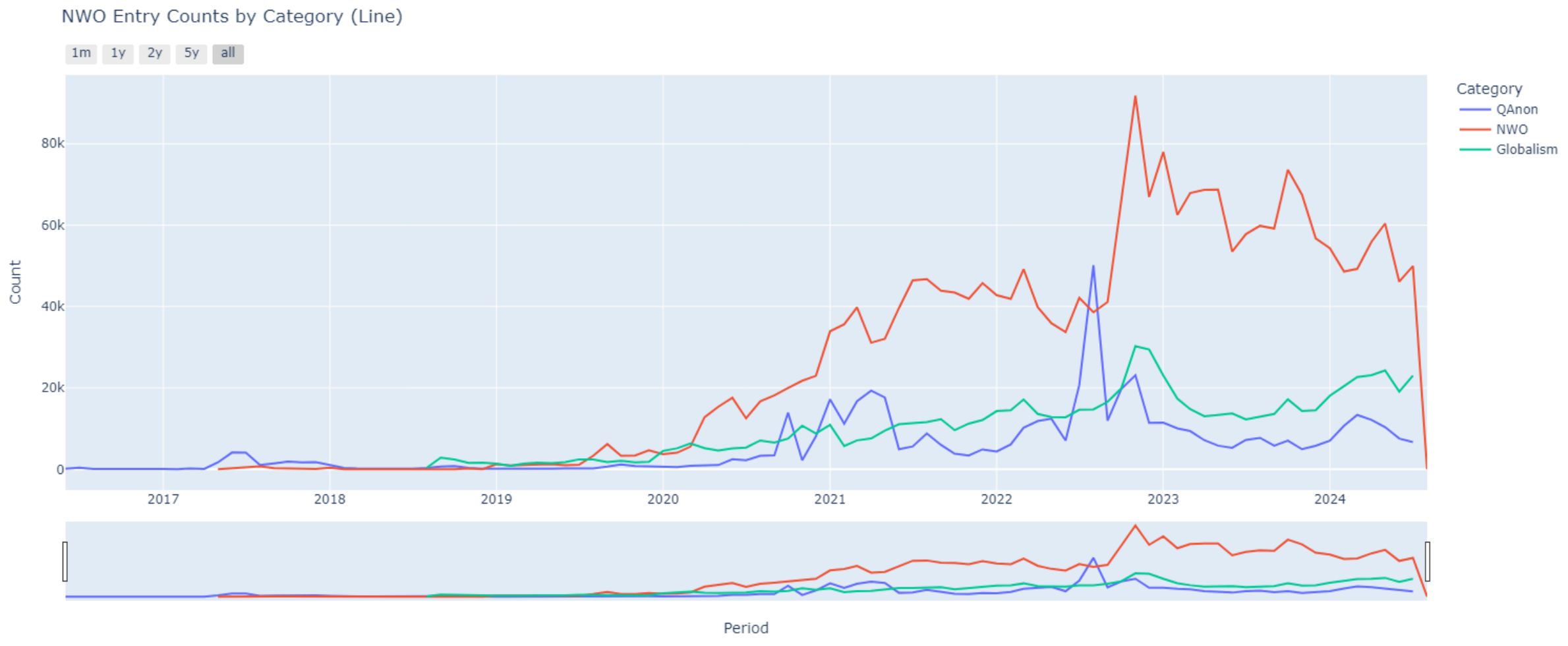

Source: Own elaboration (2024).

In the graph of categories New World Order (NWO), QAnon, Anti-Woke and Gender, and Globalism, the growth of mentions of NWO from 2019 to 2021 is impressive. Mentions rise from about 2,000 in 2019 to over 65,000 in January 2021, an increase of 3,150%. This growth can be attributed to events such as the COVID-19 pandemic and the U.S. elections, which were catalysts for the strengthening of these theories. For QAnon, the increase in mentions is equally significant, rising from about 500 to 15,000, a 2,900% increase. Discussions about Anti-Woke and Gender and Globalism follow a more gradual growth pattern, with increases of 800% and 1,000%, respectively, during the same period. The peaks of NWO and QAnon in January 2021 gradually decline over the following year, with a reduction of about 30% by the end of 2022. Nevertheless, the NWO theme remained predominant within the set even in the most recent dates, through the first half of 2024.

[ english version / português abaixo ]

**Figure 08.** Absolute area chart over the period

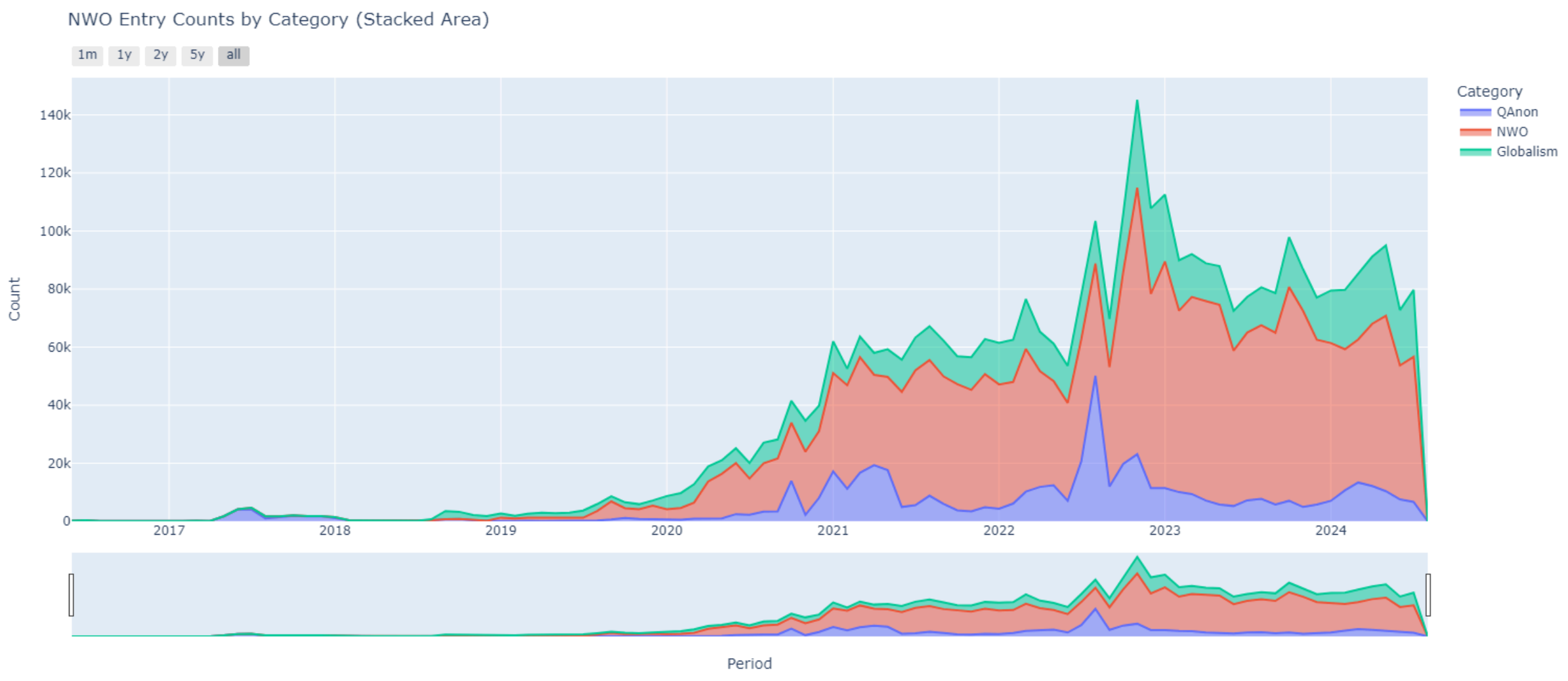

Source: Own elaboration (2024).

The absolute area graph over the period reveals that the New World Order (NWO) theme already had considerable relevance before 2020. This is not surprising, given that NWO is one of the longest-lasting and most widely disseminated conspiracy theories. From 2020 onward, there is a notable increase, with peaks in 2021 and 2022, corresponding to over 140,000 entries. These increases can be associated with the intensification of conspiratorial discourse during the COVID-19 pandemic when theories about global control gained traction. The period also coincides with the rise of the QAnon movement, which explores narratives of a "deep state" and global manipulation, as well as the growth of Globalism as a conspiracy theory. Discussions about Globalism, which peak in 2021, reflect growing concerns about globalization and the perception that international elites are controlling world events. QAnon, in turn, presents less pronounced but still significant peaks, especially after the 2020 U.S. presidential election, which was a catalyst for this conspiratorial narrative. These movements suggest an environment of distrust and radicalization, where global crises are seen as orchestrated by hidden forces.

**Figure 09.** Relative area chart over the period

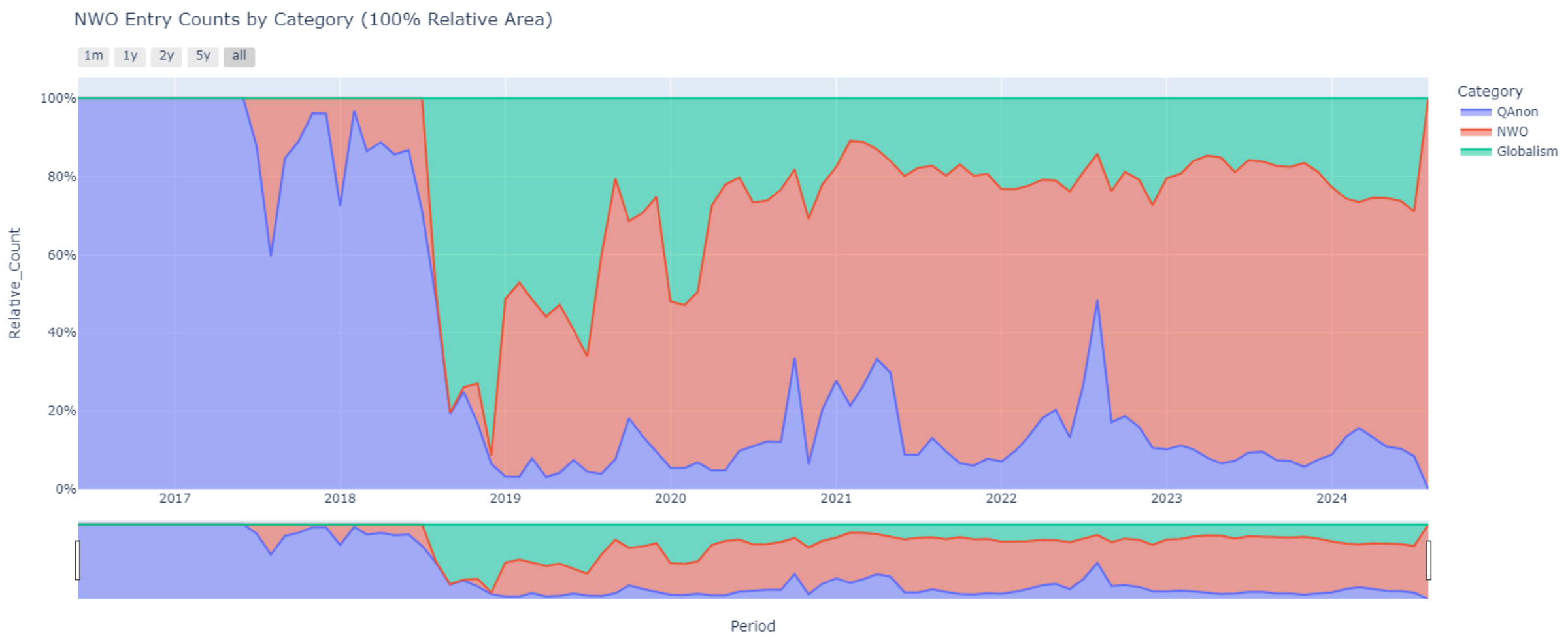

Source: Own elaboration (2024).

When analyzing the relative area graph, it becomes evident that before 2020, mentions of the NWO dominated discussions, representing the majority of entries. However, after 2020, there is a clear shift in thematic distribution, with Globalism gaining significant ground, especially in 2021, when it takes on a dominant position. This shift can be interpreted as a reflection of growing concerns about globalization policies and international interventions, which were exacerbated by the pandemic. QAnon, while less dominant in relative terms, still maintains a constant presence, particularly during periods of elections and political crises in the U.S. The relative analysis shows how the focus of conspiracy theories can shift over time depending on the global context, but also how these narratives remain interconnected, fueling a worldview where global events are interpreted as part of a grand conspiracy.

### 3.3. Content analysis

The content analysis of communities related to the New World Order (NWO), Globalism, and QAnon, using word clouds, offers a deep understanding of how these groups articulate their beliefs and narratives over time. The words that emerge most frequently in the clouds indicate the main focuses of these communities, reflecting not only the topics of interest but also the discursive strategies that sustain the cohesion and expansion of these ideas. Terms like “world”, “Brazil”, “truth”, and “now” appear prominently, suggesting a discourse that blends global perceptions with national concerns. The constant presence of “God” and “government” points to the intersection of religion and politics, often manipulated to justify complex conspiracy theories. Analyzing these clouds over time allows for the identification of changes and continuities in the narratives, offering insights into how these movements adapt to shifting political and social contexts, with their relevance and influence.

[ english version / português abaixo ]

**Figure 10.** Consolidated word cloud for new world order, globalism and QAnon

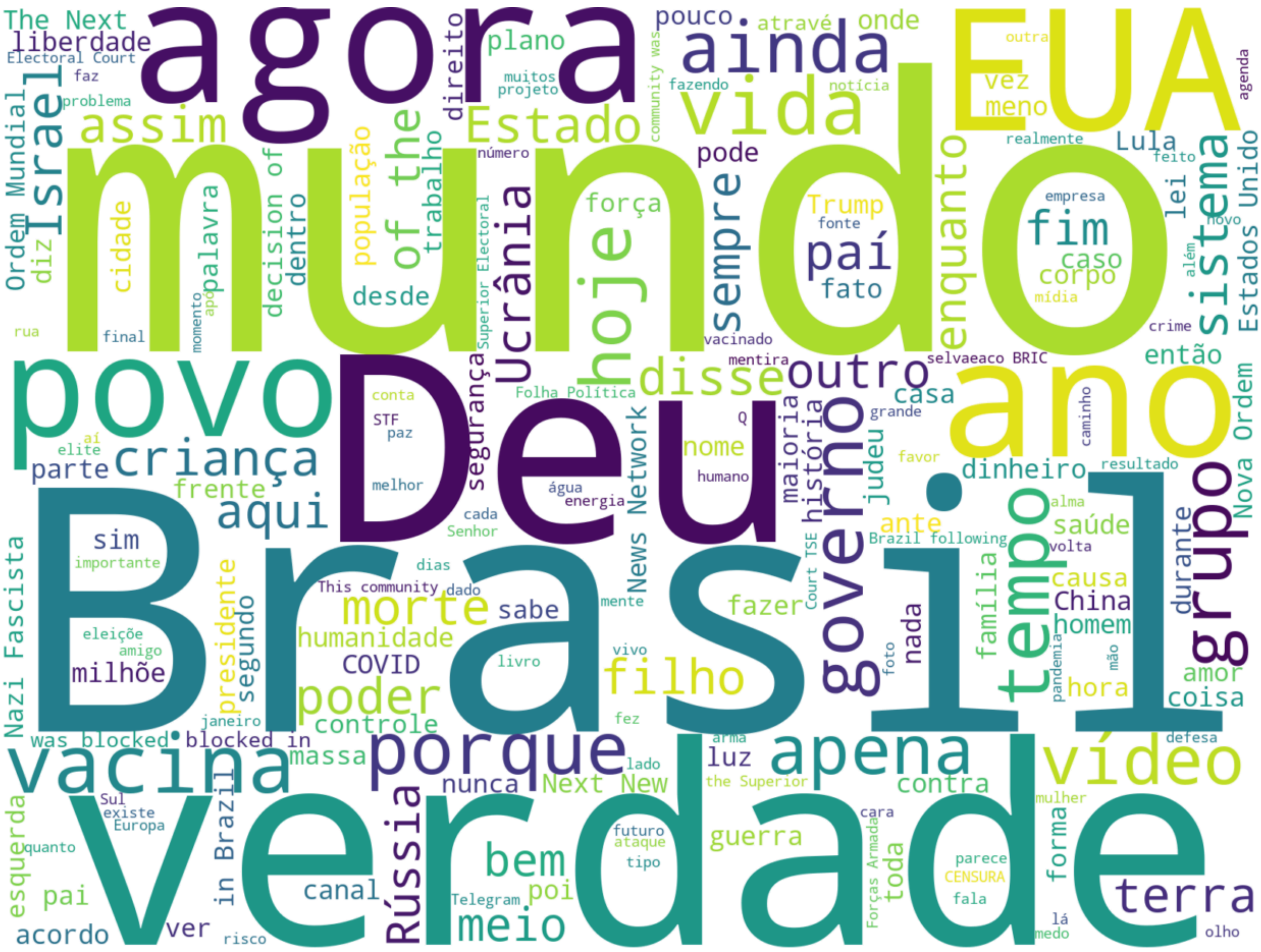

Source: Own elaboration (2024).

The consolidated word cloud that addresses the communities of the New World Order, Globalism, and QAnon reveals a strong interconnection between global themes and national issues. Terms like “world”, “Brazil”, and “truth” stand out, highlighting a narrative that seeks to connect international events to a supposed global agenda that would directly affect Brazil. The word “God” reinforces the tendency of these communities to link their theories to a religious dimension, using faith as a pillar to sustain their conspiratorial beliefs. “Government” and “people” indicate a concern with power dynamics and the idea that the people are being manipulated by hidden elites, represented by entities like the New World Order. The recurrence of “vaccine” and “pandemic” signals the incorporation of the global health crisis into the conspiracy theories, broadening the spectrum of concerns for these communities and linking public health issues with political conspiracies. In summary, the word cloud reveals the complexity and multifaceted nature of discussions within these communities, where issues of sovereignty, religion, and science intertwine to form a discourse of resistance and widespread distrust.

[ english version / português abaixo ]

**Chart 01.** Temporal word cloud series for new world order

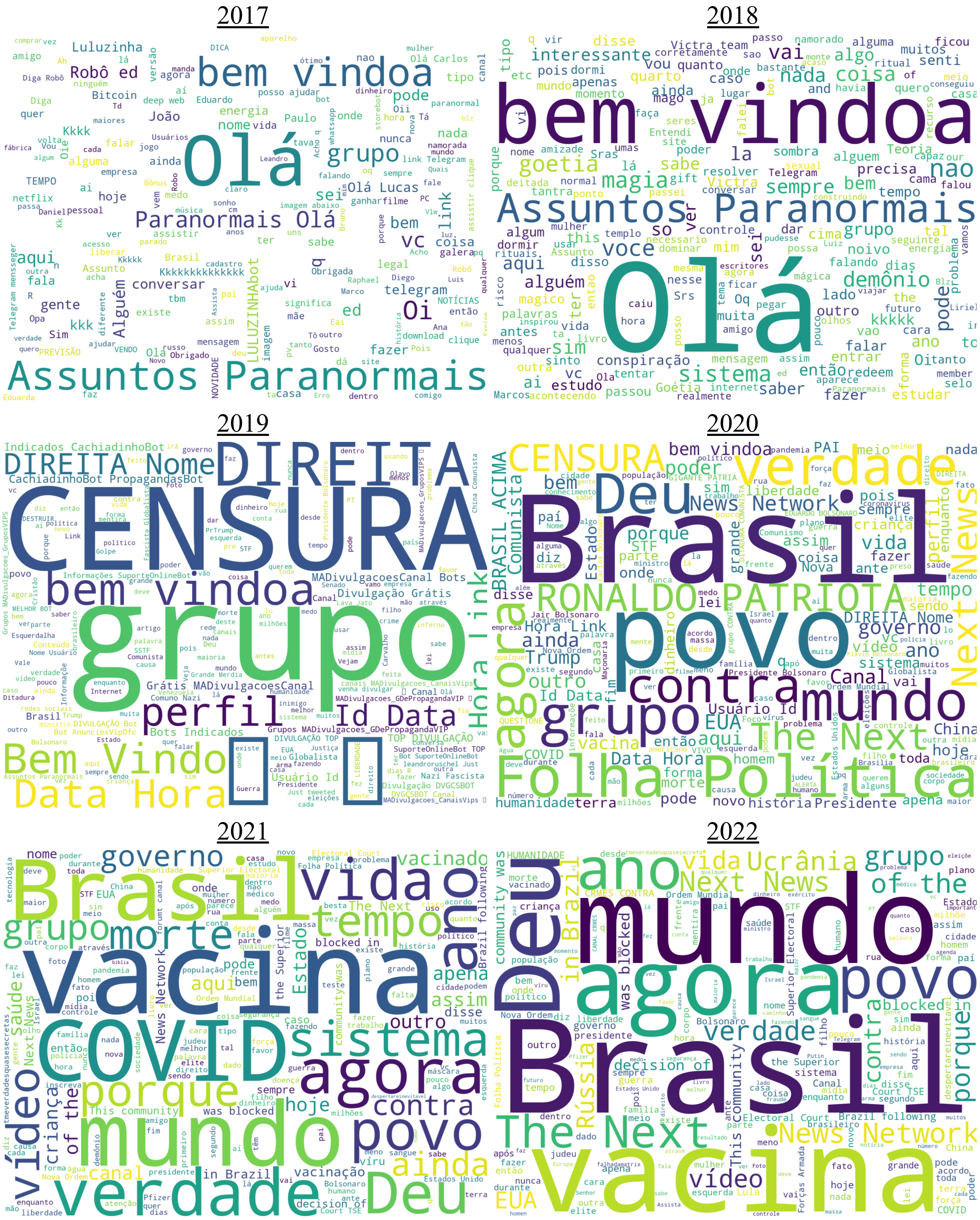

[ english version / português abaixo ]

2023 2024

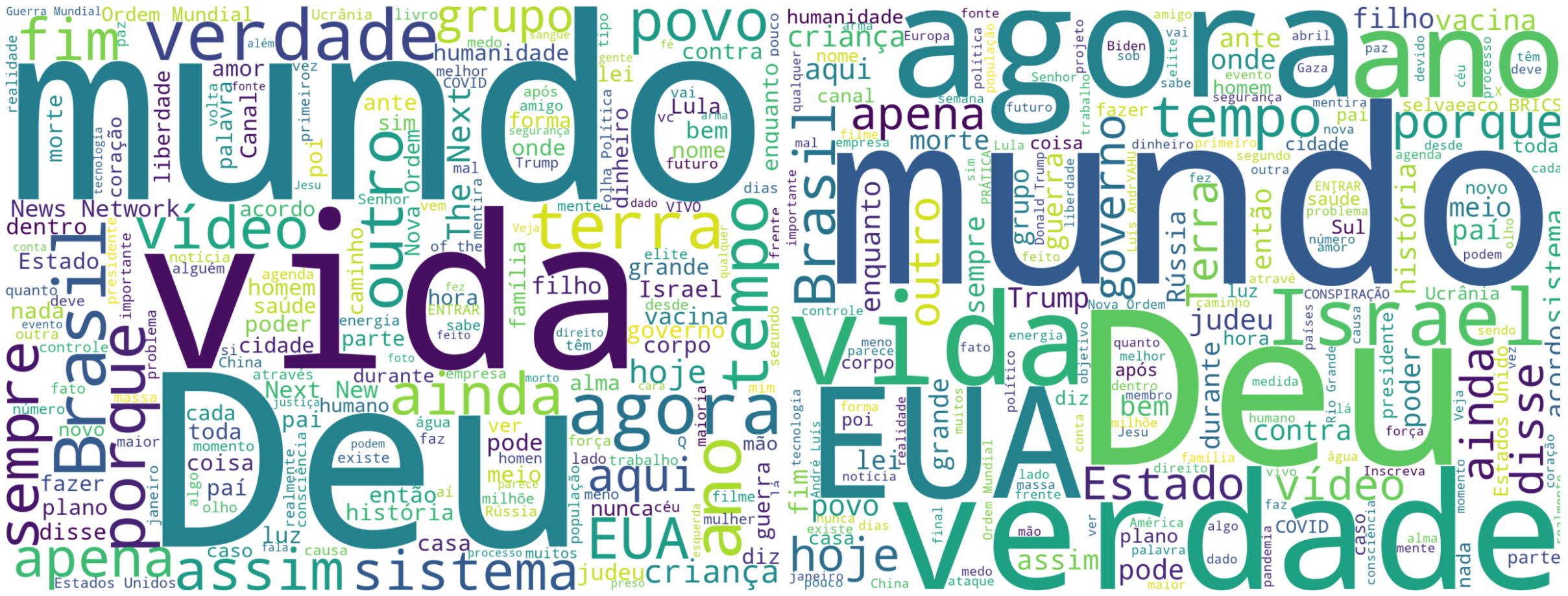

Source: Own elaboration (2024).

The time series word cloud for the New World Order highlights how the narratives and focuses of these communities have evolved over recent years. In 2017, terms like “Hello” and “group” predominate, suggesting an initial phase of formation and organization within these communities, where welcoming new members was essential. In 2018, the word “censorship” emerges, possibly in response to content moderation measures on social media, marking the beginning of a rhetoric of victimization and resistance against digital platforms. From 2020 onward, “Brazil” and “world” gain prominence, reflecting the intensification of discussions about the pandemic and how it is perceived to be part of a larger global domination plan. The term “vaccine”, emerging in 2021, reinforces this narrative, with conspiracy theories linking vaccination campaigns to a supposed population control agenda. In 2023 and 2024, the consolidation of terms like “God” and “truth” indicates a deepening of radicalization, where faith and the search for an alternative “truth” become central to the discourse, strengthening the ideological cohesion of these communities.

[ english version / português abaixo ]

**Chart 02.** Temporal word cloud series for globalism

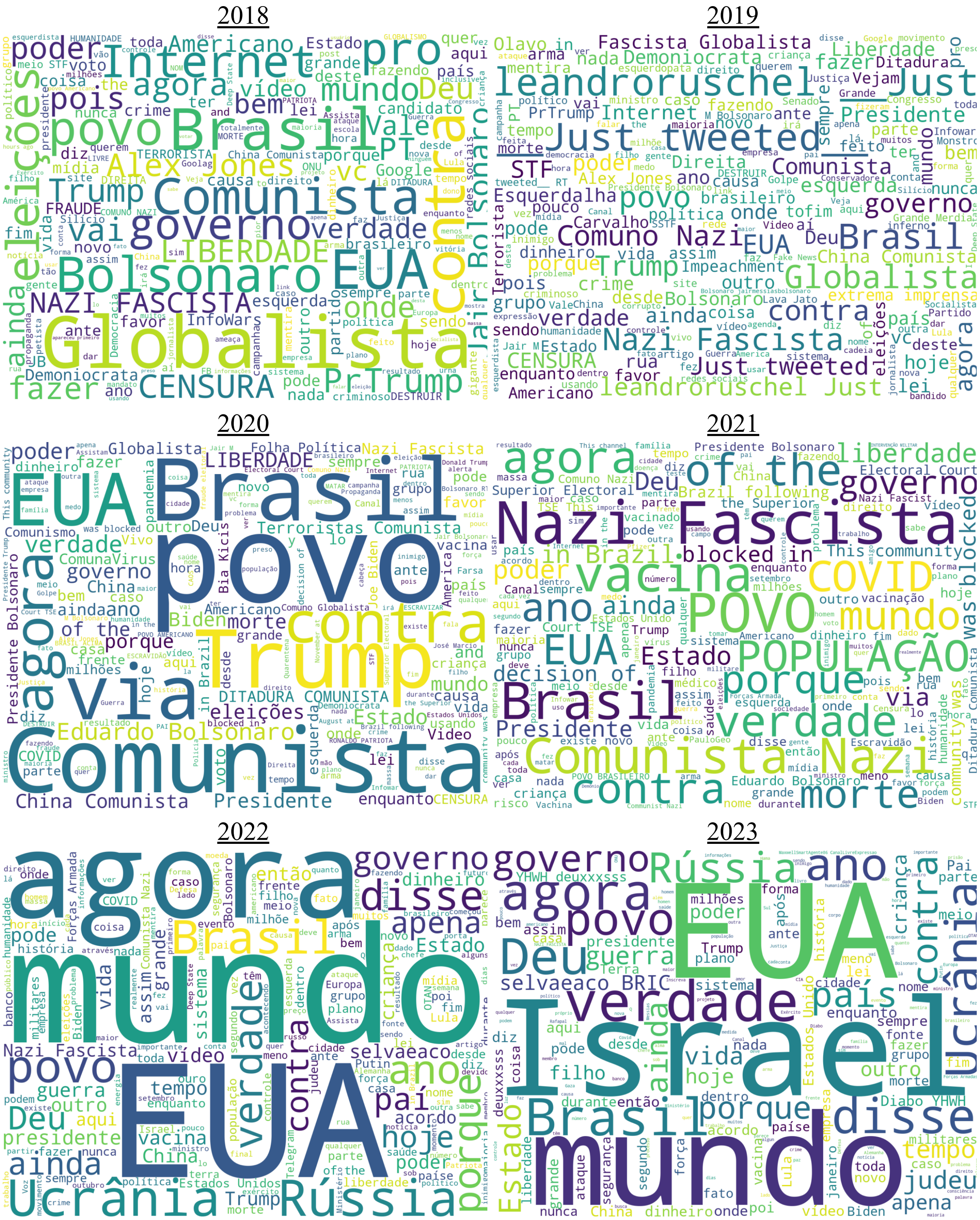

[ english version / português abaixo ]

2024

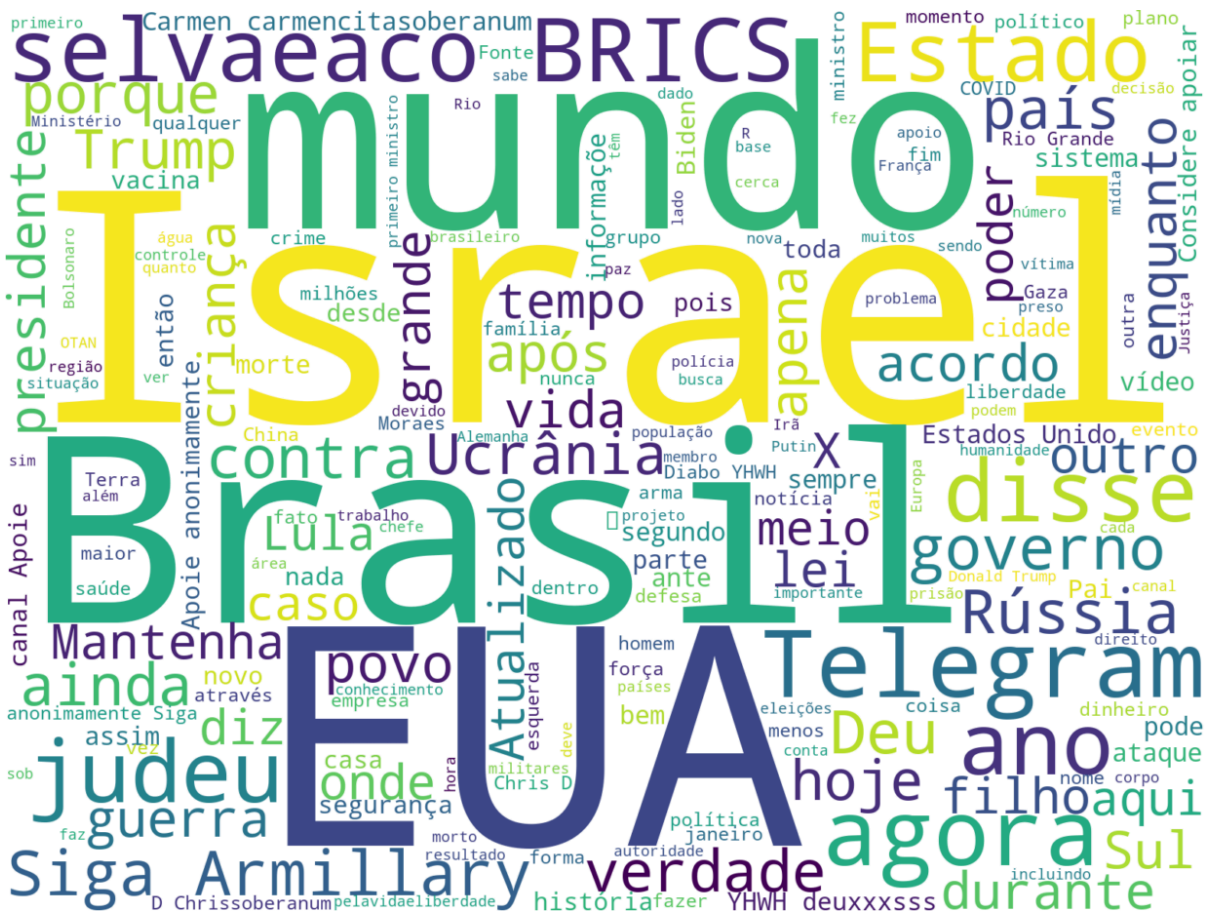

Source: Own elaboration (2024).

The time series word cloud for Globalism reveals a narrative progression that reflects political and social transformations at both global and national levels. In 2019, “USA”, “Brazil”, and “people” are predominant terms, suggesting a concern with American influence and how it is perceived within the conspiratorial narrative of Globalism. The word “communist” begins to gain traction in 2020, aligning with far-right rhetoric that views Globalism as a communist threat. In 2021, “Nazi” and “Fascist” appear more frequently, indicating an intensification of polarized discourse that equates Globalism with totalitarian regimes. From 2022 onward, “government” becomes prominent, signaling a growing concern with public policies and their supposed submission to a globalist agenda. In the following years, “world” and “now” remain central terms, suggesting a sense of urgency and a view that global events are culminating in a new world order that demands immediate resistance.

**Chart 02.** Temporal word cloud series for QAnon

2016

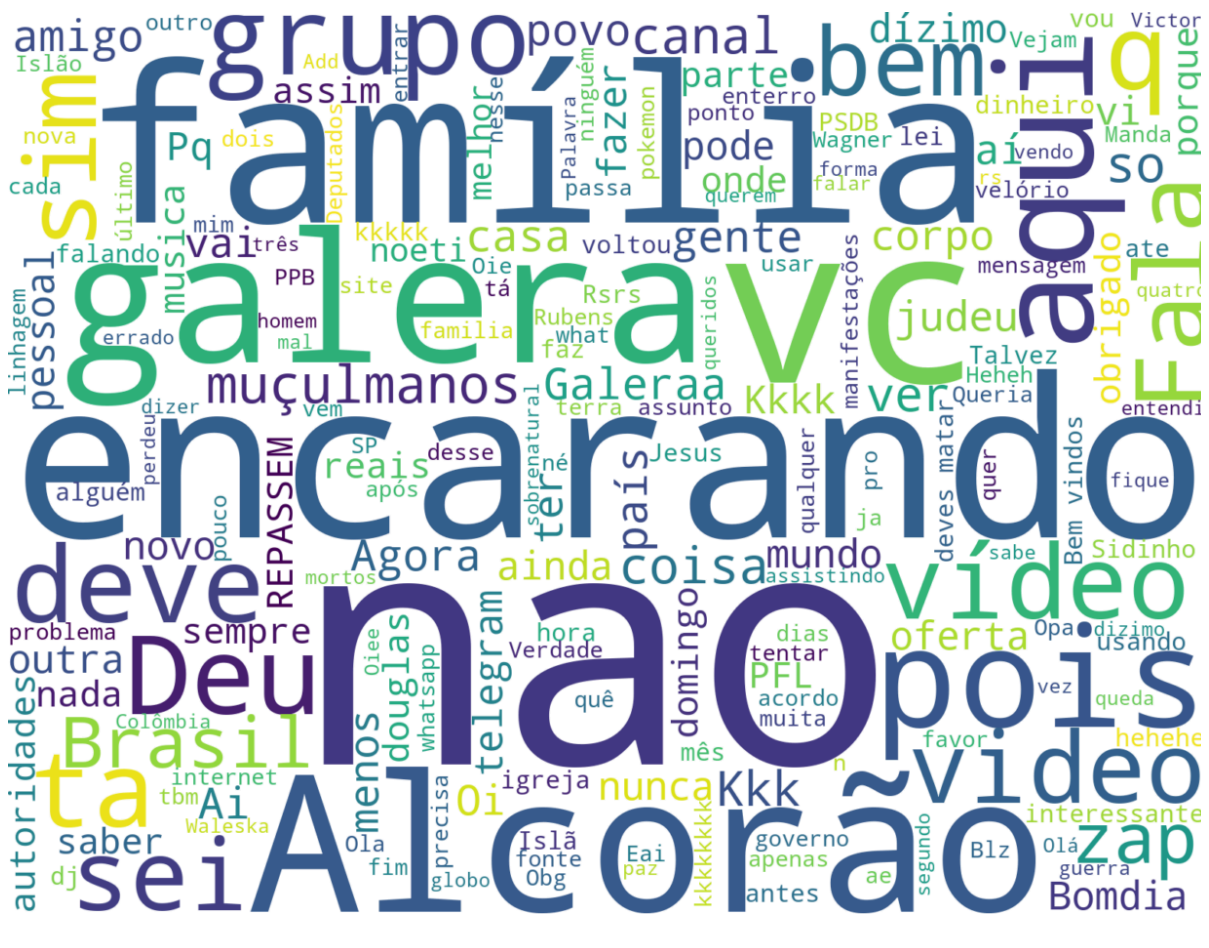

2017

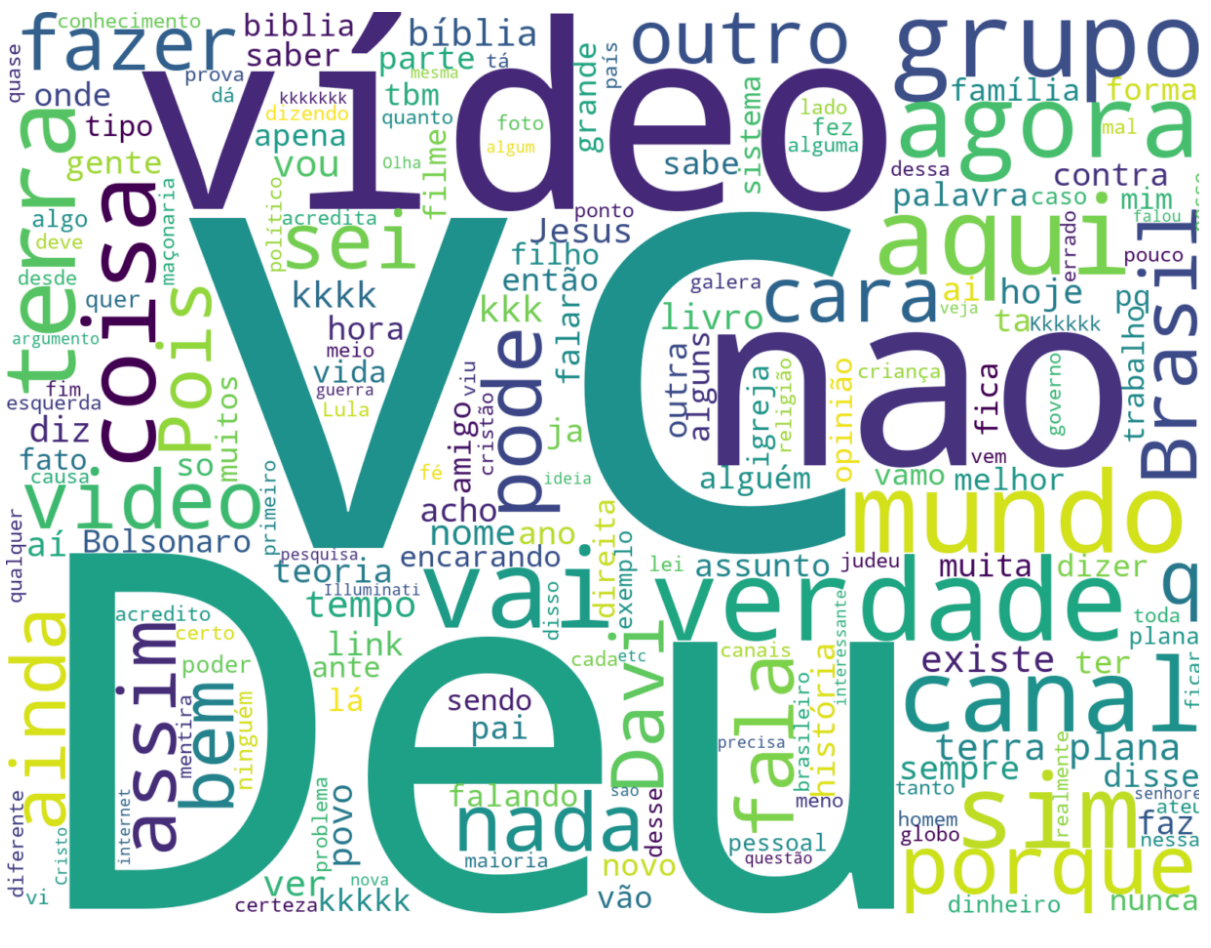

[ english version / português abaixo ]

2018

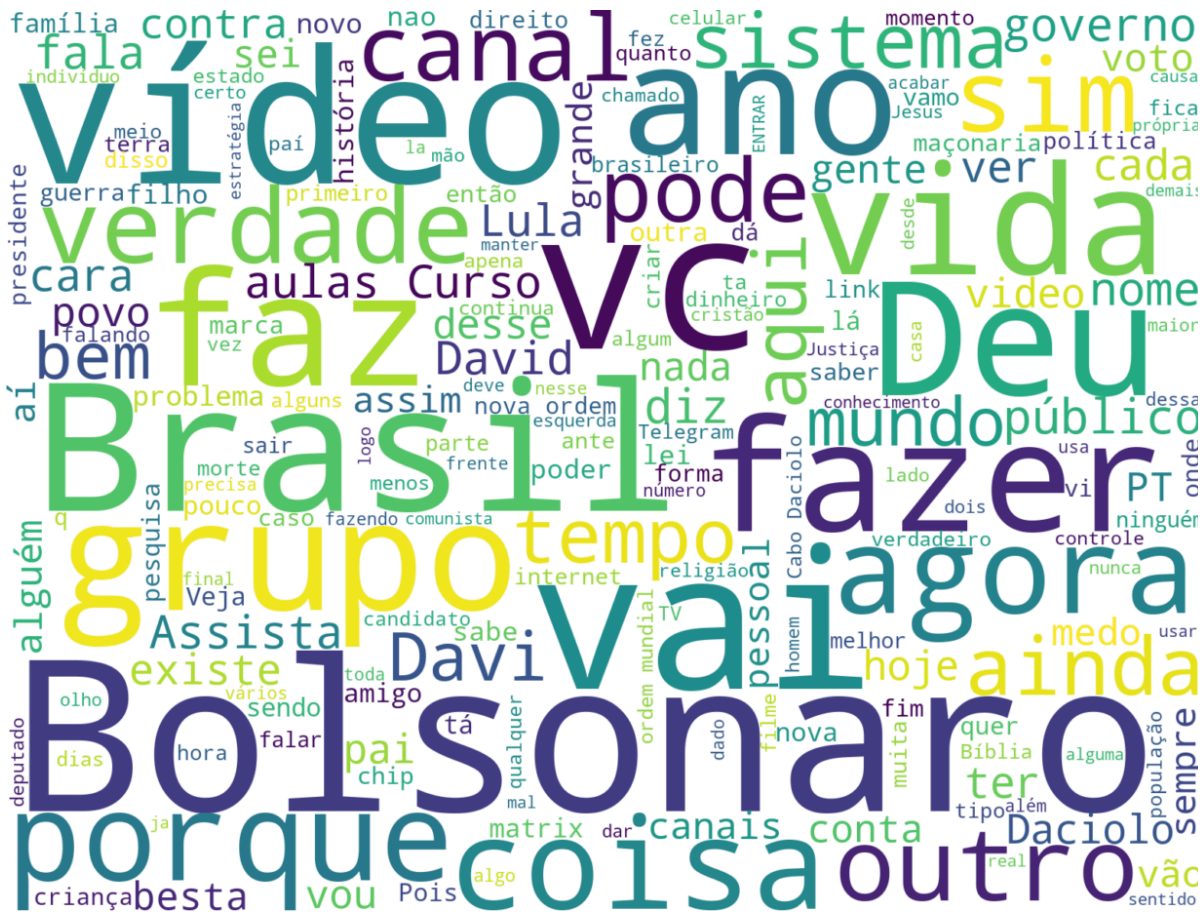

2019

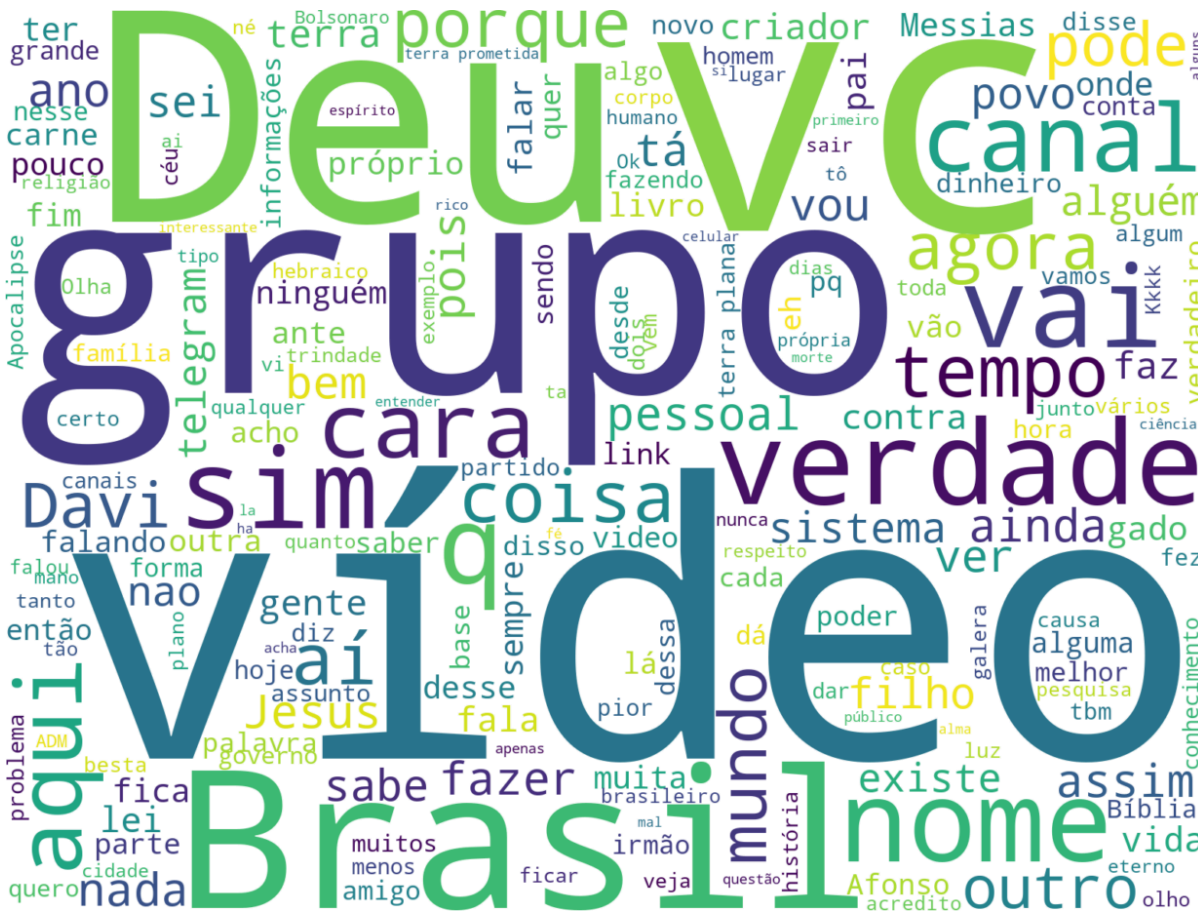

2020

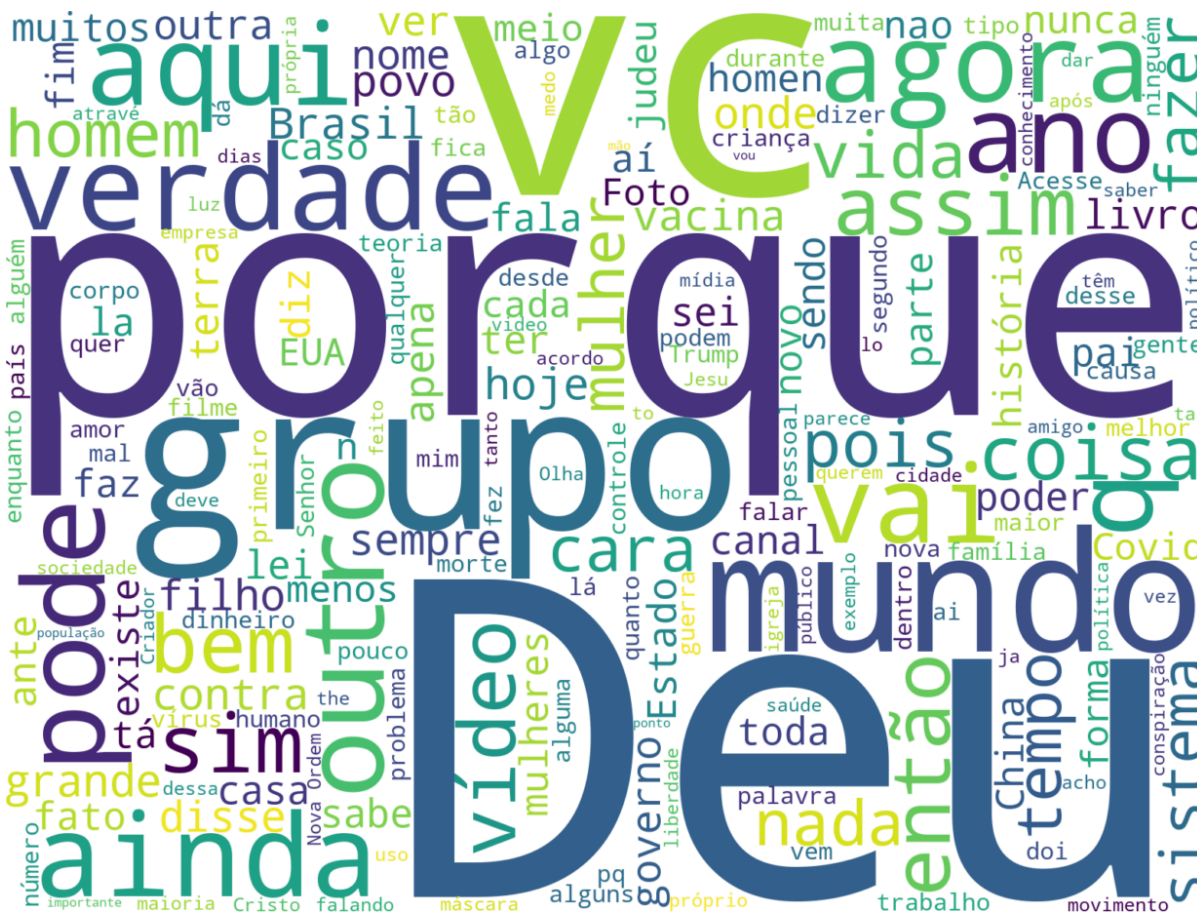

2021

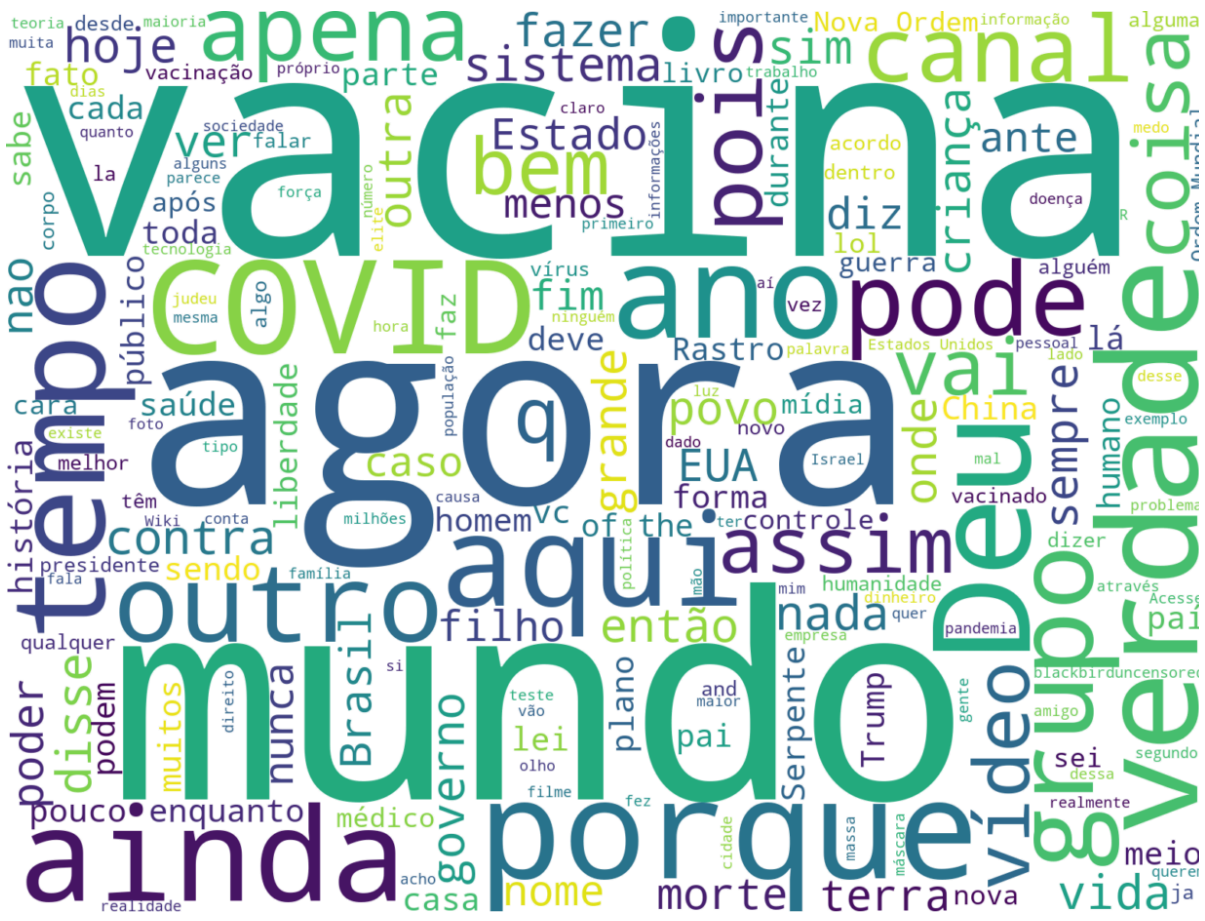

2022

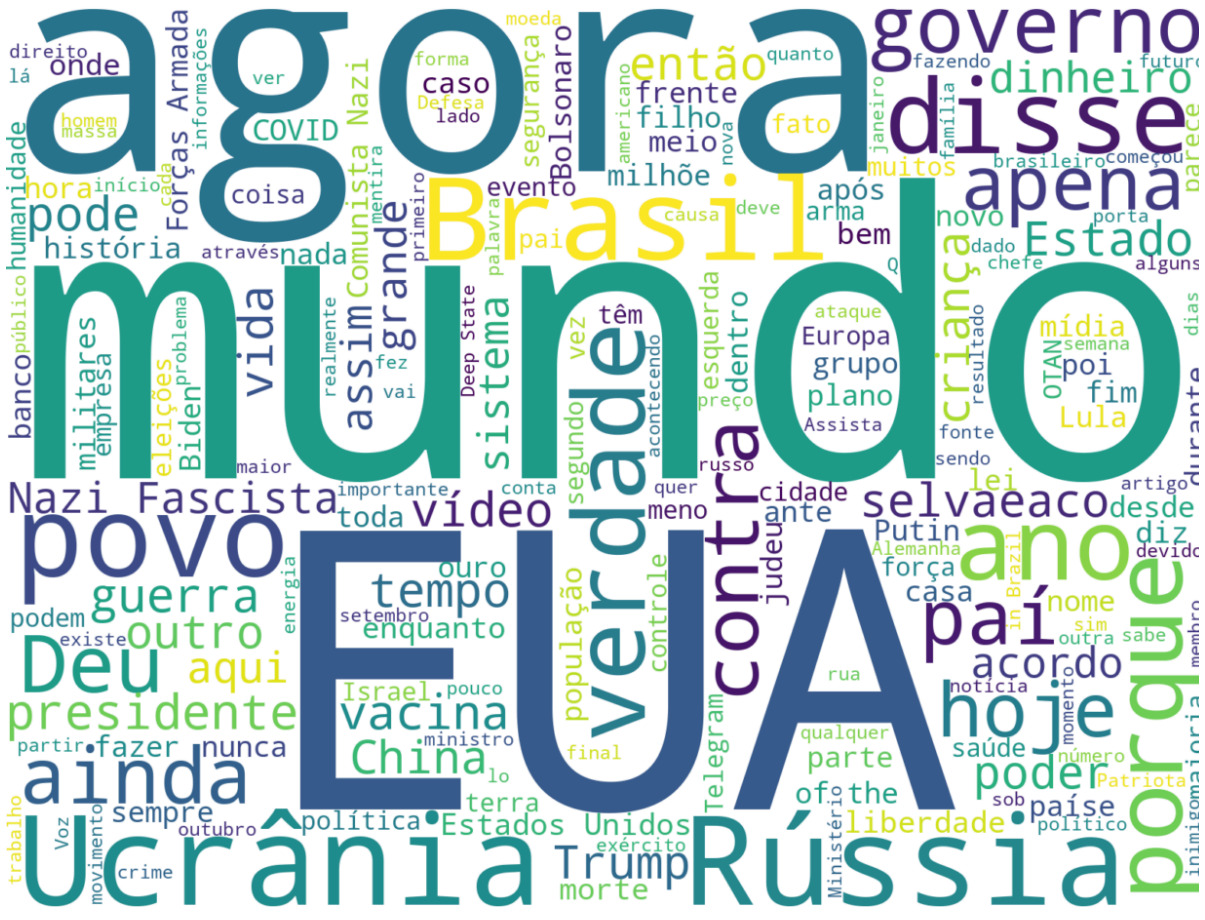

2023

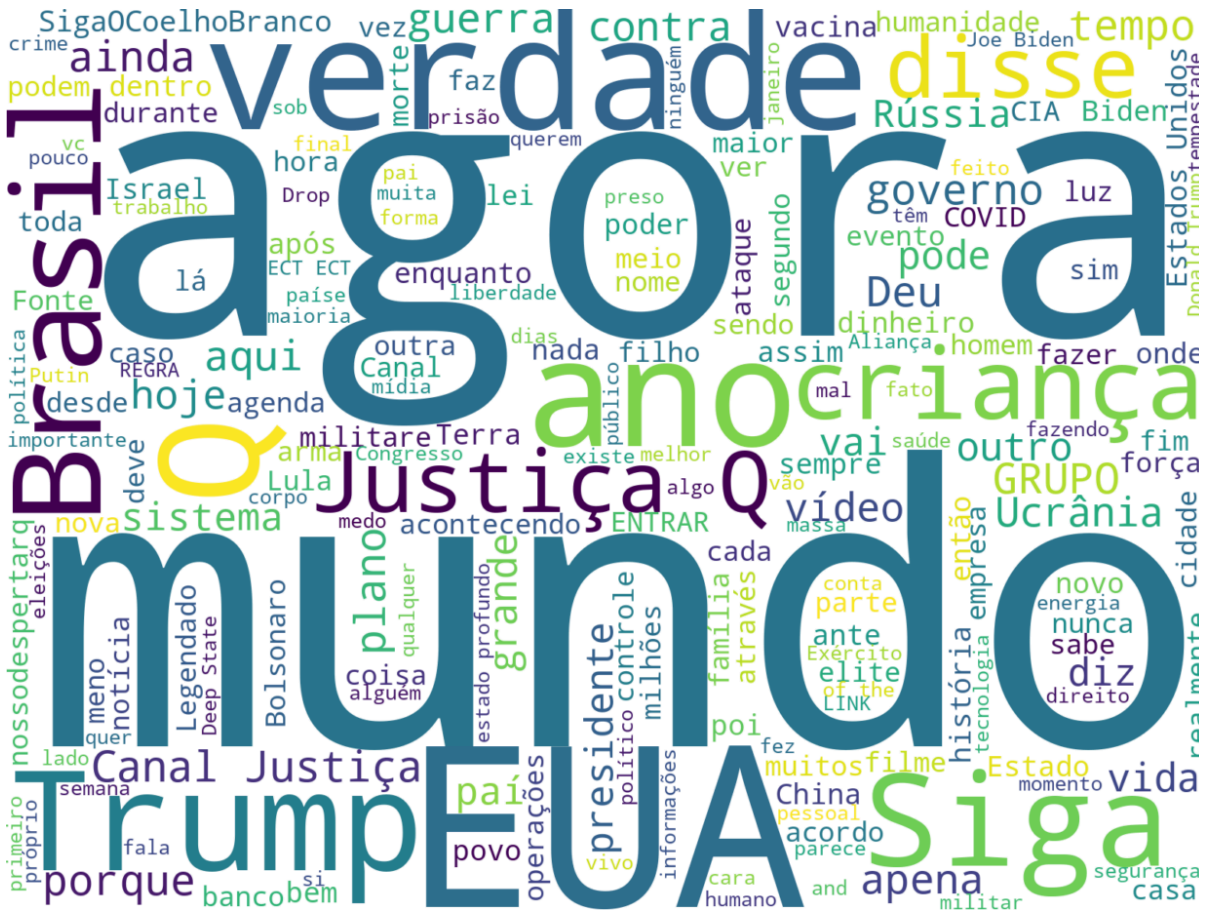

[ english version / português abaixo ]

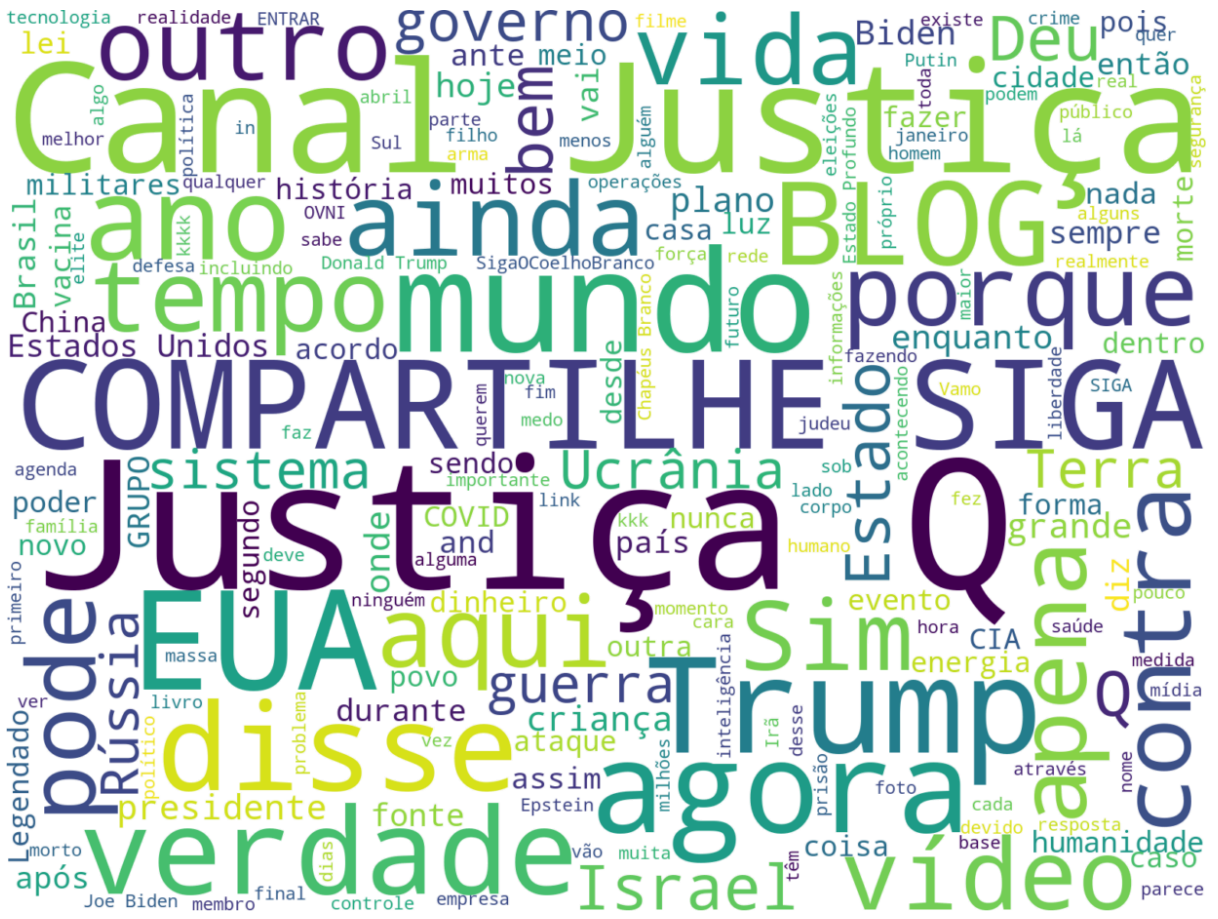

Source: Own elaboration (2024).

The time series word cloud for QAnon highlights the evolution of narratives within this community, marked by a strong religious and conspiratorial tone. In 2016 and 2017, “God”, “you”, and “family” are frequent terms, reflecting a phase in which the community was strongly focused on conservative and Christian values. In 2018, “video” begins to emerge, indicating the growing role of audiovisual content in the dissemination of QAnon ideas, which is reinforced in 2019 with the popularization of more complex theories. In 2020 and 2021, “Brazil”, “world”, and “vaccine” become central as the COVID-19 pandemic is integrated into the conspiracy theories, expanding the reach of QAnon narratives beyond the United States. In recent years, in 2023 and 2024, the combination of terms like “justice”, “truth”, and “share” indicates a phase of active mobilization, where QAnon communities are not only discussing theories but also encouraging concrete actions of dissemination and resistance against what they perceive as a global conspiracy orchestrated by elites.

### 3.4. Thematic agenda overlap

The following figures explore the thematic overlap in conspiracy theory communities, specifically those related to the New World Order (NWO), Globalism, and the QAnon movement. Through the visual analysis of the topics addressed by these communities, it is possible to observe how global, cultural, and religious themes intertwine, reinforcing conspiratorial narratives that aim to question and delegitimize political, social, and scientific structures. The intersection of these themes creates a cohesive network of disinformation that perpetuates core beliefs, making it difficult to counter their influence.

**Figure 11.** Global conflicts themes

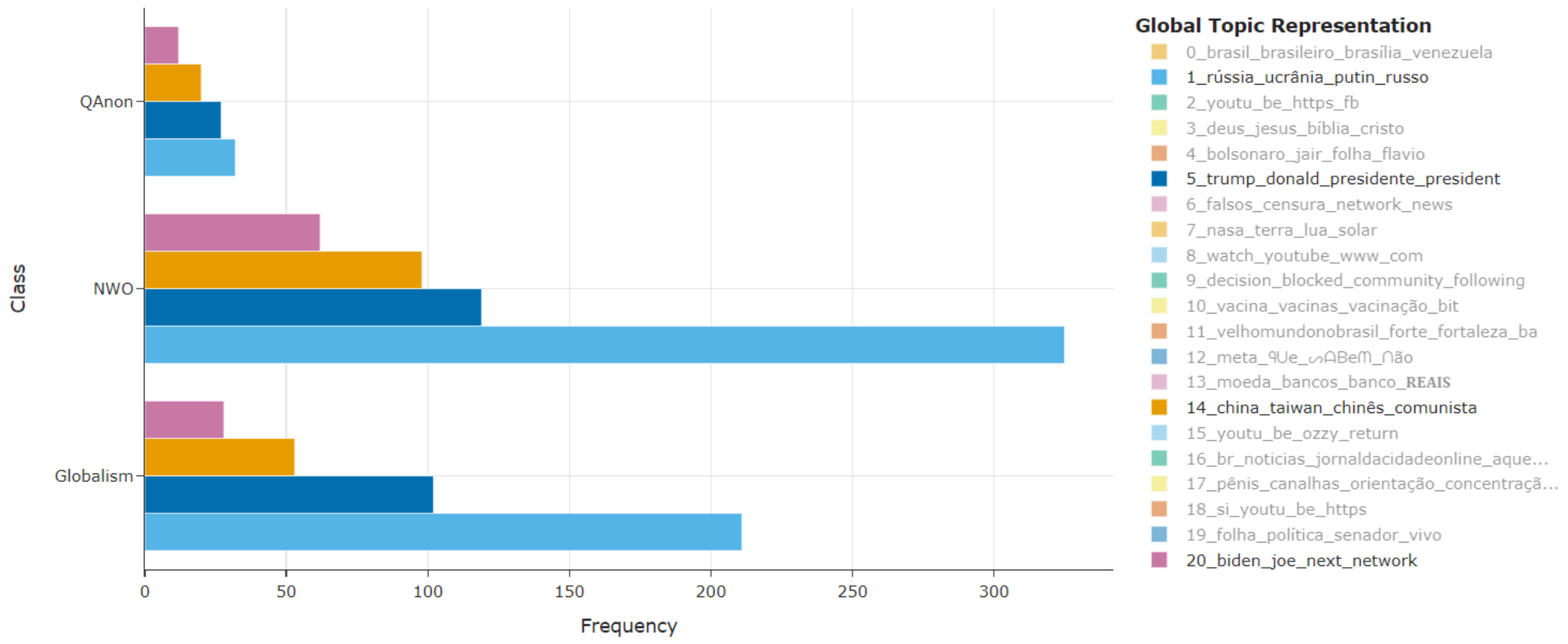

Source: Own elaboration (2024).

Figure 11 analyzes the themes of global conflicts within the QAnon, NWO, and Globalism communities. Topics such as “Russia”, “Ukraine”, and “Biden” are predominant, suggesting that these communities use recent geopolitical events to validate their narratives of a global conspiracy. The prominence of NWO indicates a strong association of these communities with the idea that international conflicts are orchestrated as part of a plan to establish a new world order. The intersection of these topics reinforces the perception that these theories are fueled by the instrumentalization of real events to sustain a discourse of global distrust and manipulation.

**Figure 12.** Faith and religion themes

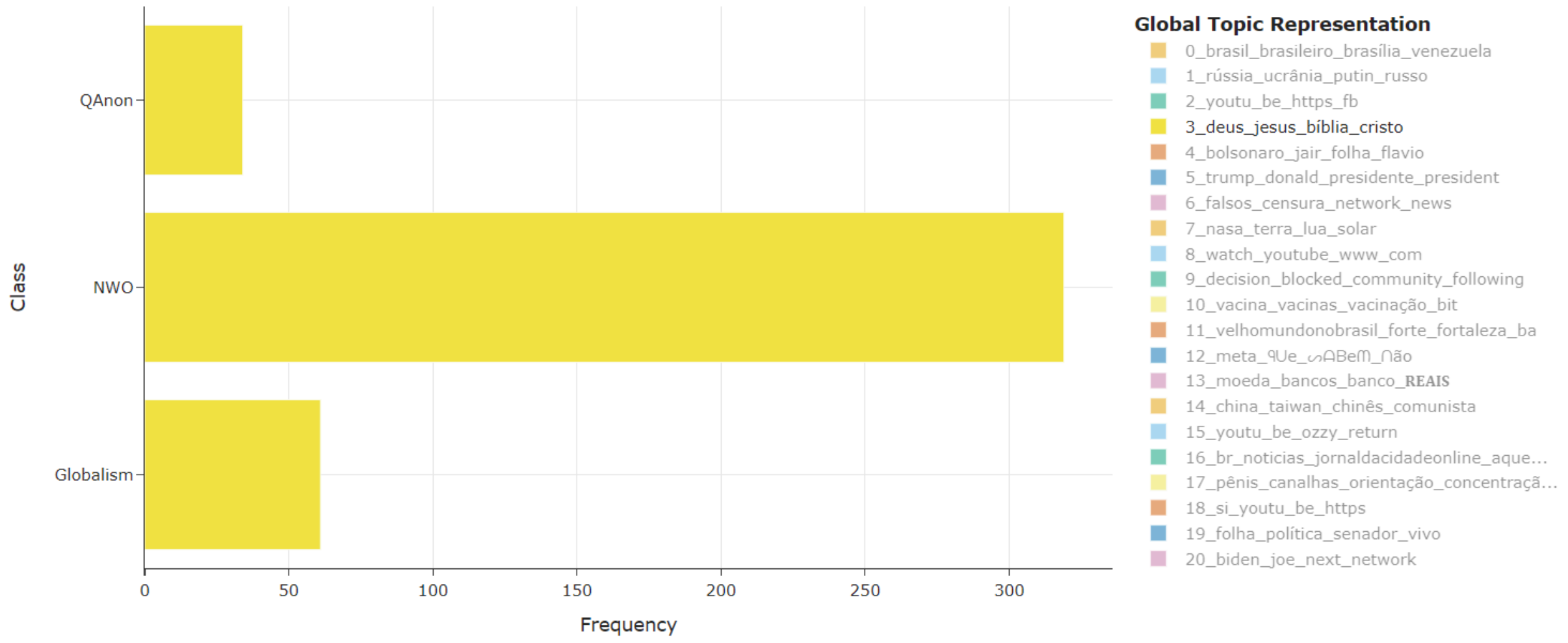

Source: Own elaboration (2024).

In Figure 12, we observe the intersection of faith and religion themes with the narratives of QAnon, NWO, and Globalism. Topics such as “God”, “Jesus”, and “Bible” are widely discussed, especially within NWO narratives, suggesting an attempt to link the fight against the supposed global conspiracy to a spiritual battle. The prevalence of these themes shows how religion is instrumentalized to strengthen belief in conspiracy theories, framing the fight against NWO as a sacred duty and justifying resistance to political and social structures from the perspective of a divine mission.

**Figure 13.** Anti-LGBT discourse and globalist agenda themes

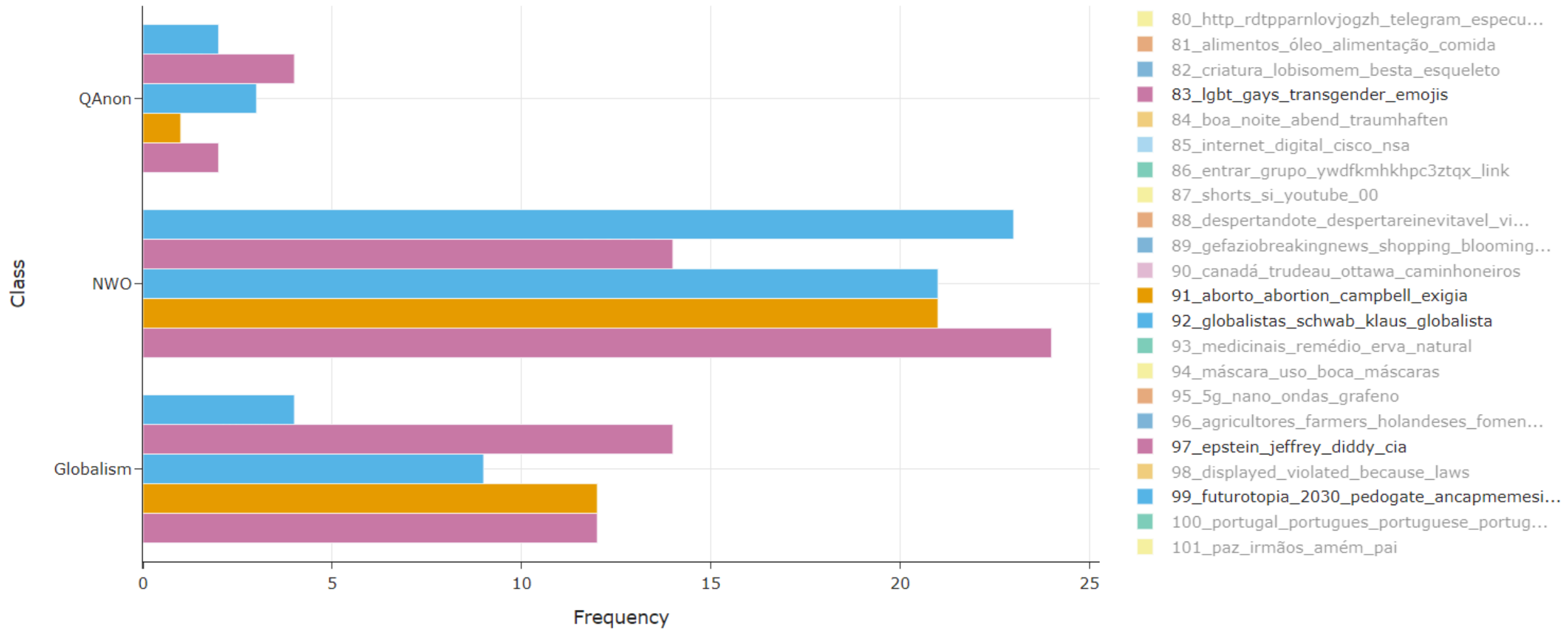

Source: Own elaboration (2024).

Figure 13 highlights the association between anti-LGBT discourse and the globalist agenda in the QAnon, NWO, and Globalism communities. Topics such as “LGBT”, “abortion”, and “globalists” indicate that these communities frequently integrate narratives that attack civil rights and minorities, linking these issues to a globalist agenda aimed at destroying traditional values. The overlap of themes suggests that anti-LGBT rhetoric is used to mobilize against what these communities perceive as a global threat, reinforcing the idea that the protection of moral values is intrinsically linked to resistance against a conspiracy.

**Figure 14.** Neo-Nazism, racism, and anti-Semitism themes

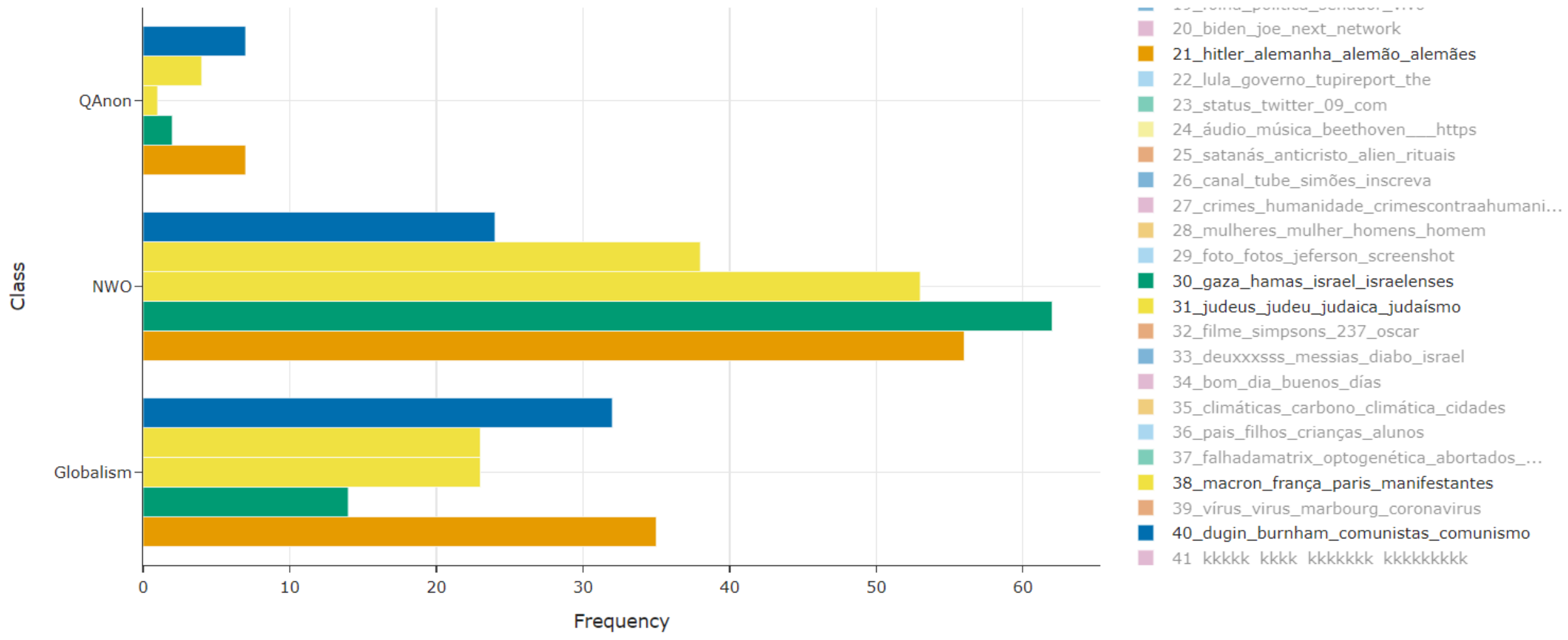

Source: Own elaboration (2024).

In Figure 14, themes of neo-Nazism, racism, and anti-Semitism are addressed within the narratives of QAnon, NWO, and Globalism. Topics such as “Hitler”, “Jews”, and “racism” are evident, especially in discussions associated with NWO, suggesting that these communities use these extremist ideologies to support their conspiracy theories. The significant presence of these themes indicates that racism and anti-Semitism are frequently used as tools to reinforce the narrative of a global conspiracy, associating minority groups with a malevolent plan to control the world.

**Figure 15.** Climate change denial themes

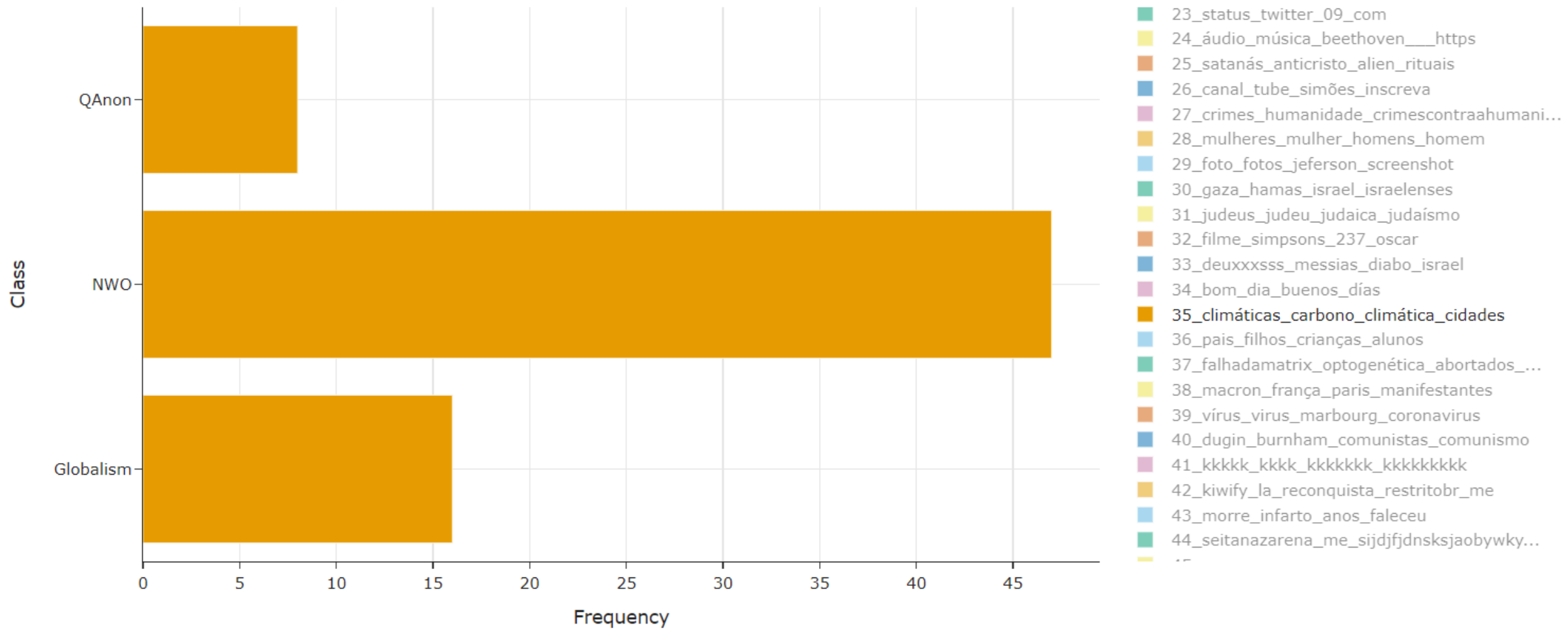

Source: Own elaboration (2024).

Figure 15 explores the relationship between climate change denial and the theories of QAnon, NWO, and Globalism. Topics such as “climate”, “carbon”, and “cities” are prominent, especially in NWO narratives, where climate change is seen as an invention to justify global control. This thematic overlap reflects a resistance to conventional science, where climate change denial is used to reinforce distrust in environmental policies, presenting them as part of a globalist agenda to subjugate populations and eliminate individual freedoms.

**Figure 16.** Vaccine denial and supposed sudden deaths from vaccines themes

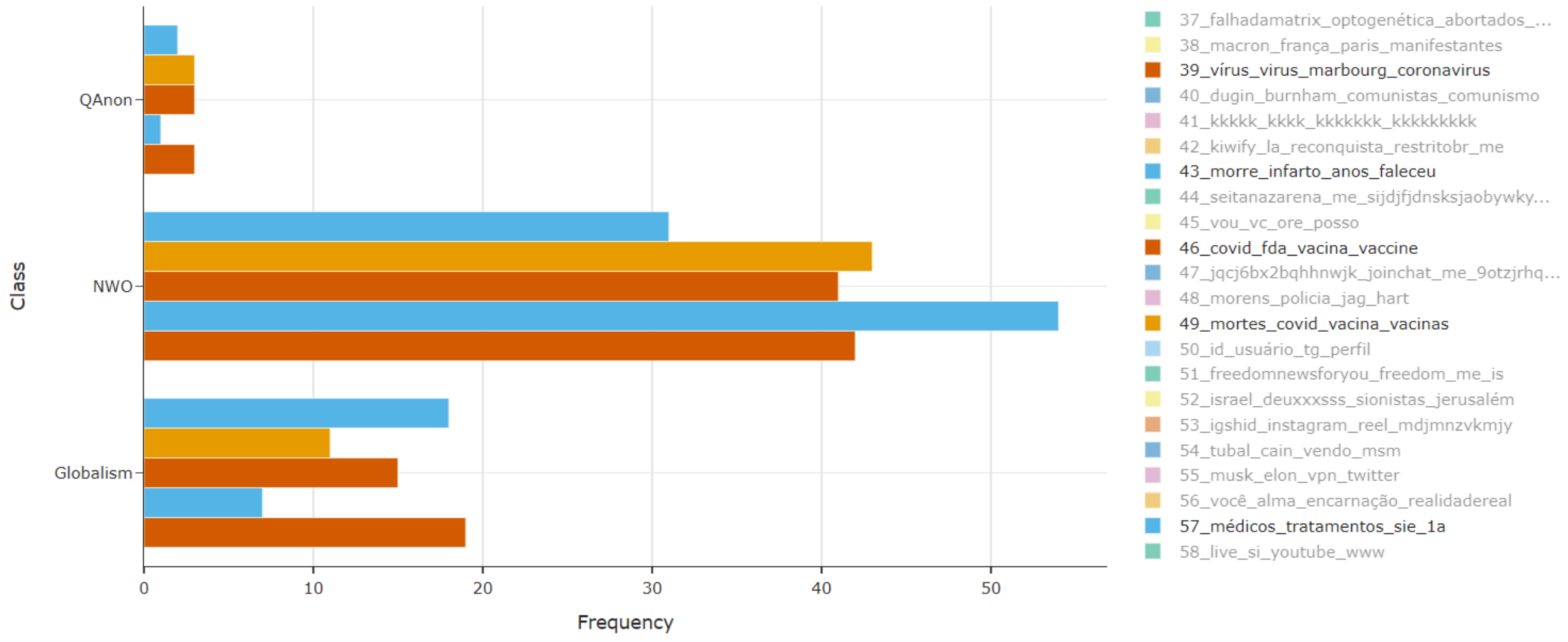

Source: Own elaboration (2024).

In Figure 16, themes of vaccine denial are discussed, with a focus on narratives of supposed sudden deaths caused by vaccines within the QAnon, NWO, and Globalism communities. Topics such as “COVID”, “vaccine”, and “deaths” are widely present, suggesting that these communities use these narratives to reinforce distrust in vaccines and the global healthcare system. The prevalence of this discourse, especially associated with NWO, reflects an attempt to discredit vaccination campaigns, presenting them as part of a plan to reduce the global population and supposedly exert control over humanity.

## 4. Reflections and future works

To answer the research question, **“how are Brazilian conspiracy theory communities on new world order, globalism and QAnon topics characterized and articulated on Telegram?”**, this study adopted mirrored techniques in a series of seven publications aimed at characterizing and describing the phenomenon of conspiracy theories on Telegram, focusing on Brazil as a case study. After months of investigation, it was possible to extract a total of 217 Brazilian conspiracy theory communities on Telegram on new world order, globalism and QAnon topics, summing up 5,545,369 content pieces published between June 2016 (initial publications) and August 2024 (date of this study), with 718,246 users aggregated from within these communities.

Four main approaches were adopted: **(i)** Network, which involved the creation of an algorithm to map connections between communities through invitations circulated among groups and channels; **(ii)** Time series, which used libraries like “Pandas” (McKinney, 2010) and “Plotly” (Plotly Technologies Inc., 2015) to analyze the evolution of publications and engagements over time; **(iii)** Content analysis, where textual analysis techniques were applied to identify patterns and word frequencies in the communities over the semesters; and **(iv)** Thematic agenda overlap, which utilized the BERTopic model (Grootendorst, 2020) to group and interpret large volumes of text, generating coherent topics from the analyzed publications. The main reflections are detailed below, followed by suggestions for future works.

[ english version / português abaixo ]

### 4.1. Main reflections

**The New World Order and Globalism as central catalysts of conspiracy theories on Brazilian Telegram:** The study reveals that the New World Order (NWO) and Globalism communities play central roles in the dissemination of conspiracy theories on Brazilian Telegram. With 2,411,003 posts and 329,304 active users, discussions about NWO are widely spread, creating an environment of global distrust. Globalism also stands out, with 768,176 posts and 326,596 users, showing that these communities not only perpetuate conspiratorial narratives but also function as hubs connecting various other theories;

**The interconnection between NWO, Globalism, and QAnon strengthens a dense network of disinformation:** The analysis of the network of communities shows that NWO, Globalism, and QAnon are interconnected in a way that creates a cohesive conspiratorial web. QAnon, although smaller in volume (531,678 posts), acts as a "narrative hub" that connects theories, while NWO and Globalism amplify these connections, reinforcing the view of a world controlled by a global elite. The centrality of these themes indicates that, upon entering one of these communities, users are quickly exposed to a broad network of disinformation;

**Significant growth in mentions of NWO and Globalism during global crises:** During critical events such as the COVID-19 pandemic and the 2020 U.S. presidential elections, mentions of NWO and Globalism grew exponentially. NWO, for example, registered a 3,150% increase in mentions between 2019 and 2021. This growth is attributed to the amplification of conspiratorial narratives in response to global crises, reflecting growing distrust in international and national institutions;

**Gateways: NWO and Globalism attract new members through other conspiracy theories:** The NWO and Globalism communities not only centralize narratives but also serve as gateways for new members already immersed in other conspiracy theories. NWO received 8,699 invitation links from General Conspiracies and 6,260 from anti-vaccine communities, showing that these communities function as epicenters that attract and retain followers from other narratives, strengthening the cohesion and reach of these theories;

**Globalism as an organizing principle of various conspiracy theories:** The study highlights that Globalism functions as an organizing principle that connects different conspiracy theories, such as Anti-Woke, Revisionism, and even Climate Change. With 32,640 links received from General Conspiracies, Globalism positions itself as an integrative narrative, suggesting that all theories are part of a larger plan, increasing the interconnectivity and resilience of conspiratorial communities;

**QAnon as a flexible metanarrative that absorbs other theories:** The QAnon narrative, despite being smaller in volume compared to NWO and Globalism, plays a crucial role by functioning as a metanarrative that absorbs and connects other conspiracy theories. With 6,308 links from General Conspiracies and 5,033 from NWO, QAnon positions itself as a convergence point for various beliefs, reinforcing its followers' conspiratorial worldview and making it difficult for contradictory information to penetrate;

**NWO as a central interpretive paradigm in the conspiratorial universe:** NWO stands out as a central paradigm that unifies various conspiracy narratives against a supposed global elite. With an impressive volume of 3,488,686 posts, NWO operates as an aggregator of disinformation, where theories such as anti-vaccine, apocalypse, and occultism find coherence and legitimacy, reinforcing a worldview where all global events are manipulated by a secret elite;

**The instrumentalization of religious themes in conspiratorial narratives:** Religious terms such as “God” and “Bible” are recurrent in discussions about NWO and QAnon, indicating a strong intersection between religion and conspiracies. These narratives use faith to justify theories about global control and mass manipulation, suggesting that the fight against NWO and Globalism is a spiritual battle, which strengthens the ideological cohesion of communities and resistance to scientific information;

**Rejection of climate change as part of a globalist agenda:** Theories about climate change are widely discussed in NWO and Globalism communities, where they are seen as part of a global conspiracy to justify control over the population. With topics such as “carbon” and “climate” frequently mentioned, these communities reject conventional science and present environmental policies as a threat to individual freedoms, contributing to polarization and disinformation;

**The perpetuation of vaccine disinformation and its link to NWO:** NWO communities are particularly active in spreading vaccine disinformation, using narratives about supposed sudden deaths caused by vaccines to reinforce distrust. These theories are often associated with the idea of a population control plan, suggesting that vaccination campaigns are part of a manipulation agenda, which perpetuates vaccine resistance and amplifies the impact of disinformation.

### 4.2. Future works

Based on the key findings of this study, several directions can be suggested for future research. The first would be to explore more deeply the connections between NWO, Globalism, and QAnon, especially in how these communities use global crises to validate their narratives. Investigating how these events are reinterpreted and amplified by these communities could provide valuable insights for developing strategies to combat disinformation during periods of crisis.

Additionally, the instrumentalization of religion within NWO communities deserves further investigation. Understanding how faith is manipulated to justify and perpetuate conspiracy theories could help identify more effective approaches to deconstruct these narratives. Research that explores the intersection between religion and disinformation could provide the foundation for factual correction campaigns that respect religious beliefs while offering an alternative to the conspiratorial narrative. Another relevant point is the analysis of the connections between anti-LGBT theories and the globalist agenda within these communities. Future studies could focus on mapping how these narratives mutually reinforce each other and how they are used to mobilize support against civil rights. Understanding these dynamics is crucial for developing interventions aimed at protecting human rights while dismantling the associated conspiratorial narratives.

The persistence of conspiracy theories in public discourse also highlights the need to investigate the mechanisms that allow these beliefs to be reintroduced and gain traction again. Studies could focus on how these narratives are maintained and adapted over time, especially in the face of external interventions. Understanding these cycles of resilience could be key to developing long-term strategies to combat disinformation.

Finally, the creation of an interconnected network of disinformation between NWO, Globalism, and QAnon suggests that there is a feedback loop of beliefs that needs to be disrupted. Future studies could investigate how these networks form, how they remain cohesive, and what the weak points are that could be exploited to destabilize these narratives. Developing methods to identify and neutralize these disinformation hubs before they gain traction is essential for reducing the impact of these theories.

## 6. Author biography

**Ergon Cugler de Moraes Silva** has a Master's degree in Public Administration and Government (FGV), Postgraduate MBA in Data Science & Analytics (USP) and Bachelor's degree in Public Policy Management (USP). He is associated with the Bureaucracy Studies Center (NEB FGV), collaborates with the Interdisciplinary Observatory of Public Policies (OIPP USP), with the Study Group on Technology and Innovations in Public Management (GETIP USP) with the Monitor of Political Debate in the Digital Environment (Monitor USP) and with the Working Group on Strategy, Data and Sovereignty of the Study and Research Group on International Security of the Institute of International Relations of the University of Brasília (GEPSI UnB). He is also a researcher at the Brazilian Institute of Information in Science and Technology (IBICT), where he works for the Federal Government on strategies against disinformation. Brasília, Federal District, Brazil. Web site: https://ergoncugler.com/.

[ versão português / english above ]

# Comunidades de nova ordem mundial, globalismo e QAnon no Telegram brasileiro: como o conspiracionismo abre portas para grupos mais nocivos

*Ergon Cugler de Moraes Silva*

Instituto Brasileiro de Informação
em Ciência e Tecnologia (IBICT)
Brasília, Distrito Federal, Brasil

contato@ergoncugler.com
www.ergoncugler.com

## Resumo

As teorias da conspiração envolvendo a Nova Ordem Mundial (NOM), o Globalismo e QAnon têm se tornado centrais nas discussões do Telegram brasileiro, especialmente durante crises globais como a Pandemia da COVID-19. Dessa forma, esse estudo busca responder à pergunta de pesquisa: **como são caracterizadas e articuladas as comunidades de teorias da conspiração brasileiras sobre temáticas de nova ordem mundial, globalismo e QAnon no Telegram?** Vale ressaltar que este estudo faz parte de uma série de um total de sete estudos que possuem como objetivo principal compreender e caracterizar as comunidades brasileiras de teorias da conspiração no Telegram. Esta série de sete estudos está disponibilizada abertamente e originalmente no arXiv da Cornell University, aplicando um método espelhado nos sete estudos, mudando apenas o objeto temático de análise e provendo uma replicabilidade de investigação, inclusive com códigos próprios e autorais elaborados, somando-se à cultura de software livre e de código aberto. No que diz respeito aos principais achados deste estudo, observa-se: NOM e Globalismo tornaram-se catalisadores centrais para a disseminação de teorias conspiratórias; QAnon atua como um “narrativa-hub” que conecta NOM e Globalismo; Durante crises, menções a NOM cresceram exponencialmente, refletindo a desconfiança nas instituições; NOM e Globalismo atraem seguidores de outras teorias conspiratórias, como antivacinas, servindo como o principal *gatekeeper* de toda a rede de teoria da conspiração; Narrativas religiosas são frequentemente usadas para legitimar a NOM, reforçando a coesão ideológica.

### Principais descobertas

➔ Especialmente durante e após a Pandemia da COVID-19, a Nova Ordem Mundial (NOM) e o Globalismo se tornaram catalisadores centrais para a disseminação de teorias conspiratórias no Telegram brasileiro, com 2.411.003 publicações sobre NOM e 768.176 sobre Globalismo, criando um ambiente de desconfiança global que reforça a coesão entre essas narrativas;

➔ A interconexão entre NOM, Globalismo e QAnon forma uma rede densa de desinformação, onde QAnon atua como um “narrativa-*hub*” conectando 531.678 publicações, enquanto NOM e Globalismo amplificam essas conexões, expondo rapidamente os membros a uma vasta rede;

➔ Durante crises globais, como a Pandemia da COVID-19 e as eleições de 2020 nos EUA, as menções a NOM e Globalismo cresceram exponencialmente, com a NOM registrando um aumento de 3.150% entre 2019 e 2021, refletindo uma desconfiança crescente nas instituições;

- ➔ NOM e Globalismo atuam como portas de entrada para novos membros já imersos em outras teorias conspiratórias, recebendo 8.699 links de Conspirações Gerais e 6.260 de comunidades antivacinas, funcionando como epicentros que atraem e retêm seguidores de outras narrativas;
- ➔ Globalismo se destaca como um princípio organizador de diversas teorias conspiratórias, conectando temas como Anti-*Woke*, Revisionismo e Mudanças Climáticas, com 32.640 links recebidos de Conspirações Gerais, sugerindo que todas as teorias são parte de um plano maior;
- ➔ QAnon opera como uma metanarrativa flexível que absorve e conecta outras teorias conspiratórias, recebendo 6.308 links de Conspirações Gerais e 5.033 de NOM, reforçando a visão conspiratória de seus seguidores e dificultando a entrada de informações contraditórias;
- ➔ A NOM se consolida como um paradigma interpretativo central no universo conspiratório, unificando diversas narrativas contra uma suposta elite global, com 3.488.686 publicações, funcionando como um aglutinador de desinformação e legitimando uma visão conspiratória;
- ➔ Temas religiosos como "Deus" e "bíblia" são instrumentalizados nas narrativas sobre NOM e QAnon, sugerindo que a luta contra a NOM e o Globalismo é uma batalha espiritual, o que fortalece a coesão ideológica das comunidades e a resistência às informações científicas;
- ➔ Nas comunidades de NOM e Globalismo, as mudanças climáticas são vistas como parte de uma conspiração global para controlar a população, com tópicos como "carbono" e "climáticas" frequentemente citados, contribuindo para a rejeição da ciência e a polarização;
- ➔ As comunidades de NOM perpetuam a desinformação sobre vacinas, associando-as a um plano de controle populacional, sugerindo que as campanhas de vacinação são parte de uma agenda de manipulação, o que perpetua a resistência às vacinas e amplia a desinformação.

## 1. Introdução

Após percorrer milhares de comunidades brasileiras de teorias da conspiração no Telegram, extrair dezenas de milhões de conteúdos dessas comunidades, elaborados e/ou compartilhados por milhões de usuários que as compõem, este estudo tem o objetivo de compor uma série de um total de sete estudos que tratam sobre o fenômeno das teorias da conspiração no Telegram, adotando o Brasil como estudo de caso. Com as abordagens de identificação implementadas, foi possível alcançar um total de 217 comunidades de teorias da conspiração brasileiras no Telegram sobre temáticas de nova ordem mundial, globalismo e QAnon, estas somando 5.545.369 de conteúdos publicados entre junho de 2016 (primeiras publicações) até agosto de 2024 (realização deste estudo), com 718.246 usuários somados dentre as comunidades. Dessa forma, este estudo tem como objetivo compreender e caracterizar as comunidades sobre temáticas de nova ordem mundial, globalismo e QAnon presentes nessa rede brasileira de teorias da conspiração identificada no Telegram.

Para tal, será aplicado um método espelhado em todos os sete estudos, mudando apenas o objeto temático de análise e provendo uma replicabilidade de investigação. Assim, abordaremos técnicas para observar as conexões, séries temporais, conteúdos e sobreposições temáticas das comunidades de teorias da conspiração. Além desse estudo, é possível encontrar os seis demais disponibilizados abertamente e originalmente no arXiv da Cornell University.

Essa série contou com a atenção redobrada para garantir a integridade dos dados e o respeito à privacidade dos usuários, conforme a legislação brasileira prevê (Lei nº 13.709/2018).

Portanto questiona-se: **como são caracterizadas e articuladas as comunidades de teorias da conspiração brasileiras sobre temáticas de nova ordem mundial, globalismo e QAnon no Telegram?**

## 2. Materiais e métodos

A metodologia deste estudo está organizada em três subseções, sendo: **2.1. Extração de dados**, que descreve o processo e as ferramentas utilizadas para coletar as informações das comunidades no Telegram; **2.2. Tratamento de dados**, onde são abordados os critérios e métodos aplicados para classificar e anonimizar os dados coletados; e **2.3. Abordagens para análise de dados**, que detalha as técnicas utilizadas para investigar as conexões, séries temporais, conteúdos e sobreposições temáticas das comunidades de teorias da conspiração.

### 2.1. Extração de dados

Este projeto teve início em fevereiro de 2023, com a publicação da primeira versão do TelegramScrap (Silva, 2023), uma ferramenta própria e autoral, de software livre e de código aberto, que faz uso da Application Programming Interface (API) da plataforma Telegram por meio da biblioteca Telethon e organiza ciclos de extração de dados de grupos e canais abertos no Telegram. Ao longo dos meses, a base de dados pôde ser ampliada e qualificada fazendo uso de quatro abordagens de identificação de comunidades de teorias da conspiração:

**(i) Uso de palavras chave:** no início do projeto, foram elencadas palavras-chave para identificação diretamente no buscador de grupos e canais brasileiros no Telegram, tais como "apocalipse", "sobrevivencialismo", "mudanças climáticas", "terra plana", "teoria da conspiração", "globalismo", "nova ordem mundial", "ocultismo", "esoterismo", "curas alternativas", "qAnon", "reptilianos", "revisionismo", "alienígenas", dentre outras. Essa primeira abordagem forneceu algumas comunidades cujos títulos e/ou descrições dos grupos e canais contassem com os termos explícitos relacionados a teorias da conspiração. Contudo, com o tempo foi possível identificar outras diversas comunidades cujas palavras-chave elencadas não davam conta de abarcar, algumas inclusive propositalmente com caracteres trocados para dificultar a busca de quem a quisesse encontrar na rede;

**(ii) Mecanismo de recomendação de canais do Telegram:** com o tempo, foi identificado que canais do Telegram (exceto grupos) contam com uma aba de recomendação chamada de "canais similares", onde o próprio Telegram sugere dez canais que tenham alguma similaridade com o canal que se está observando. A partir desse mecanismo de recomendação do próprio Telegram, foi possível encontrar mais comunidades de teorias da conspiração brasileiras, com estas sendo recomendadas pela própria plataforma;

**(iii) Abordagem de bola de neve para identificação de convites:** após algumas comunidades iniciais serem acumuladas para a extração, foi elaborado um algoritmo próprio autoral de identificação de urls que contivessem “t.me/”, sendo o prefixo de qualquer convite para grupos e canais do Telegram. Acumulando uma frequência de centenas de milhares de links que atendessem a esse critério, foram elencados os endereços únicos e contabilizadas as suas repetições. Dessa forma, foi possível fazer uma investigação de novos grupos e canais brasileiros mencionados nas próprias mensagens dos já investigados, ampliando a rede. Esse processo foi sendo repetido periodicamente, buscando identificar novas comunidades que tivessem identificação com as temáticas de teorias da conspiração no Telegram;

**(iv) Ampliação para tweets publicados no X que mencionassem convites:** com o objetivo de diversificar ainda mais a fonte de comunidades de teorias da conspiração brasileiras no Telegram, foi elaborada uma query de busca própria que pudesse identificar as palavras-chave de temáticas de teorias da conspiração, porém usando como fonte tweets publicados no X (antigo Twitter) e que, além de conter alguma das palavras-chave, contivesse também o “t.me/”, sendo o prefixo de qualquer convite para grupos e canais do Telegram, “https://x.com/search?q=lang%3Apt%20%22t.me%2F%22%20TERMO-DE-BUSCA”.

Com as abordagens de identificação de comunidades de teorias da conspiração implementadas ao longo de meses de investigação e aprimoramento de método, foi possível construir uma base de dados do projeto com um total de 855 comunidades de teorias da conspiração brasileiras no Telegram (considerando as demais temáticas também não incluídas nesse estudo), estas somando 27.227.525 de conteúdos publicados entre maio de 2016 (primeiras publicações) até agosto de 2024 (realização deste estudo), com 2.290.621 usuários somados dentre as comunidades brasileiras. Há de se considerar que este volume de usuários conta com dois elementos, o primeiro é que trata-se de uma variável, pois usuários podem entrar e sair diariamente, portanto este valor representa o registrado na data de extração de publicações da comunidade; além disso, é possível que um mesmo usuário esteja em mais de um grupo e, portanto, é contabilizado mais de uma vez. Nesse sentido, o volume ainda sinaliza ser expressivo, mas pode ser levemente menor quando considerado o volume de cidadãos deduplicados dentro dessas comunidades brasileiras de teorias da conspiração.

### 2.2. Tratamento de dados

Com todos os grupos e canais brasileiros de teorias da conspiração no Telegram extraídos, foi realizada uma classificação manual considerando o título e a descrição da comunidade. Caso houvesse menção explícita no título ou na descrição da comunidade a alguma temática, esta foi classificada entre: (i) “Anticiência”; (ii) “Anti-Woke e Gênero”; (iii) “Antivax”; (iv) “Apocalipse e Sobrevivencialismo”; (v) “Mudanças Climáticas”; (vi) Terra Plana; (vii) “Globalismo”; (viii) “Nova Ordem Mundial”; (ix) “Ocultismo e Esoterismo”; (x) “Off Label e Charlatanismo”; (xi) “QAnon”; (xii) “Reptilianos e Criaturas”; (xiii) “Revisionismo e Discurso de Ódio”; (xiv) “OVNI e Universo”. Caso não houvesse nenhuma menção explícita relacionada às temáticas no título ou na descrição da comunidade, esta foi classificada como (xv) “Conspiração Geral”. No Quadro a seguir, podemos observar as métricas relacionadas à classificação dessas comunidades de teorias da conspiração no Brasil.

**Tabela 01.** Comunidades de teorias da conspiração no Brasil (métricas até agosto de 2024)

| Categorias | Grupos | Usuários | Publicações | Comentários | Total |
|---|---|---|---|---|---|
| **Anticiência** | 22 | 58.138 | 187.585 | 784.331 | 971.916 |
| **Anti-*Woke* e Gênero** | 43 | 154.391 | 276.018 | 1.017.412 | 1.293.430 |
| **Antivacinas (*Antivax*)** | 111 | 239.309 | 1.778.587 | 1.965.381 | 3.743.968 |
| **Apocalipse e Sobrevivência** | 33 | 109.266 | 915.584 | 429.476 | 1.345.060 |
| **Mudanças Climáticas** | 14 | 20.114 | 269.203 | 46.819 | 316.022 |
| **Terraplanismo** | 33 | 38.563 | 354.200 | 1.025.039 | 1.379.239 |
| **Conspirações Gerais** | 127 | 498.190 | 2.671.440 | 3.498.492 | 6.169.932 |
| **Globalismo** | 41 | 326.596 | 768.176 | 537.087 | 1.305.263 |
| **Nova Ordem Mundial (NOM)** | 148 | 329.304 | 2.411.003 | 1.077.683 | 3.488.686 |
| **Ocultismo e Esoterismo** | 39 | 82.872 | 927.708 | 2.098.357 | 3.026.065 |
| **Medicamentos *off label*** | 84 | 201.342 | 929.156 | 733.638 | 1.662.794 |
| **QAnon** | 28 | 62.346 | 531.678 | 219.742 | 751.420 |
| **Reptilianos e Criaturas** | 19 | 82.290 | 96.262 | 62.342 | 158.604 |
| **Revisionismo e Ódio** | 66 | 34.380 | 204.453 | 142.266 | 346.719 |
| **OVNI e Universo** | 47 | 58.912 | 862.358 | 406.049 | 1.268.407 |
| **Total** | **855** | **2.296.013** | **13.183.411** | **14.044.114** | **27.227.525** |

Fonte: Elaboração própria (2024).

Com esse volume de dados extraídos, foi possível segmentar para apresentar neste estudo apenas comunidades e conteúdos referentes às temáticas de nova ordem mundial, globalismo e QAnon. Em paralelo, as demais temáticas de comunidades brasileiras de teorias da conspiração também contaram com estudos elaborados para a caracterização da extensão e da dinâmica da rede, estes sendo disponibilizados abertamente e originalmente no arXiv da Cornell University.

Além disso, cabe citar que apenas foram extraídas comunidades abertas, isto é, não apenas identificáveis publicamente, mas também sem necessidade de solicitação para acessar ao conteúdo, estando aberto para todo e qualquer usuário com alguma conta do Telegram sem que este necessite ingressar no grupo ou canal. Além disso, em respeito à legislação brasileira e especialmente da Lei Geral de Proteção de Dados Pessoais (LGPD), ou Lei nº 13.709/2018, que trata do controle da privacidade e do uso/tratamento de dados pessoais, todos os dados extraídos foram anonimizados para a realização de análises e investigações. Dessa forma, nem mesmo a identificação das comunidades é possível por meio deste estudo, estendendo aqui a privacidade do usuário ao anonimizar os seus dados para além da própria comunidade à qual ele se submeteu ao estar em um grupo ou canal público e aberto no Telegram.

### 2.3. Abordagens para análise de dados

Totalizando 217 comunidades selecionadas nas temáticas de nova ordem mundial, globalismo e QAnon, contendo 5.545.369 publicações e 718.246 usuários somados, quatro abordagens serão utilizadas para investigar as comunidades de teorias da conspiração selecionadas para o escopo do estudo. Tais métricas são detalhadas no Quadro a seguir:

**Tabela 02.** Comunidades selecionadas para análise (métricas até agosto de 2024)

| Categorias | Grupos | Usuários | Publicações | Comentários | Total |
|---|---|---|---|---|---|
| **Nova Ordem Mundial (NOM)** | 148 | 329.304 | 2.411.003 | 1.077.683 | 3.488.686 |
| **QAnon** | 28 | 62.346 | 531.678 | 219.742 | 751.420 |
| **Globalismo** | 41 | 326.596 | 768.176 | 537.087 | 1.305.263 |
| **Total** | **217** | **718.246** | **3.710.857** | **1.834.512** | **5.545.369** |

Fonte: Elaboração própria (2024).

**(i) Rede:** com a elaboração de um algoritmo próprio para a identificação de mensagens que contenham o termo de "t.me/" (de convite para entrarem em outras comunidades), propomos apresentar volumes e conexões observadas sobre como **(a)** as comunidades da temática de nova ordem mundial, globalismo e QAnon circulam convites para que os seus usuários conheçam mais grupos e canais da mesma temática, reforçando os sistemas de crença que comungam; e como **(b)** essas mesmas comunidades circulam convites para que os seus usuários conheçam grupos e canais que tratem de outras teorias da conspiração, distintas de seu propósito explícito. Esta abordagem é interessante para observar se essas comunidades utilizam a si próprias como fonte de legitimação e referência e/ou se embasam-se em demais temáticas de teorias da conspiração, inclusive abrindo portas para que seus usuários conheçam outras conspirações. Além disso, cabe citar o estudo de Rocha *et al.* (2024) em que uma abordagem de identificação de rede também foi aplicada em comunidades do Telegram, porém observando conteúdos similares a partir de um ID gerado para cada mensagem única e suas similares;

**(ii) Séries temporais:** utiliza-se da biblioteca "Pandas" (McKinney, 2010) para organizar os data frames de investigação, observando **(a)** o volume de publicações ao longo dos meses; e **(b)** o volume de engajamento ao longo dos meses, considerando metadados de visualizações, reações e comentários coletados na extração; Além da volumetria, a biblioteca "Plotly" (Plotly Technologies Inc., 2015) viabilizou a representação gráfica dessa variação;

**(iii) Análise de conteúdo:** além da análise geral de palavras com identificação das frequências, são aplicadas séries temporais na variação das palavras mais frequentes ao longo dos semestres — observando entre junho de 2016 (primeiras publicações) até agosto de 2024 (realização deste estudo). E com as bibliotecas "Pandas" (McKinney, 2010) e "WordCloud" (Mueller, 2020), os resultados são apresentados tanto volumetricamente quanto graficamente;

**(iv) Sobreposição de agenda temática:** seguindo a abordagem proposta por Silva & Sátiro (2024) para identificação de sobreposição de agenda temática em comunidades do Telegram, utilizamos o modelo “BERTopic” (Grootendorst, 2020). O BERTopic é um algoritmo de modelagem de tópicos que facilita a geração de representações temáticas a partir de grandes quantidades de textos. Primeiramente, o algoritmo extrai embeddings dos documentos usando modelos transformadores de sentenças, como o “all-MiniLM-L6-v2”. Em seguida, essas embeddings têm sua dimensionalidade reduzida por técnicas como “UMAP”, facilitando o processo de agrupamento. A clusterização é realizada usando “HDBSCAN”, uma técnica baseada em densidade que identifica clusters de diferentes formas e tamanhos, além de detectar outliers. Posteriormente, os documentos são tokenizados e representados em uma estrutura de bag-of-words, que é normalizada (L1) para considerar as diferenças de tamanho entre os clusters. A representação dos tópicos é refinada usando uma versão modificada do “TF-IDF”, chamada “Class-TF-IDF”, que considera a importância das palavras dentro de cada cluster (Grootendorst, 2020). Cabe destacar que, antes de aplicar o BERTopic, realizamos a limpeza da base removendo “stopwords” em português, por meio da biblioteca “NLTK” (Loper & Bird, 2002). Para a modelagem de tópicos, utilizamos o backend “loky” para otimizar o desempenho durante o ajuste e a transformação dos dados.

Em síntese, a metodologia aplicada compreendeu desde a extração de dados com a ferramenta própria autoral TelegramScrap (Silva, 2023), até o tratamento e a análise de dados coletados, utilizando diversas abordagens para identificar e classificar comunidades de teorias da conspiração brasileiras no Telegram. Cada uma das etapas foi cuidadosamente implementada para garantir a integridade dos dados e o respeito à privacidade dos usuários, conforme a legislação brasileira prevê. A seguir, serão apresentados os resultados desses dados, com o intuito de revelar as dinâmicas e as características das comunidades estudadas.

## 3. Resultados

A seguir, os resultados são detalhados na ordem prevista na metodologia, iniciando com a caracterização da rede e suas fontes de legitimação e referência, avançando para as séries temporais, incorporando a análise de conteúdo e concluindo com a identificação de sobreposição de agenda temática dentre as comunidades de teorias da conspiração.

### 3.1. Rede

A análise da rede de comunidades ligadas à Nova Ordem Mundial (NOM), Globalismo e QAnon revela a complexidade e interconectividade desse ecossistema conspiratório. A primeira figura, que retrata a rede interna dessas temáticas, destaca como essas comunidades se entrelaçam de maneira densa, criando um ciclo contínuo de reforço mútuo entre suas crenças. A interdependência entre essas narrativas é evidenciada pelos grandes nós que atuam como epicentros de disseminação, onde uma vez que os seguidores adentram nesse universo conspiratório, eles são rapidamente envolvidos por uma multiplicidade de narrativas. Essas interações fazem com que a separação entre realidade e teoria conspiratória se torne cada vez mais difícil para os indivíduos envolvidos, levando a um ambiente onde as ideias sobre controle global, elites ocultas e QAnon estão em constante retroalimentação. A segunda figura explora as comunidades que servem como porta de entrada para essas temáticas, mostrando como NOM, Globalismo e QAnon atraem novos seguidores ao expor essas comunidades a um vasto universo de teorias. Essas comunidades atuam como *hubs* centrais na rede conspiratória, facilitando a transição e o aprofundamento dos seguidores em outras teorias interconectadas. A sobreposição significativa entre essas comunidades destaca que a entrada em uma delas pode levar rapidamente à adoção de múltiplas outras teorias, criando um ciclo de reforço contínuo dessas crenças.

Na terceira figura, observamos como essas temáticas se interconectam com outras áreas conspiratórias, atuando como portas de saída para novas narrativas. A centralidade das comunidades de NOM na rede, por exemplo, mostra que elas desempenham um papel crucial na conexão entre diferentes teorias, funcionando como núcleos de disseminação que incentivam a exploração de novas áreas conspiratórias. Isso indica que os seguidores de NOM, Globalismo e QAnon não apenas reforçam suas crenças, mas também são constantemente expostos a novas narrativas, ampliando seu engajamento dentro desse ecossistema. Os gráficos de fluxo de links de convites, que analisam a centralidade de NOM, Globalismo e QAnon, reforçam a ideia de que essas temáticas não são isoladas, mas profundamente interligadas a uma vasta rede de teorias conspiratórias. NOM, por exemplo, recebe um volume expressivo de links de Conspirações Gerais e Antivacinas, funcionando como um paradigma interpretativo central que unifica diferentes narrativas contra uma suposta elite global. O Globalismo, por sua vez, atua como um princípio organizador de outras teorias, fortalecendo a coesão interna dessas comunidades ao sugerir que todas as conspirações fazem parte de um plano maior. Já QAnon, ao se conectar com diversas outras teorias, funciona como um "narrativa-hub", sintetizando diferentes crenças em um todo coerente e reforçando a visão conspiratória original de seus seguidores.

Em suma, as redes e os fluxos de links entre NOM, Globalismo e QAnon mostram que essas temáticas operam como motores centrais dentro do universo conspiratório, agindo tanto como portas de entrada quanto de saída para outras teorias, ao mesmo tempo em que reforçam a coesão e a expansão desse ecossistema de crenças interligadas.

[ versão português / english above ]

**Figura 01.** Rede interna entre nova ordem mundial, globalismo e QAnon

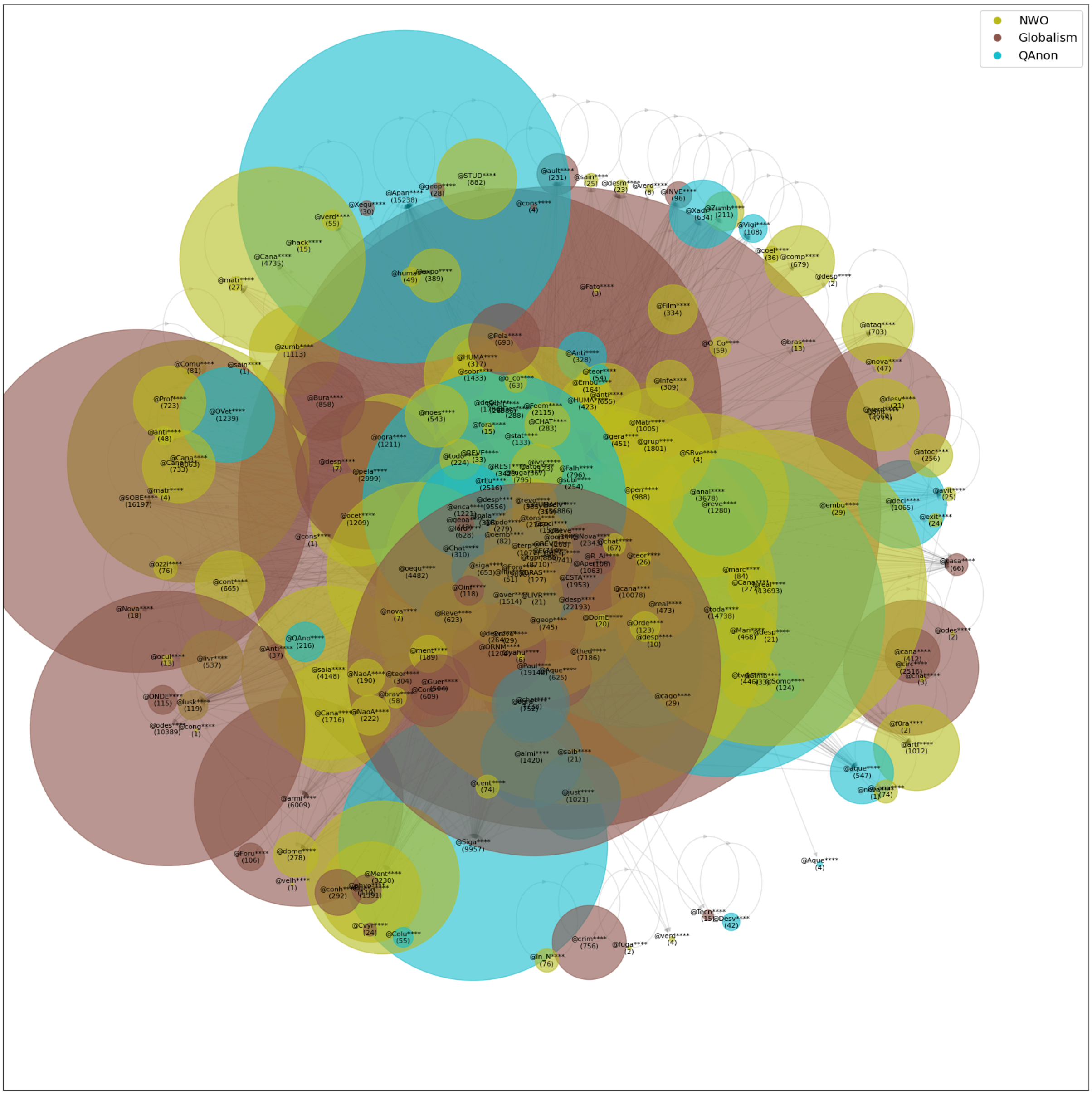

Fonte: Elaboração própria (2024).

Este gráfico revela uma rede complexa e densamente conectada entre comunidades que discutem temas como nova ordem mundial, globalismo e QAnon. A sobreposição significativa entre os grupos sugere uma forte interdependência entre essas narrativas conspiratórias, com as comunidades frequentemente se referenciando umas às outras. Essa rede reflete um ecossistema conspiratório onde ideias sobre controle global, elites ocultas e teorias relacionadas a QAnon são constantemente interligadas, criando um ciclo contínuo de reforço dessas crenças. A densidade e a amplitude da rede indicam que uma vez que os seguidores entram nesse círculo, eles são rapidamente envolvidos por uma multiplicidade de narrativas, tornando difícil para eles discernir a realidade das teorias conspiratórias. Os grandes nós representam os epicentros dessas interações, com uma alta capacidade de amplificação de conteúdo que se propaga por toda a rede, reforçando a coesão entre as diferentes narrativas.

[ versão português / english above ]

**Figura 02.** Rede de comunidades que abrem portas para a temática (porta de entrada)

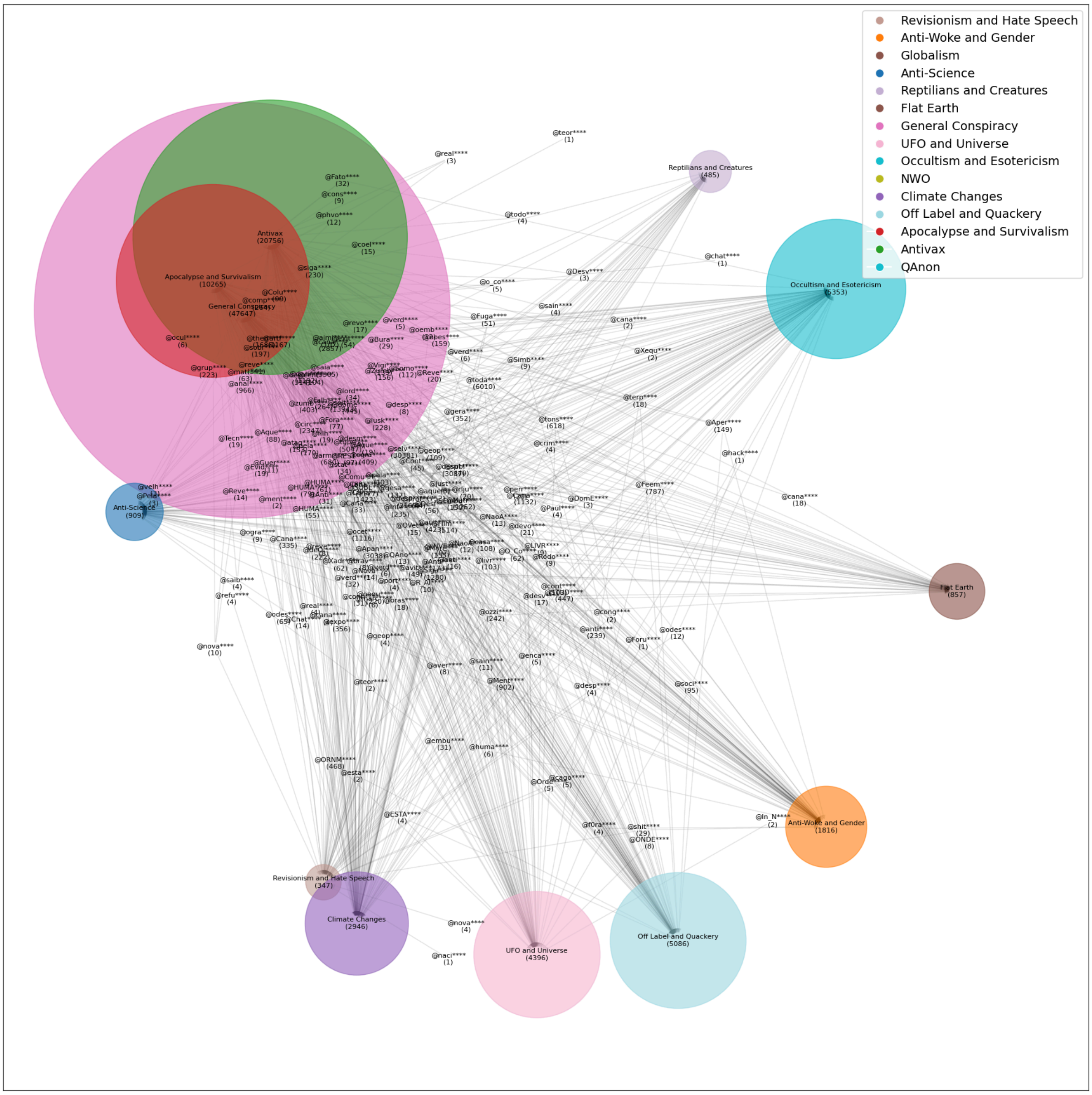

Fonte: Elaboração própria (2024).

A figura mostra a rede de comunidades que são portas de entrada para discussões sobre a nova ordem mundial, globalismo e QAnon. As grandes esferas que representam essas comunidades destacam a sua importância como *hubs* centrais na rede conspiratória. Esses temas atuam como imãs para seguidores de teorias conspiratórias, que ao entrar em contato com essas comunidades, são expostos a uma vasta rede de outras teorias correlatas. A sobreposição significativa entre essas comunidades sugere que a transição entre as diferentes teorias é comum, criando um ciclo contínuo de reforço dessas crenças. Além disso, essas comunidades não apenas centralizam essas narrativas, mas também ampliam a capacidade de conexão entre diferentes áreas conspiratórias, aumentando o engajamento dos indivíduos.

[ versão português / english above ]

**Figura 03.** Rede de comunidades cuja temática abre portas (porta de saída)

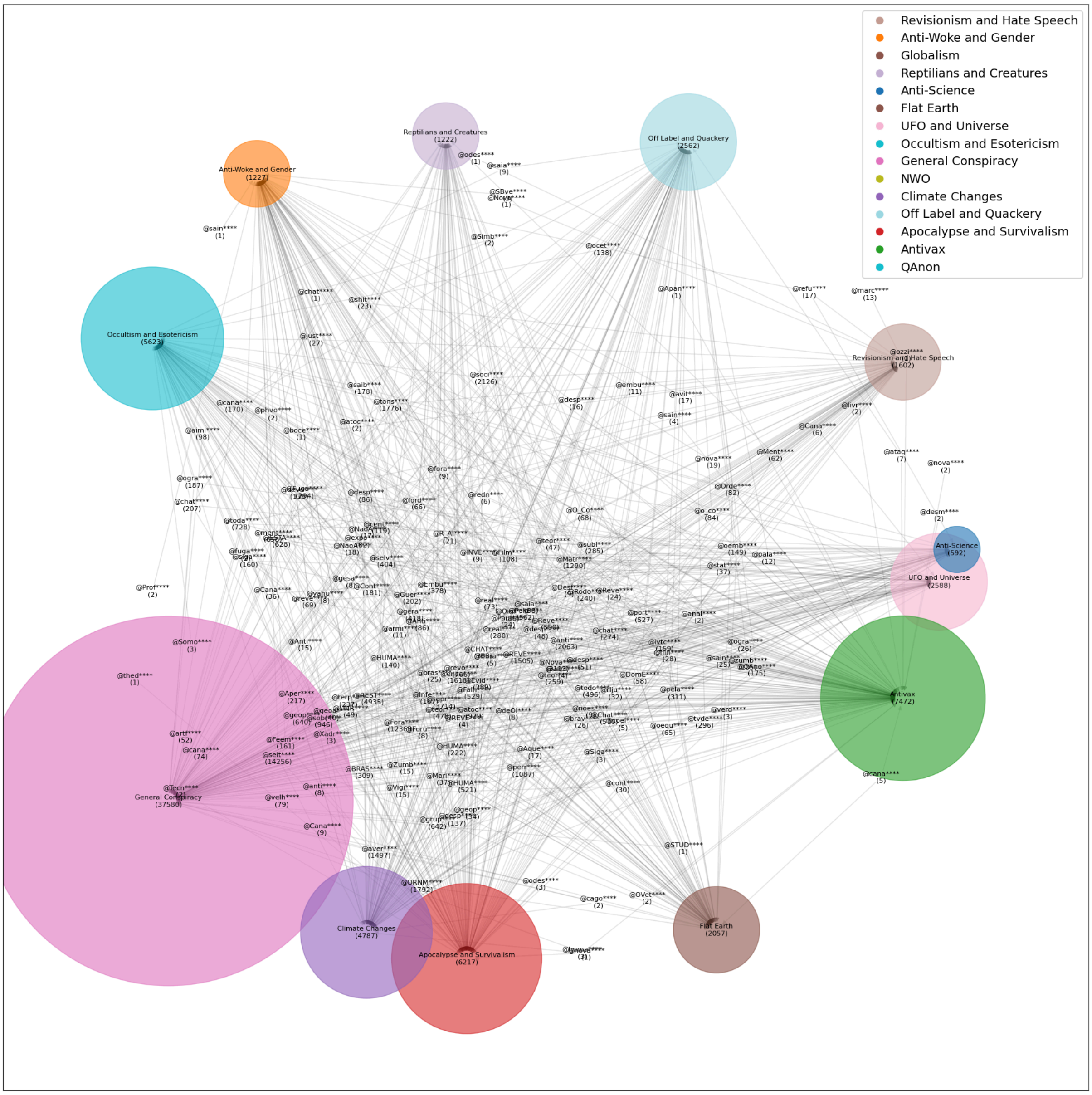

Fonte: Elaboração própria (2024).

Este gráfico destaca a rede de comunidades relacionadas à Nova Ordem Mundial (NOM), Globalismo e QAnon, demonstrando como elas se entrelaçam com diversas outras temáticas conspiratórias. A centralidade das comunidades de NOM na rede indica que elas desempenham um papel crucial em conectar diferentes teorias da conspiração. As interações entre NOM e temáticas como Globalismo, QAnon e Ocultismo são especialmente notáveis, sugerindo que essas comunidades funcionam como núcleos de disseminação de desinformação, que não apenas reforçam as crenças dos seus membros, mas também incentivam a exploração de novas áreas conspiratórias. A densa teia de conexões evidencia que os seguidores de NOM têm maior probabilidade de se engajar em uma gama variada de teorias conspiratórias, indicando que essas comunidades agem como um elo vital para a propagação de uma visão de mundo conspiratória amplificada e multifacetada.

**Figura 04.** Fluxo de links de convites entre comunidades de nova ordem mundial

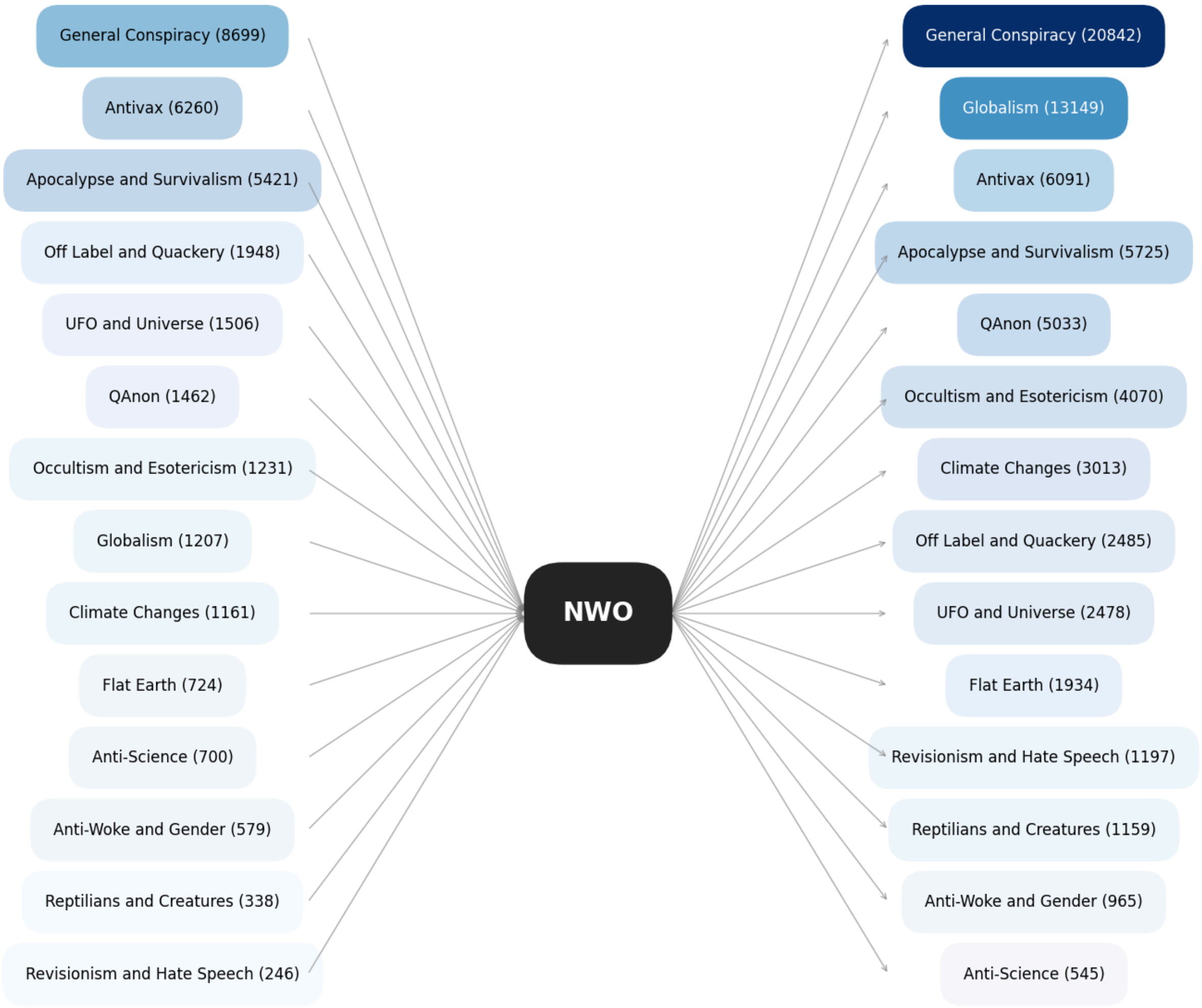

Fonte: Elaboração própria (2024).

O gráfico de fluxos de convites na temática da Nova Ordem Mundial (NOM) demonstra claramente sua centralidade no universo conspiratório. Com um volume expressivo de links oriundos de Conspirações Gerais (8.699) e Antivacinas (6.260), a NOM se configura não apenas como um tema central, mas como o principal aglutinador de narrativas que se entrelaçam em torno da desconfiança contra instituições globais. A NOM funciona como um paradigma interpretativo, onde diversas narrativas encontram coerência e sentido ao se posicionarem contra uma suposta elite global que manipula eventos globais. Essa interconexão não é trivial. Ela implica que a NOM atua como uma plataforma de reforço mútuo entre diferentes teorias conspiratórias. As teorias que circulam dentro dessa rede são mutuamente fortalecedoras, criando um ciclo onde o indivíduo que adentra por uma dessas portas (como Antivacinas ou Apocalipse e Sobrevivência) é progressivamente levado a adotar uma visão de mundo onde tudo está interconectado por um fio comum de dominação e controle. Esse ciclo, portanto, não é apenas uma adição de crenças, mas uma reconfiguração do pensamento crítico, que passa a interpretar todos os eventos mundiais como peças de um mesmo quebra-cabeça manipulativo.

**Figura 05.** Fluxo de links de convites entre comunidades de globalismo

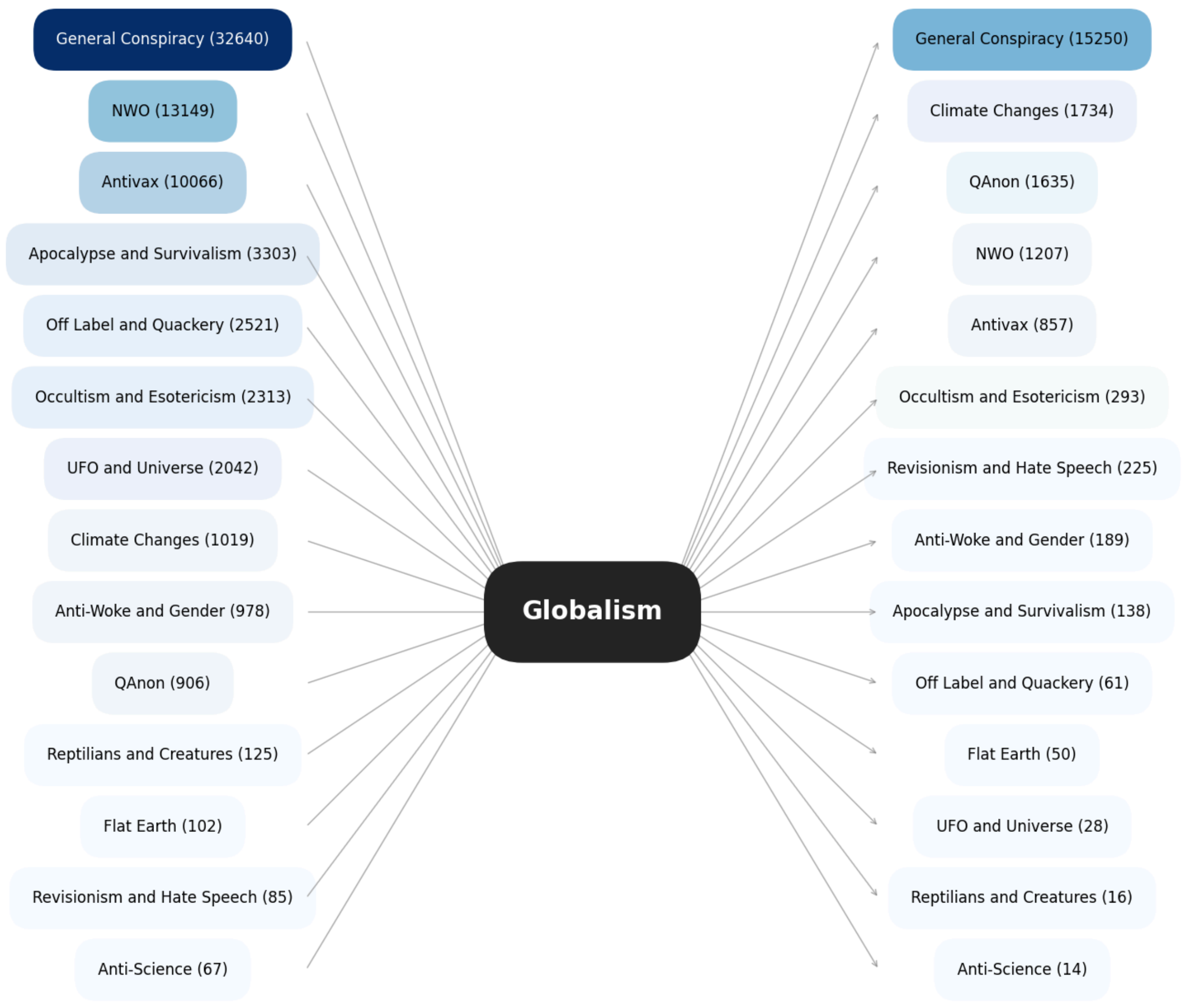

Fonte: Elaboração própria (2024).

O gráfico de Globalismo revela uma relação intrínseca entre a temática e Conspirações Gerais, que envia 32.640 links, evidenciando que o Globalismo é um conceito que se fortalece e se retroalimenta das mais diversas teorias conspiratórias. Essa conexão indica que o Globalismo não é apenas uma narrativa, mas um princípio organizador de outras teorias, uma espécie de "cola" que une diferentes discursos de resistência contra a ordem mundial estabelecida. A narrativa de Globalismo funciona como uma forma de legitimação das outras teorias ao sugerir que todas as supostas conspirações fazem parte de um plano maior. À medida que Globalismo distribui convites para Conspirações Gerais (15.250 links) e Mudanças Climáticas (1.734 links), fica evidente que essa teoria não só atrai, mas também dissemina um conjunto diversificado de crenças. Esse papel duplo sugere que o Globalismo opera tanto como uma entrada quanto como um canalizador de desinformação, construindo pontes entre narrativas que, à primeira vista, podem parecer desconexas. A análise aqui aponta para o Globalismo como uma narrativa integradora, que não apenas facilita a entrada de novos adeptos ao universo conspiratório, mas também reforça a coesão interna dessas comunidades ao oferecer uma explicação unificadora para uma miríade de teorias.

**Figura 06.** Fluxo de links de convites entre comunidades de QAnon

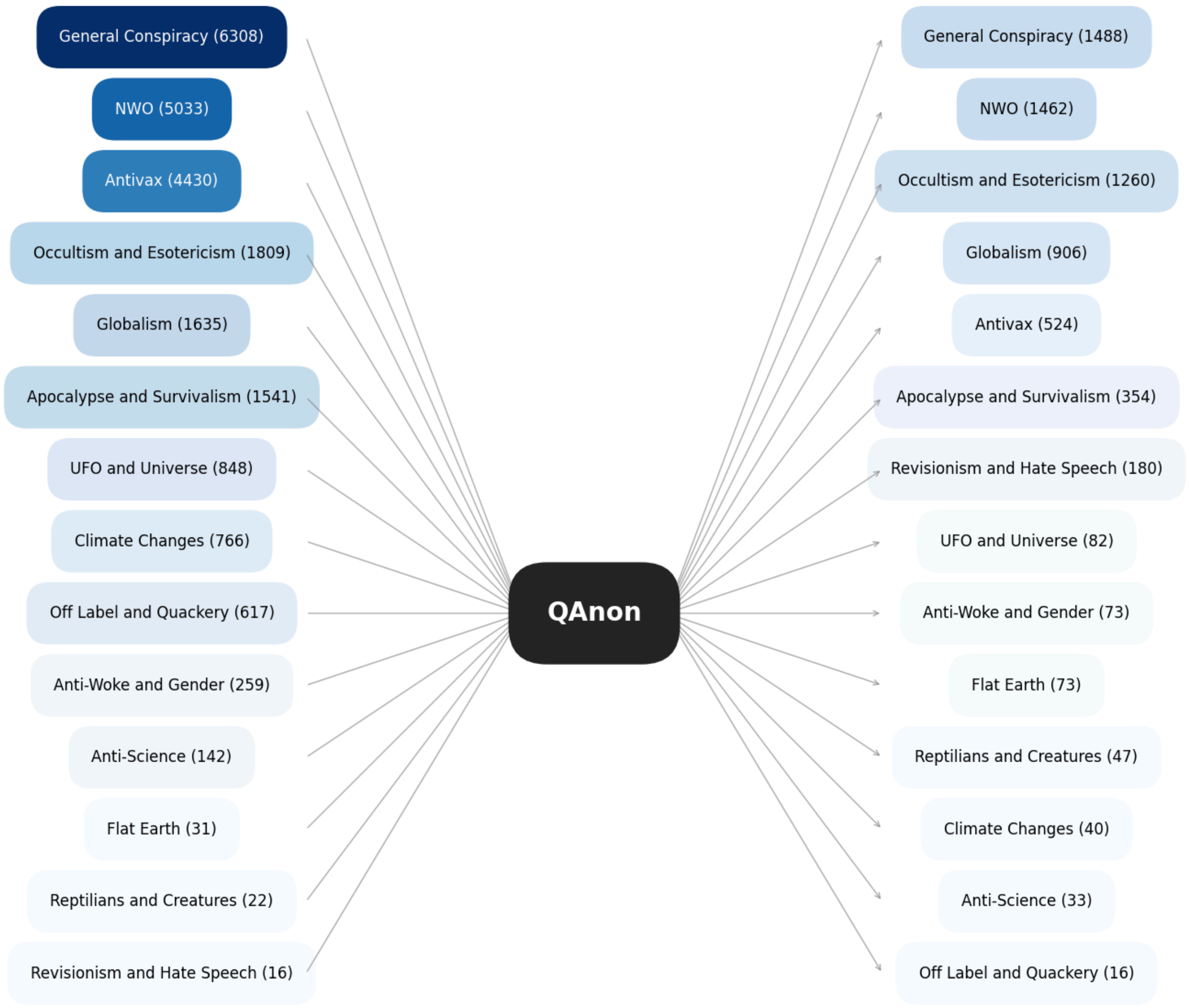

Fonte: Elaboração própria (2024).

QAnon, como mostrado no gráfico de fluxos, está profundamente interligado a uma vasta rede de teorias que vão desde Conspirações Gerais (6.308 links) até NOM (5.033 links). Esse posicionamento revela que QAnon não é apenas uma teoria entre muitas, mas uma espécie de "narrativa-hub", que conecta e sintetiza diversas outras teorias em um todo coerente para seus seguidores. O papel de QAnon na difusão de outras narrativas é evidente pelos convites que envia para Conspirações Gerais (1.488 links) e NOM (1.462 links), evidenciando que, para os adeptos de QAnon, essas teorias não são apenas paralelas, mas complementares. Refletindo sobre a função de QAnon no contexto mais amplo das teorias da conspiração, podemos ver que ele atua como um "sistema de crenças flexível". Seus adeptos podem integrar novas informações de forma a sempre reforçar a visão conspiratória original, tornando-a altamente resistente a fatos contraditórios. Isso sugere que QAnon é mais do que uma simples teoria; é um metanarrativa que oferece um modelo de como entender e organizar as outras crenças. Assim, QAnon funciona como uma via rápida para a radicalização, ao absorver e reconectar várias teorias dentro de uma moldura interpretativa abrangente.

### 3.2. Séries temporais

A seguir, exploraremos as séries temporais que destacam o crescimento expressivo de teorias conspiratórias como Nova Ordem Mundial (NOM) e QAnon, e como estas se entrelaçam com narrativas sobre Anti-*Woke* e Globalismo. O gráfico a seguir evidencia como a Pandemia da COVID-19 e as eleições presidenciais dos EUA em 2020 atuaram como catalisadores para a explosão dessas menções. A NOM, que já tinha relevância, ampliou seu alcance, refletindo uma desconfiança generalizada nas instituições globais. QAnon seguiu um padrão semelhante, solidificando sua presença. Ao longo do tempo, notamos uma gradual diminuição, mas as teorias permanecem significativamente enraizadas no discurso público, com a NOM mantendo uma posição predominante até 2024, sugerindo que as crises globais continuam a alimentar essas narrativas.

**Figura 07.** Gráfico de linhas do período

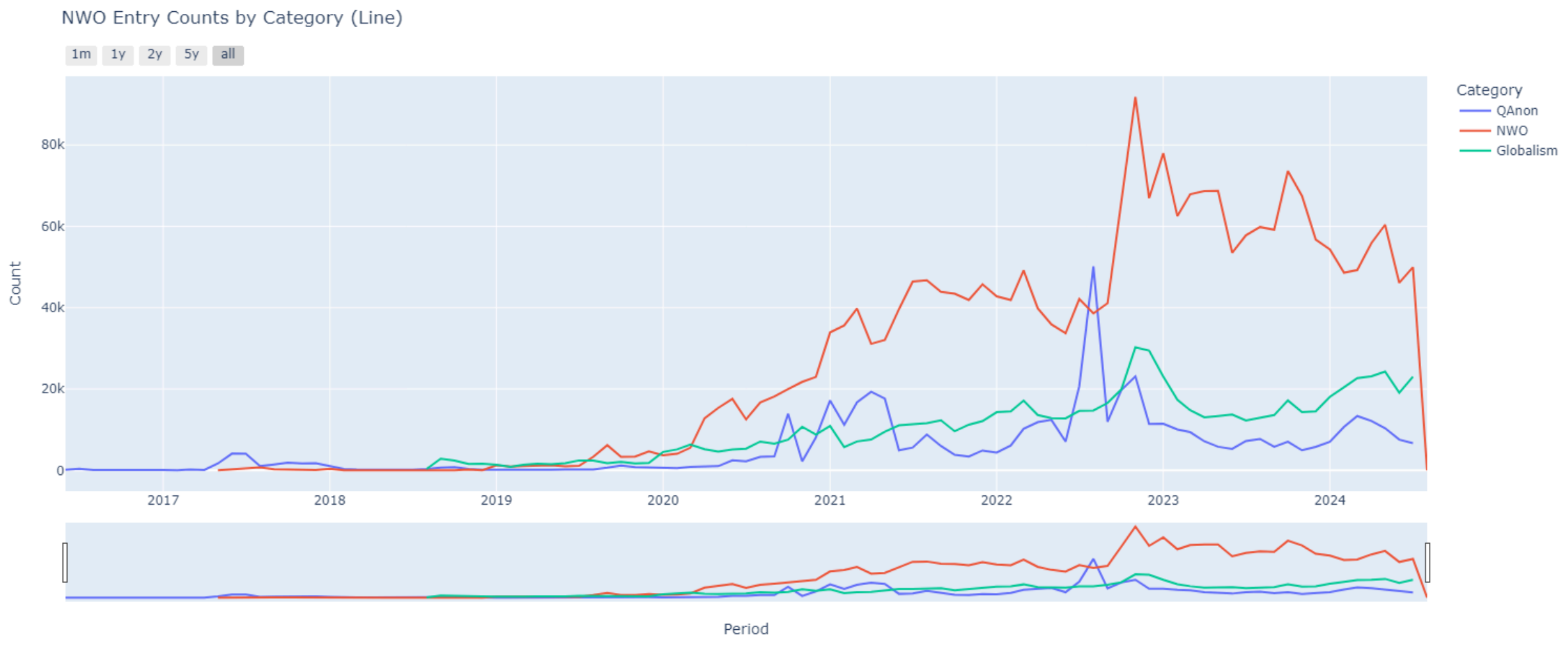

Fonte: Elaboração própria (2024).

No gráfico das categorias Nova Ordem Mundial (NOM), QAnon, Anti-*Woke* e Gênero, e Globalismo, o crescimento das menções a NOM a partir de 2019 até 2021 é impressionante. As menções sobem de cerca de 2.000 em 2019 para mais de 65.000 em janeiro de 2021, um aumento de 3.150%. Este crescimento pode ser atribuído a eventos como a Pandemia da COVID-19 e as eleições dos EUA, que foram catalisadores para o fortalecimento dessas teorias. Para QAnon, o aumento de menções é igualmente significativo, subindo de cerca de 500 para 15.000, um aumento de 2.900%. As discussões sobre Anti-*Woke* e Gênero e Globalismo seguem um padrão de crescimento mais gradual, com aumentos de 800% e 1.000%, respectivamente, durante o mesmo período. Os picos de NOM e QAnon em janeiro de 2021 caem gradualmente ao longo do ano seguinte, com uma redução de cerca de 30% até o final de 2022. Ainda assim, as temáticas de NOM se manteve como predominante do conjunto até mesmo nas datas mais recentes, quanto ao primeiro semestre de 2024.

**Figura 08.** Gráfico de área absoluta do período

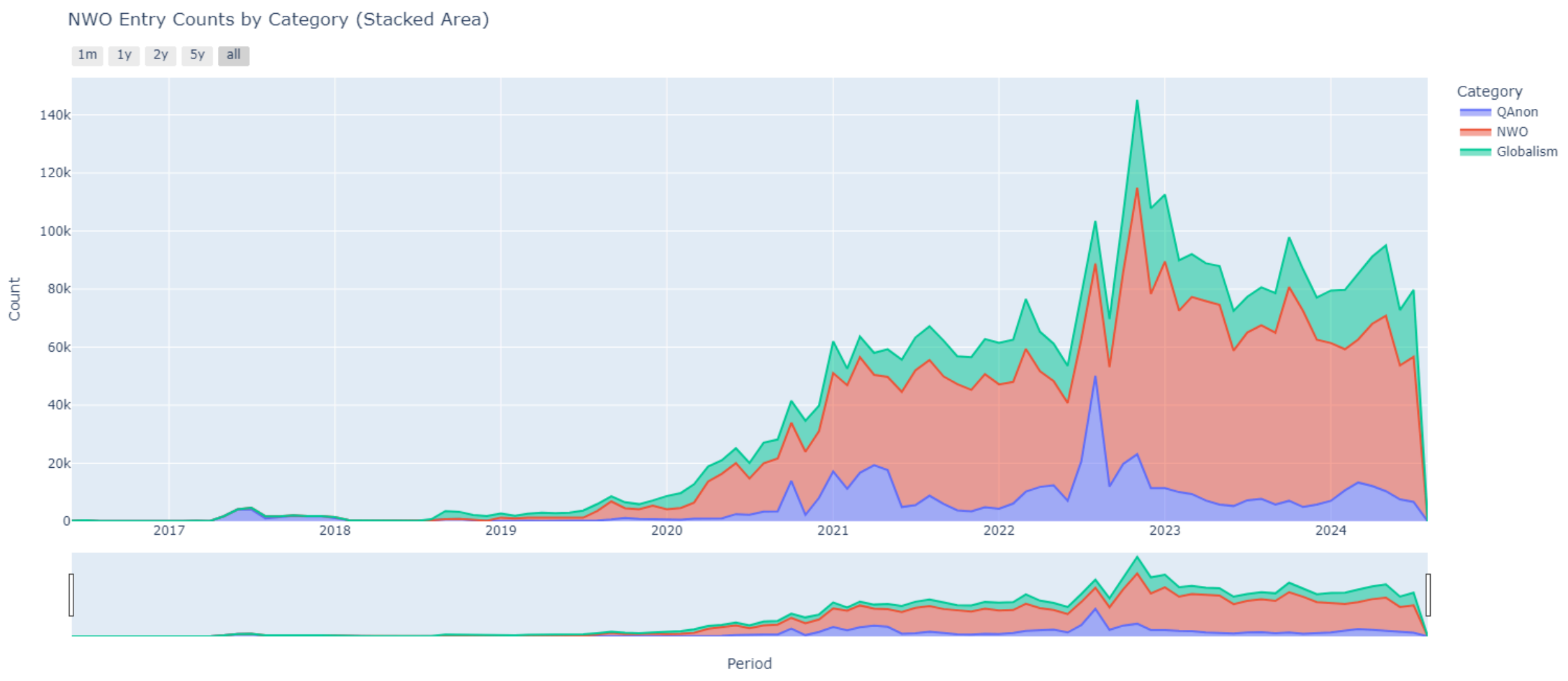

Fonte: Elaboração própria (2024).

O gráfico de área absoluta do período revela que a temática Nova Ordem Mundial (NOM) iniciou com relevância considerável já antes de 2020. Isso não é surpreendente, visto que a NOM é uma das teorias conspiratórias mais duradouras e amplamente disseminadas. A partir de 2020, há um crescimento notável, com picos em 2021 e 2022, correspondendo a mais de 140 mil entradas. Esses aumentos podem ser associados à intensificação de discursos conspiratórios durante a Pandemia da COVID-19, quando teorias sobre controle global ganharam força. O período também coincide com a ascensão do movimento QAnon, que explora narrativas de um "estado profundo" e manipulação global, assim como o crescimento do Globalismo como uma teoria conspiratória. As discussões sobre Globalismo, que atingem seu pico em 2021, refletem preocupações crescentes com a globalização e a percepção de que elites internacionais estão controlando eventos mundiais. QAnon, por sua vez, apresenta picos menos acentuados, mas ainda assim significativos, especialmente após a eleição presidencial dos EUA em 2020, que foi um catalisador para essa narrativa conspiratória. Esses movimentos sugerem um ambiente de desconfiança e radicalização, onde as crises globais são vistas como orquestradas por forças ocultas.

**Figura 09.** Gráfico de área relativa do período

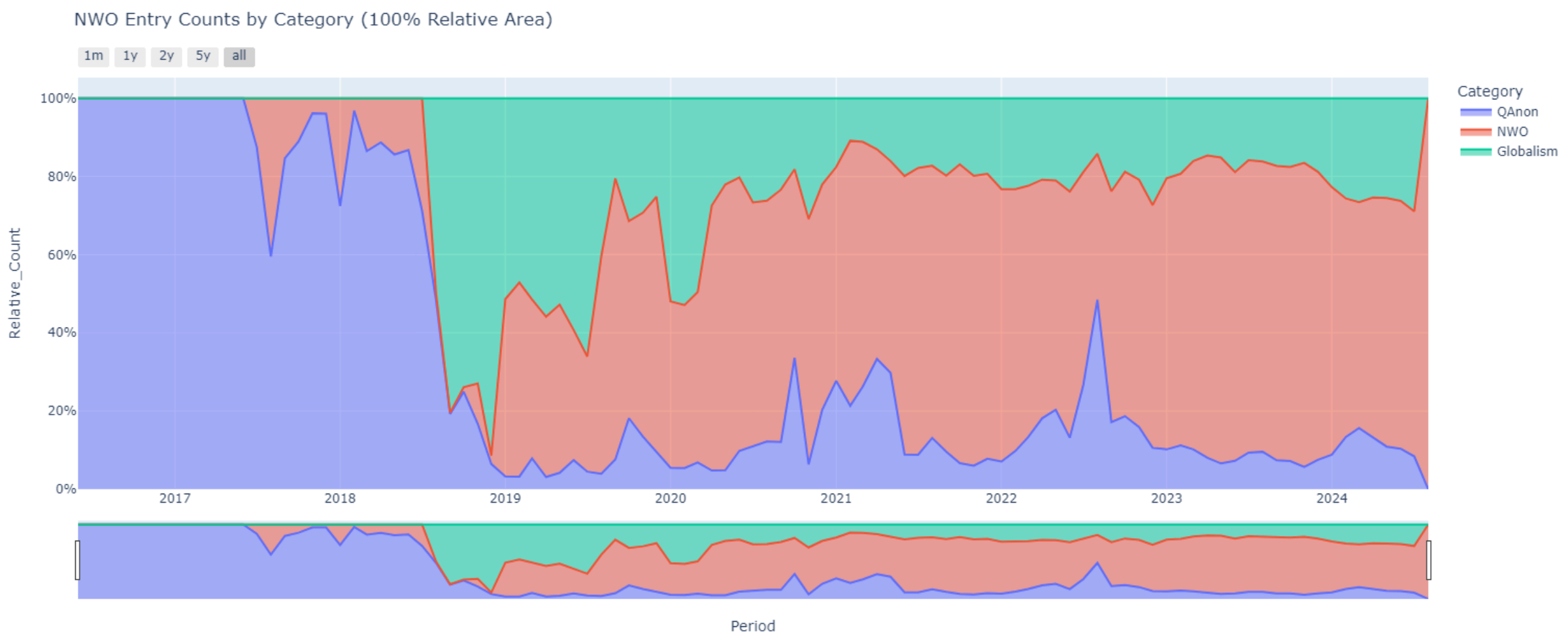

Fonte: Elaboração própria (2024).

Ao analisar o gráfico de área relativa, percebe-se que antes de 2020, as menções à NOM dominavam as discussões, representando a maior parte das entradas. No entanto, após 2020, há uma mudança evidente na distribuição temática, com o Globalismo ganhando terreno de forma significativa, especialmente em 2021, quando passa a ocupar uma posição dominante. Esse deslocamento pode ser interpretado como um reflexo das crescentes preocupações com políticas de globalização e intervenções internacionais, que foram exacerbadas pela Pandemia. QAnon, embora menos dominante em termos relativos, ainda mantém uma presença constante, especialmente durante os períodos de eleições e crises políticas nos EUA. A análise relativa mostra como o foco das teorias conspiratórias pode se deslocar ao longo do tempo, dependendo do contexto global, mas também como essas narrativas continuam interconectadas, alimentando uma visão de mundo onde eventos globais são interpretados como parte de uma grande conspiração.

### 3.3. Análise de conteúdo

A análise de conteúdo das comunidades relacionadas à Nova Ordem Mundial (NOM), globalismo e QAnon, utilizando nuvens de palavras, oferece uma compreensão profunda de como esses grupos articulam suas crenças e narrativas ao longo do tempo. As palavras que emergem com maior frequência nas nuvens indicam os principais focos dessas comunidades, refletindo não apenas os tópicos de interesse, mas também as estratégias discursivas que sustentam a coesão e a expansão dessas ideias. Termos como ”mundo”, ”Brasil”, ”verdade” e ”agora” aparecem de maneira proeminente, sugerindo um discurso que mescla percepções globais com preocupações nacionais. A presença constante de ”Deus” e ”governo” aponta para a interseção entre religião e política, frequentemente manipulada para justificar teorias conspiratórias complexas. Analisar essas nuvens ao longo do tempo permite identificar mudanças e continuidades nas narrativas, oferecendo insights sobre como esses movimentos se adaptam a contextos políticos e sociais mutantes, mantendo-se relevantes e influentes.

**Figura 10.** Nuvem de palavras consolidadas de nova ordem mundial, globalismo e QAnon

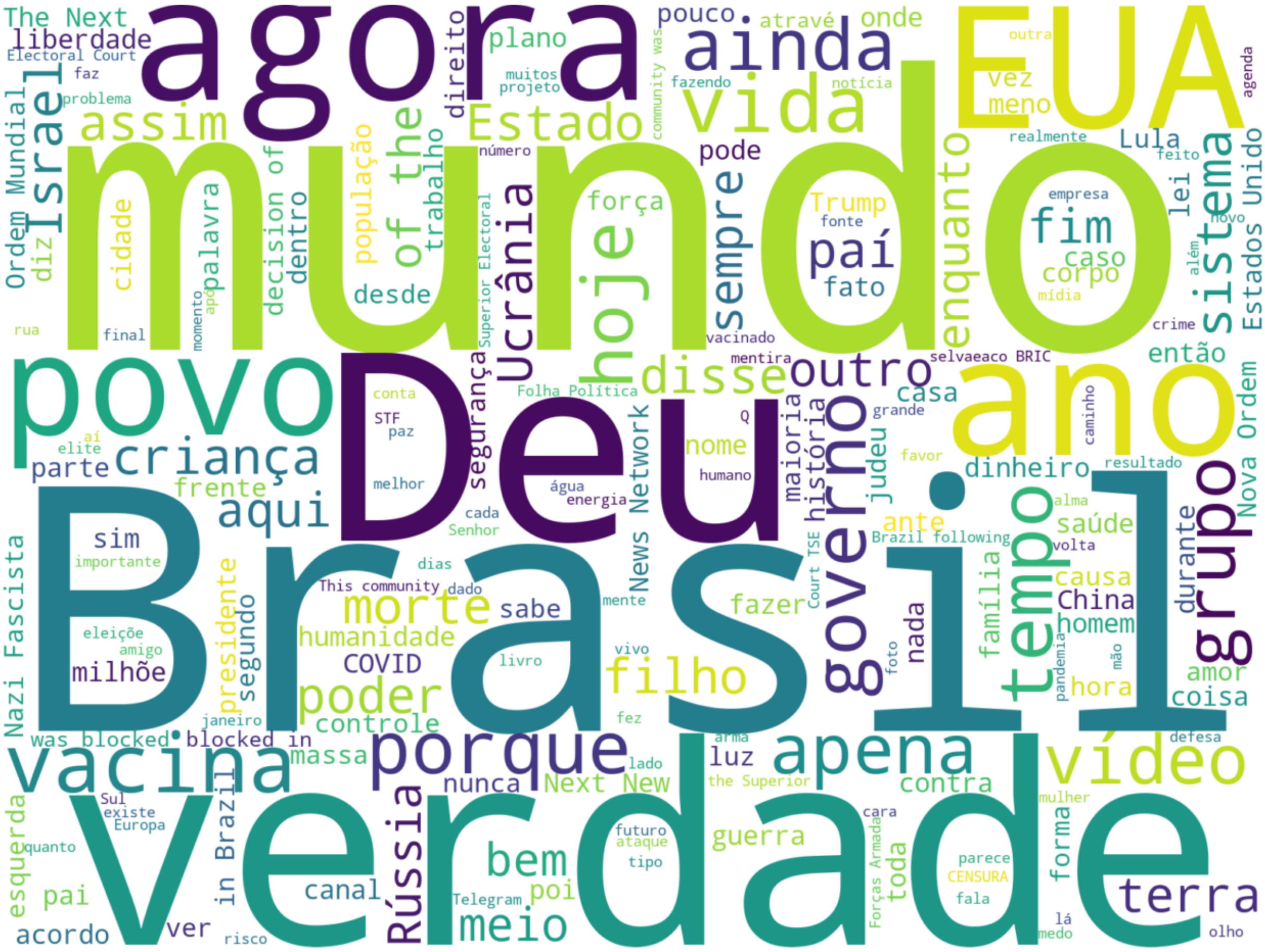

Fonte: Elaboração própria (2024).

A nuvem de palavras consolidada que aborda as comunidades da Nova Ordem Mundial, globalismo e QAnon revela uma forte interconexão entre temas globais e questões nacionais. Termos como ”mundo”, ”Brasil” e ”verdade” destacam-se, evidenciando uma narrativa que busca conectar acontecimentos internacionais a uma suposta agenda global que afetaria diretamente o Brasil. A palavra ”Deus” reforça a tendência dessas comunidades em vincular suas teorias a uma dimensão religiosa, utilizando a fé como um pilar para sustentar suas crenças conspiratórias. ”Governo” e ”povo” indicam uma preocupação com as dinâmicas de poder e a ideia de que o povo estaria sendo manipulado por elites ocultas, representadas por entidades como a Nova Ordem Mundial. A recorrência de ”vacina” e ”Pandemia” sinaliza a incorporação da crise sanitária global nas teorias conspiratórias, ampliando o espectro de preocupações dessas comunidades e interligando temas de saúde pública com conspirações políticas. Em resumo, a nuvem de palavras revela a complexidade e a multifacetada natureza das discussões dentro dessas comunidades, onde questões de soberania, religião e ciência se entrelaçam para formar um discurso de resistência e desconfiança generalizada.

[ versão português / english above ]

**Quadro 01.** Nuvem de palavras em série temporal de nova ordem mundial

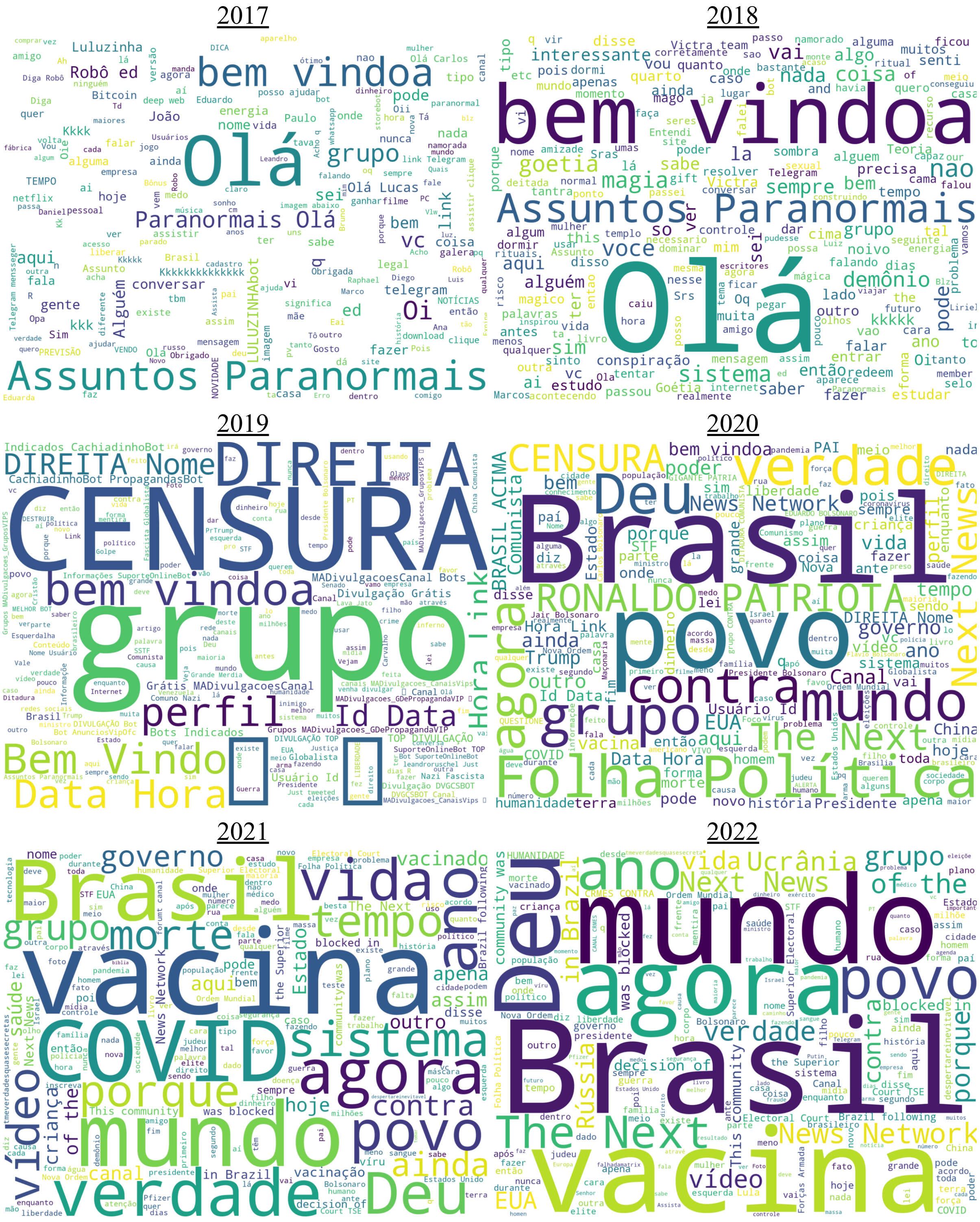

[ versão português / english above ]

2023 2024

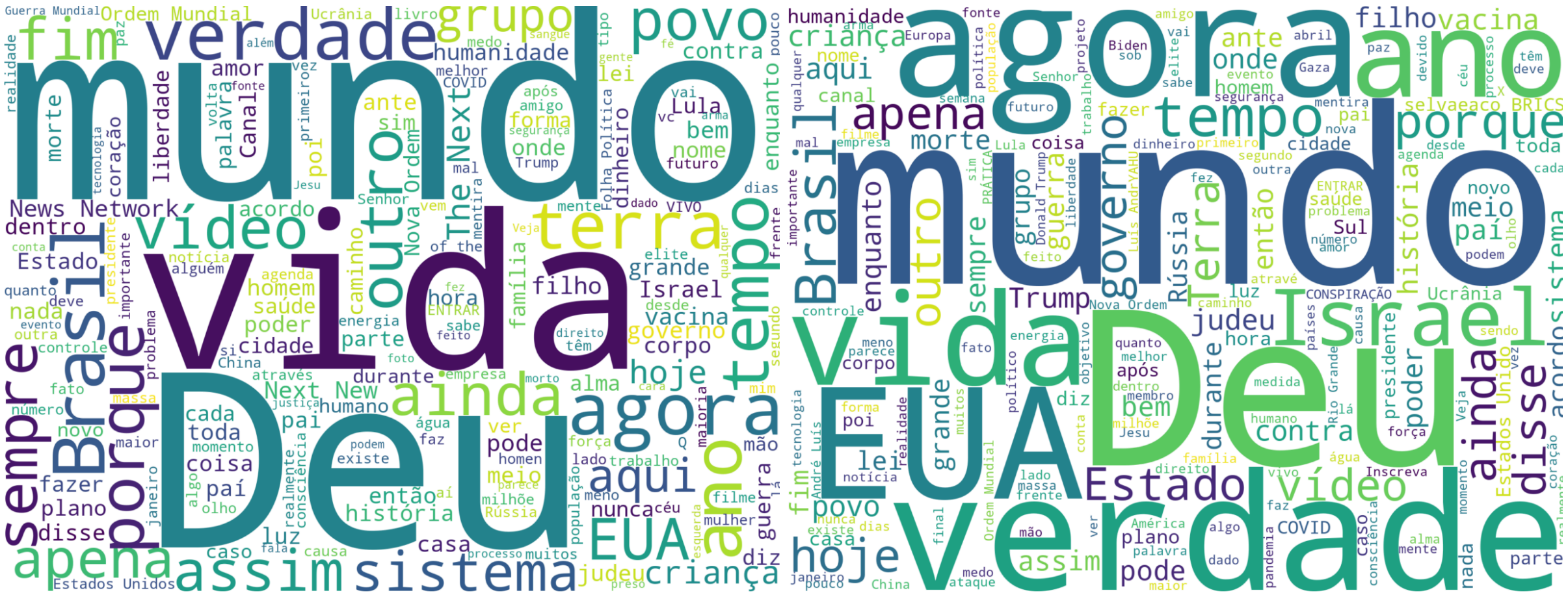

Fonte: Elaboração própria (2024).

O quadro de nuvens de palavras em série temporal para a Nova Ordem Mundial destaca como as narrativas e focos dessas comunidades evoluíram nos últimos anos. Em 2017, termos como ”Olá” e ”grupo” predominam, sugerindo uma fase inicial de formação e organização dessas comunidades, onde o acolhimento de novos membros era essencial. Em 2018, a palavra ”censura” surge, possivelmente em resposta a medidas de moderação de conteúdo nas redes sociais, marcando o início de uma retórica de vitimização e resistência contra plataformas digitais. A partir de 2020, ”Brasil” e ”mundo” ganham destaque, refletindo a intensificação das discussões sobre a Pandemia e como ela estaria inserida em um plano maior de dominação global. O termo ”vacina”, emergente em 2021, reforça essa narrativa, com as teorias conspiratórias ligando as campanhas de vacinação a uma suposta agenda de controle populacional. Em 2023 e 2024, a consolidação de termos como ”Deus” e ”verdade” indica um aprofundamento da radicalização, onde a fé e a busca por uma ”verdade” alternativa se tornam centrais no discurso, fortalecendo a coesão ideológica das comunidades.

**Quadro 02.** Nuvem de palavras em série temporal de globalismo

2018 2019

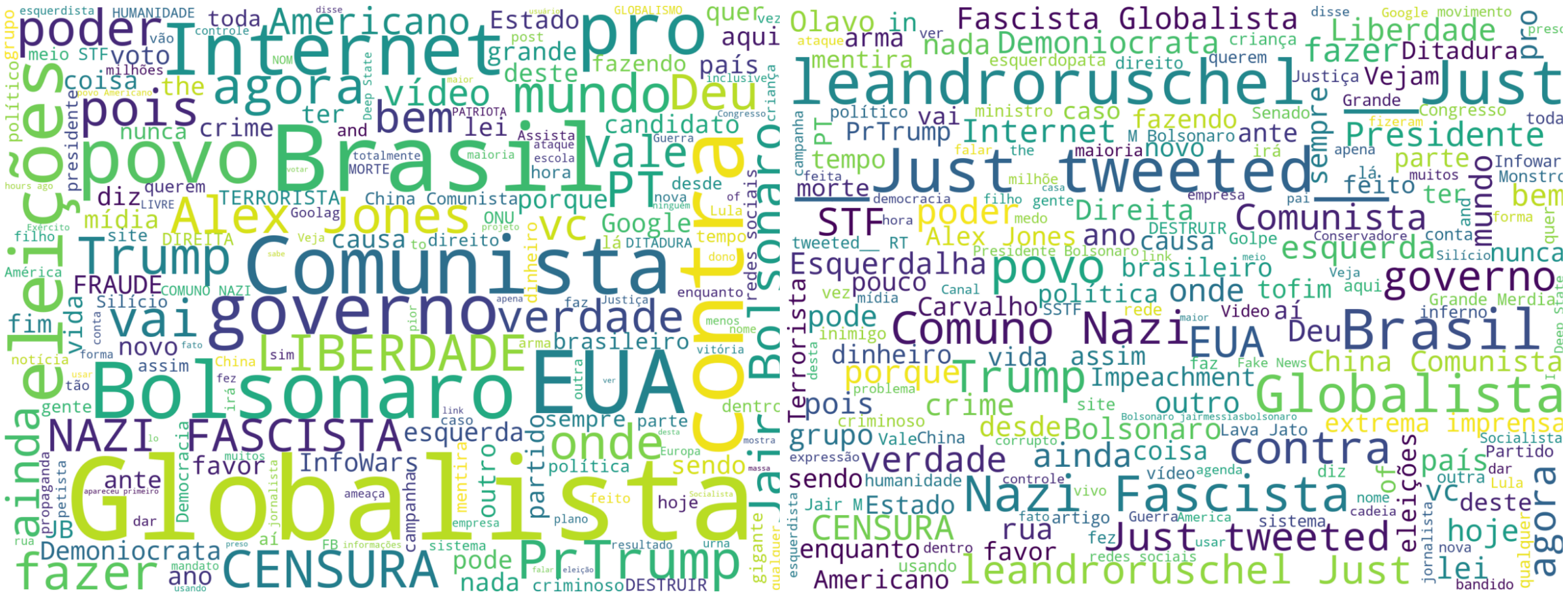

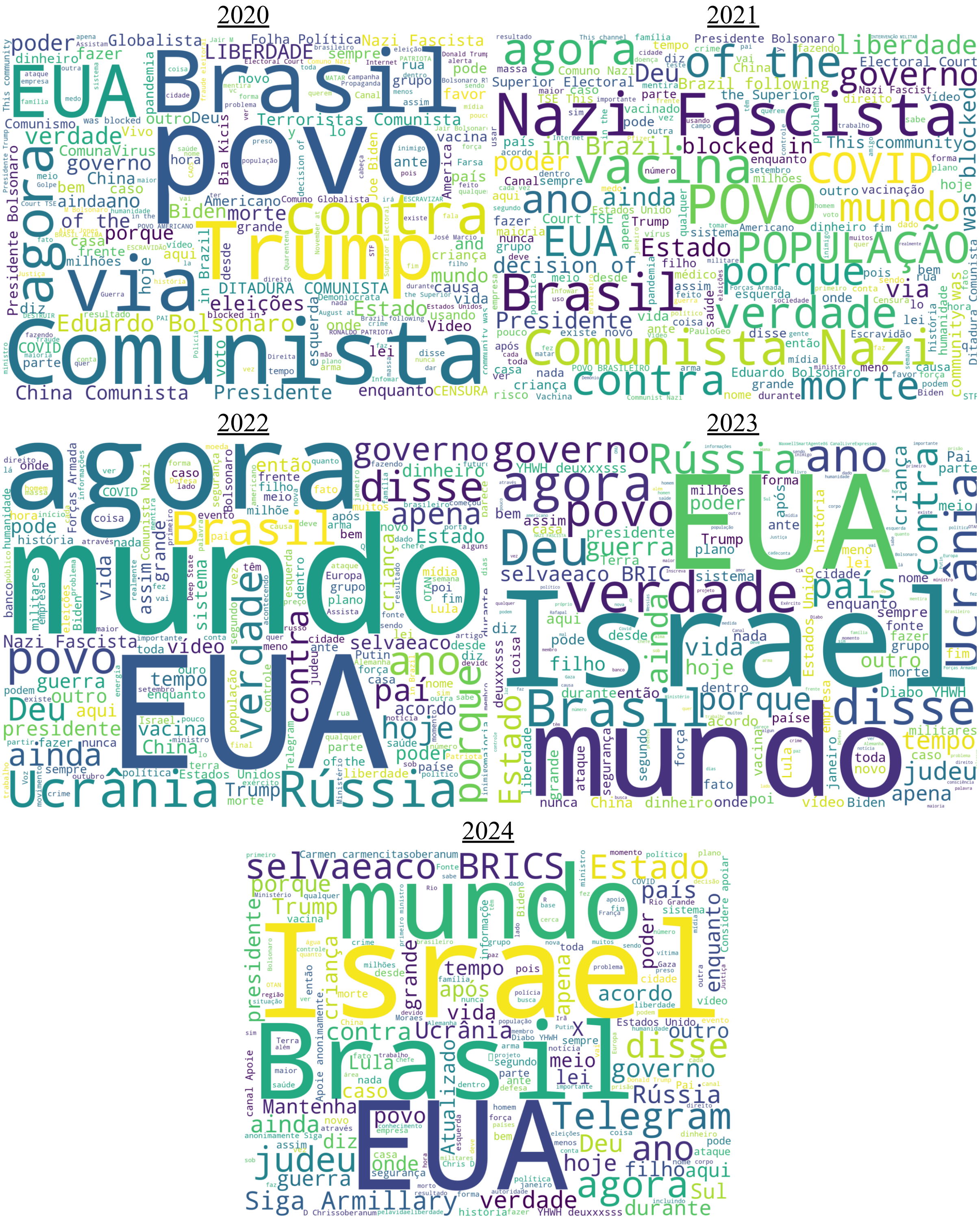

Fonte: Elaboração própria (2024).

O quadro de nuvens de palavras em série temporal para o globalismo revela uma progressão narrativa que reflete as transformações políticas e sociais em nível global e nacional. Em 2019, ”EUA”, ”Brasil” e ”povo” são termos predominantes, sugerindo uma preocupação com a influência americana e como ela seria percebida dentro da narrativa conspiratória do globalismo. A palavra ”comunista” começa a ganhar tração em 2020, alinhando-se com a retórica de extrema direita que vê o globalismo como uma ameaça comunista. Em 2021, ”Nazi” e ”Fascista” aparecem com maior frequência, indicando uma intensificação do discurso polarizado que equipara o globalismo a regimes totalitários. A partir de 2022, a palavra ”governo” torna-se mais proeminente, sinalizando uma preocupação crescente com as políticas públicas e sua suposta submissão a uma agenda globalista. Nos anos seguintes, ”mundo” e ”agora” continuam a ser termos centrais, sugerindo um senso de urgência e uma visão de que os eventos globais estão culminando em uma nova ordem mundial que exigiria resistência imediata.

**Quadro 03.** Nuvem de palavras em série temporal de QAnon

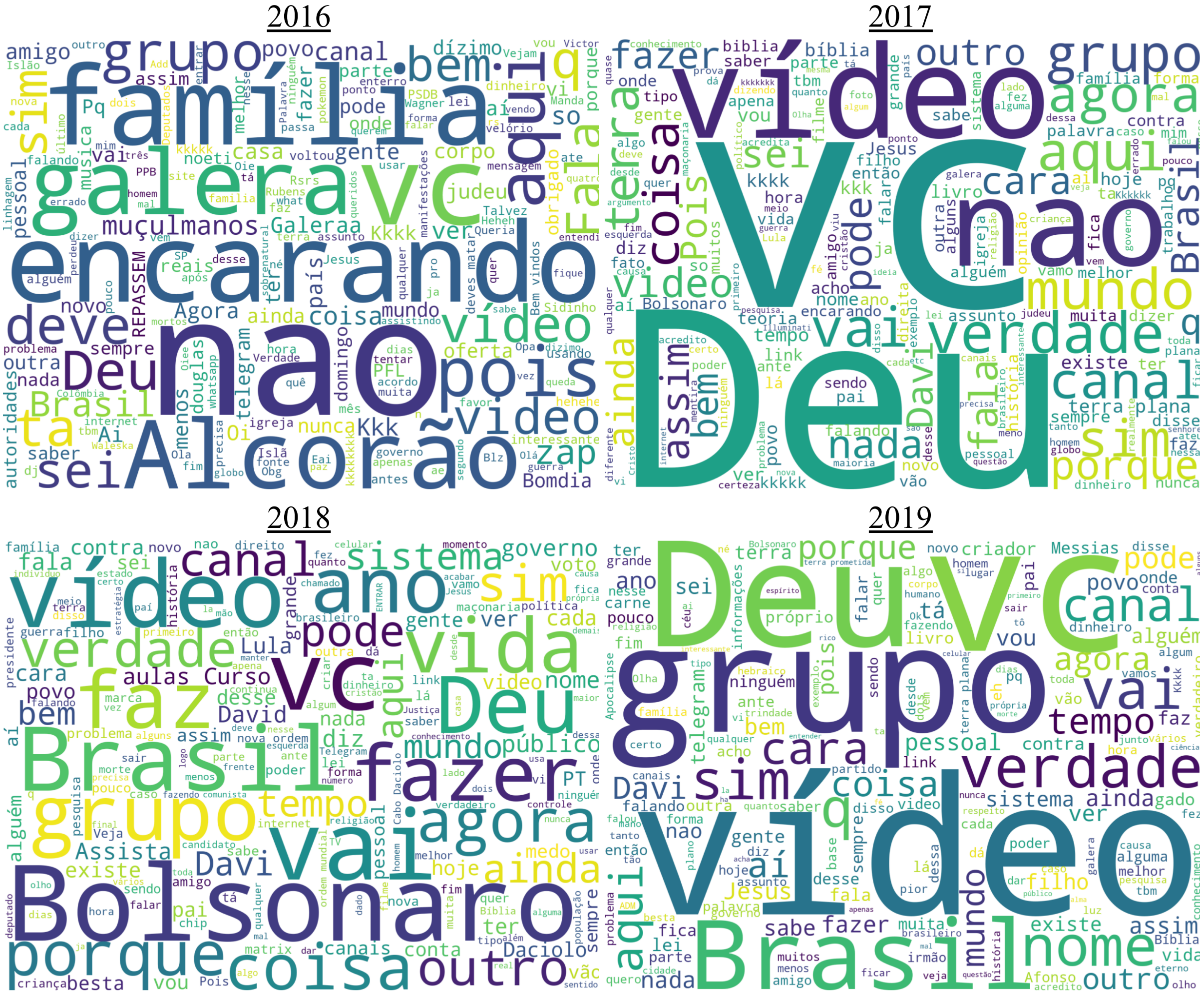

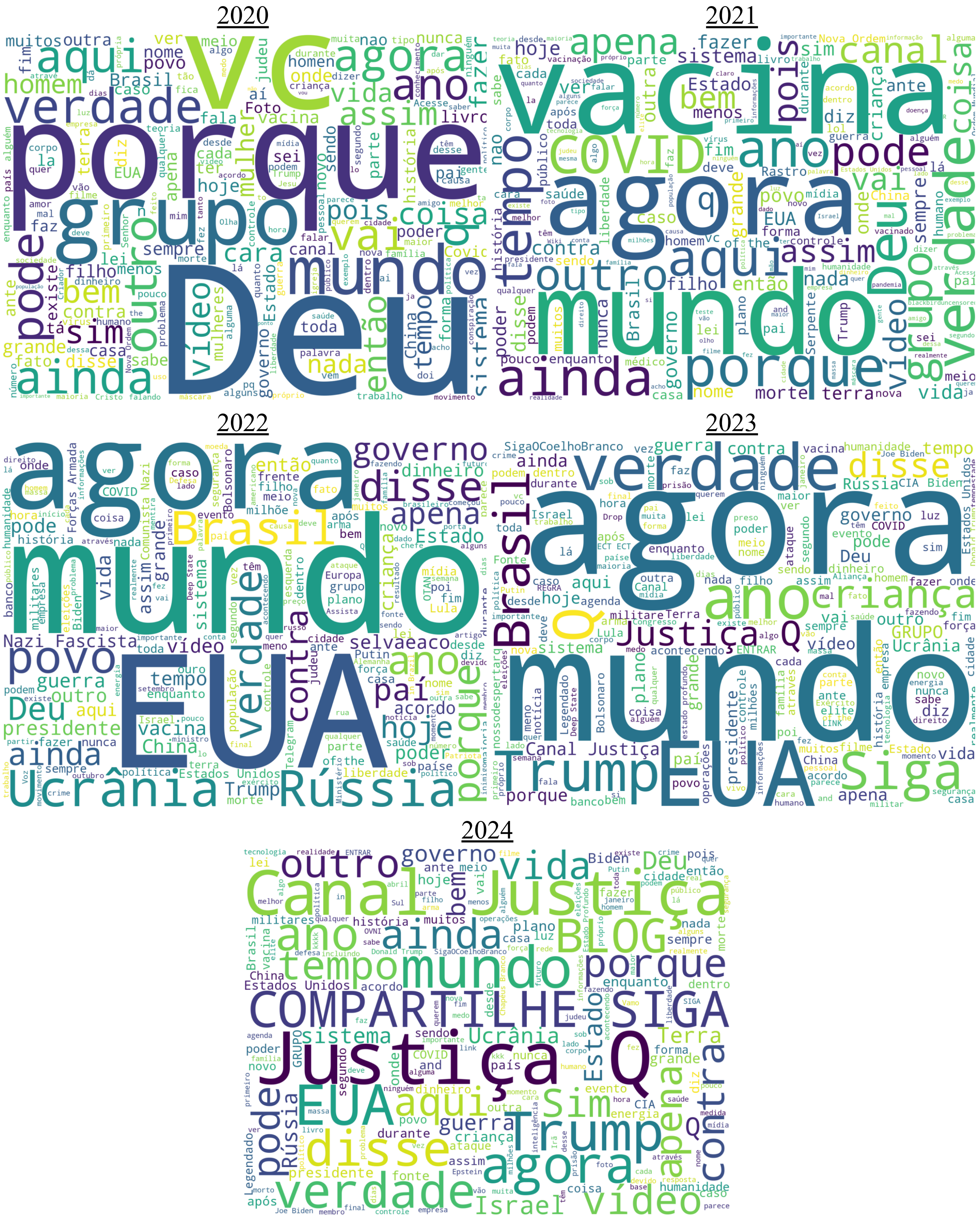

Fonte: Elaboração própria (2024).

O quadro de nuvens de palavras em série temporal para QAnon evidencia a evolução das narrativas dentro dessa comunidade, marcada por uma forte carga religiosa e conspiratória. Em 2016 e 2017, ”Deus”, ”vc” e ”família” são termos frequentes, refletindo uma fase em que a comunidade estava fortemente focada em valores conservadores e cristãos.

Em 2018, ”vídeo” começa a emergir, indicando o papel crescente do conteúdo audiovisual na disseminação das ideias QAnon, o que é reforçado em 2019 com a popularização de teorias mais complexas. Em 2020 e 2021, ”Brasil”, ”mundo” e ”vacina” tornam-se centrais, à medida que a Pandemia da COVID-19 é integrada nas teorias conspiratórias, ampliando o alcance das narrativas QAnon para além dos Estados Unidos. Nos últimos anos, especialmente em 2023 e 2024, a combinação de termos como ”justiça”, ”verdade” e ”compartilhe” indica uma fase de mobilização ativa, onde as comunidades QAnon não apenas discutem teorias, mas também incentivam ações concretas de disseminação e resistência contra o que percebem como uma conspiração global orquestrada pelas elites.

### 3.4. Sobreposição de agenda temática

As figuras a seguir exploram a sobreposição de temáticas em comunidades de teorias da conspiração, especificamente aquelas relacionadas à Nova Ordem Mundial (NWO), ao Globalismo e ao movimento QAnon. Através da análise visual dos tópicos abordados por essas comunidades, é possível perceber como temas globais, culturais e religiosos se entrelaçam, reforçando narrativas conspiratórias que visam questionar e deslegitimar estruturas políticas, sociais e científicas. A interseção desses temas cria uma rede de desinformação coesa que perpetua crenças centrais, dificultando o combate às influências.

**Figura 11.** Temáticas de conflitos globais

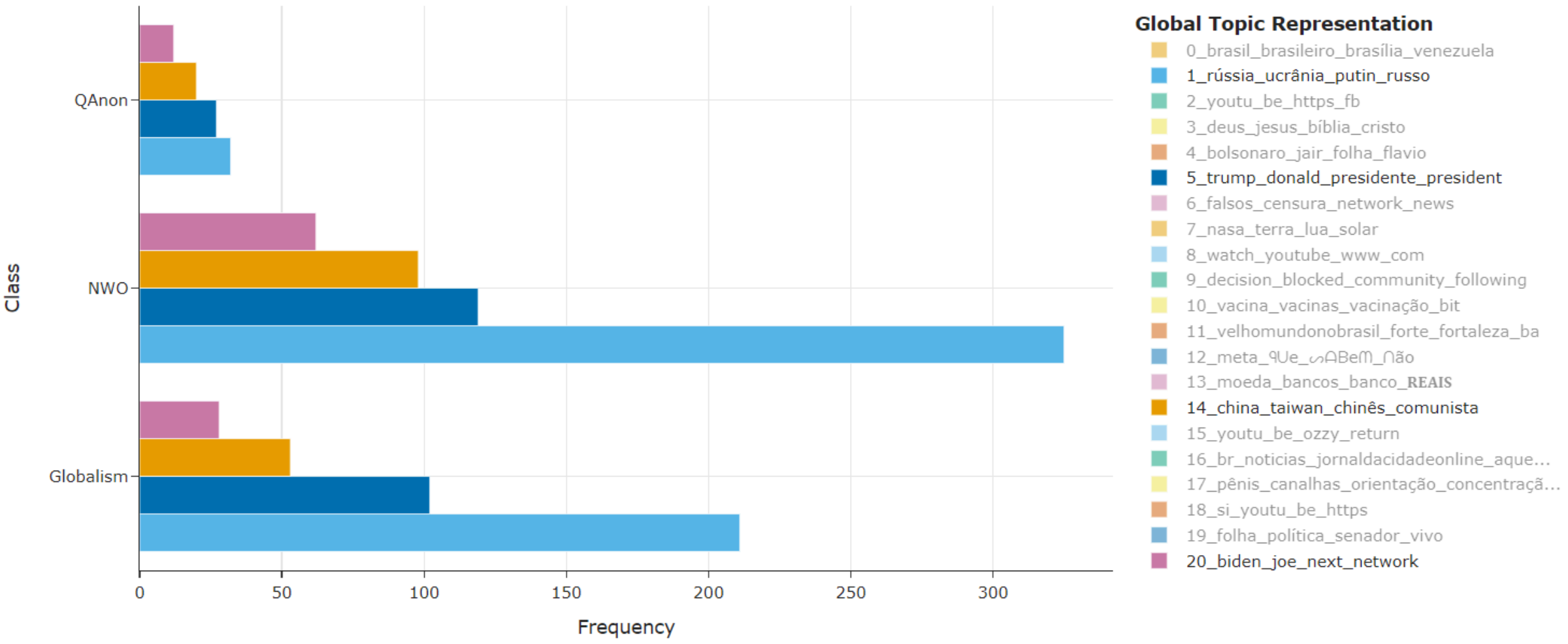

Fonte: Elaboração própria (2024).

A Figura 11 analisa as temáticas de conflitos globais dentro das comunidades de QAnon, NWO e Globalismo. Tópicos como “Rússia”, “Ucrânia”, e “Biden” são predominantes, sugerindo que essas comunidades utilizam eventos geopolíticos recentes para validar suas narrativas de uma conspiração global. O destaque para a NWO indica uma forte associação dessas comunidades com a ideia de que conflitos internacionais são orquestrados como parte de um plano para estabelecer uma nova ordem mundial. A interseção entre esses tópicos reforça a percepção de que essas teorias são alimentadas pela instrumentalização de eventos reais para sustentar um discurso de desconfiança e manipulação global.

**Figura 12.** Temáticas de fé e religião

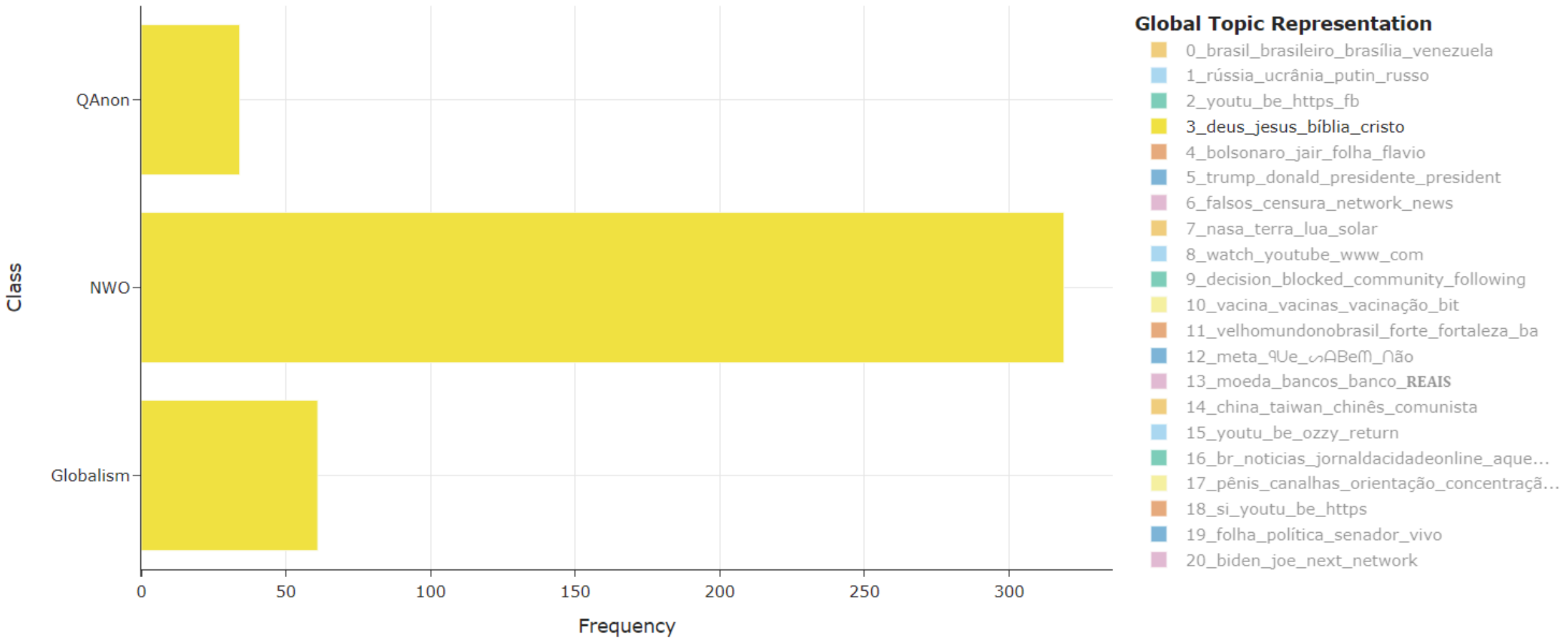

Fonte: Elaboração própria (2024).

Na Figura 12, observamos a interseção de temáticas de fé e religião com as narrativas de QAnon, NWO e Globalismo. Tópicos como “Deus”, “Jesus” e “bíblia” são amplamente discutidos, especialmente dentro das narrativas da NWO, sugerindo uma tentativa de vincular a luta contra a suposta conspiração global a uma batalha espiritual. A prevalência desses temas mostra como a religião é instrumentalizada para fortalecer a crença em teorias da conspiração, colocando a luta contra a NWO como um dever sagrado e justificando a resistência a estruturas políticas e sociais sob a ótica de uma missão divina.

**Figura 13.** Temáticas de discursos anti-LGBT e agenda globalista

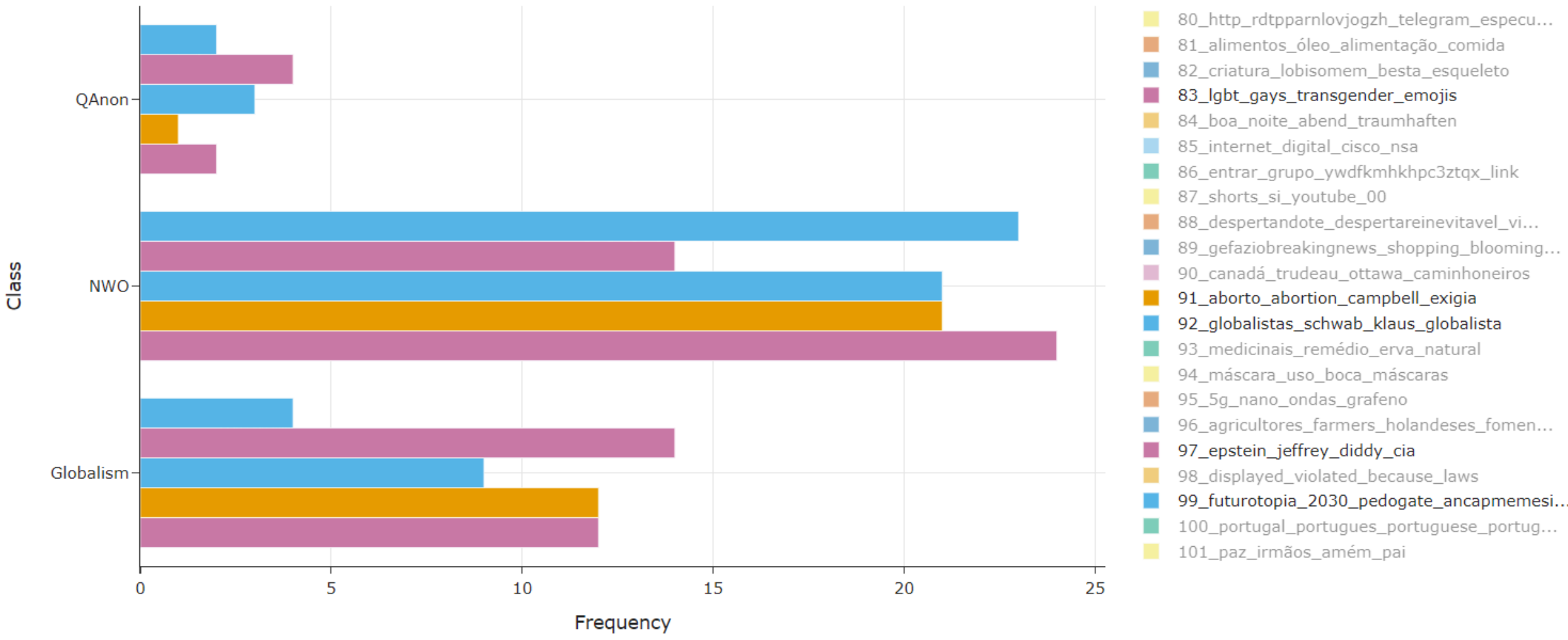

Fonte: Elaboração própria (2024).

A Figura 13 destaca a associação entre discursos anti-LGBT e a agenda globalista nas comunidades de QAnon, NWO e Globalismo. Tópicos como “LGBT”, “aborto”, e “globalistas” indicam que essas comunidades frequentemente integram narrativas que atacam direitos civis e minorias, relacionando essas questões a uma agenda globalista que visa destruir valores tradicionais. A sobreposição de temas sugere que a retórica anti-LGBT é utilizada para mobilizar apoio contra o que essas comunidades percebem como uma ameaça global, reforçando a ideia de que a proteção dos valores morais está intrinsecamente ligada à resistência contra uma conspiração global.

**Figura 14.** Temáticas de neonazismo, racismo e antissemitismo

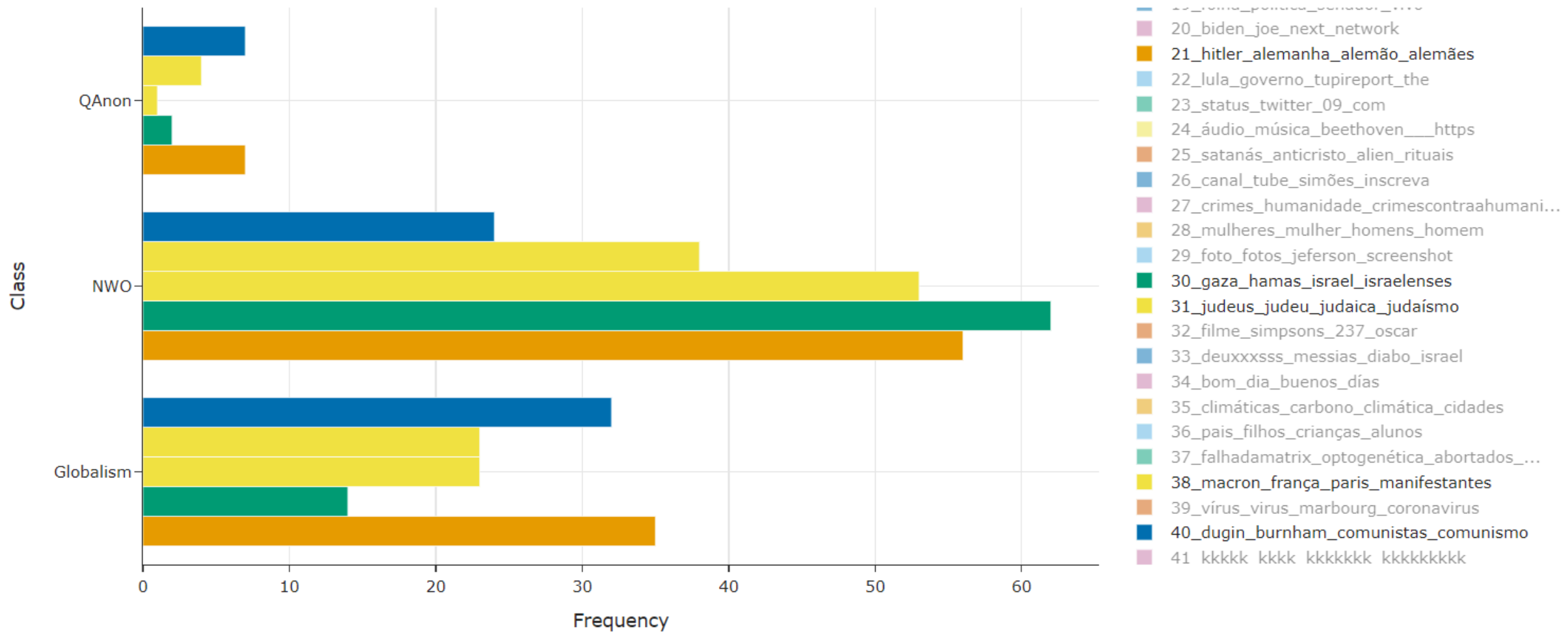

Fonte: Elaboração própria (2024).

Na Figura 14, são abordadas temáticas de neonazismo, racismo e antissemitismo dentro das narrativas de QAnon, NWO e Globalismo. Tópicos como "Hitler", "judeus", e "racismo" são evidentes, especialmente nas discussões associadas à NWO, sugerindo que essas comunidades utilizam essas ideologias extremistas para sustentar suas teorias conspiratórias. A presença significativa desses temas indica que o racismo e o antissemitismo são frequentemente usados como ferramentas para reforçar a narrativa de uma conspiração global, associando grupos minoritários a um plano maligno para controlar o mundo.

**Figura 15.** Temáticas de mudanças climáticas

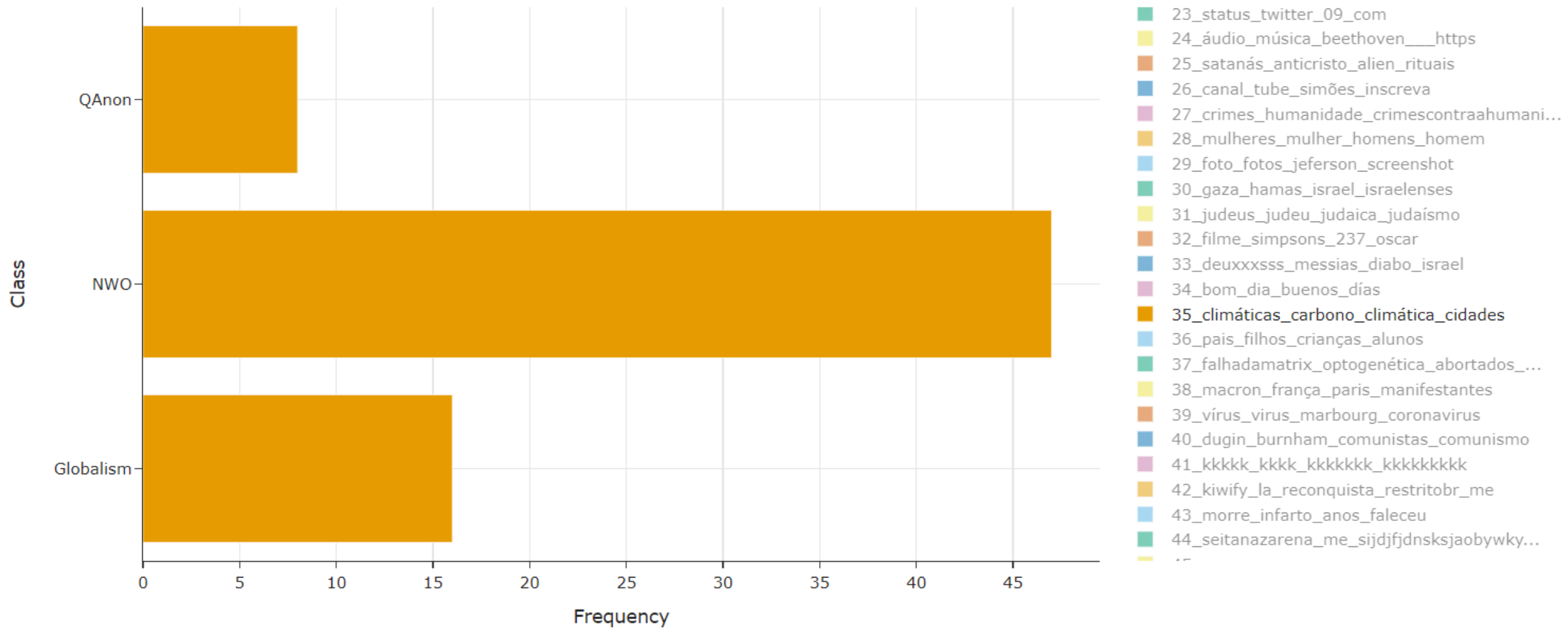

Fonte: Elaboração própria (2024).

A Figura 15 explora a relação entre a negação das mudanças climáticas e as teorias de QAnon, NWO e Globalismo. Tópicos como "climáticas", "carbono", e "cidades" são proeminentes, especialmente nas narrativas da NWO, onde as mudanças climáticas são vistas como uma invenção para justificar o controle global. Essa sobreposição temática reflete uma resistência à ciência convencional, onde a negação das mudanças climáticas é utilizada para reforçar a desconfiança em relação a políticas ambientais, apresentando-as como parte de uma agenda globalista para subjugar populações e eliminar liberdades individuais.

**Figura 16.** Temáticas de negação às vacinas com supostas mortes súbitas por vacinas

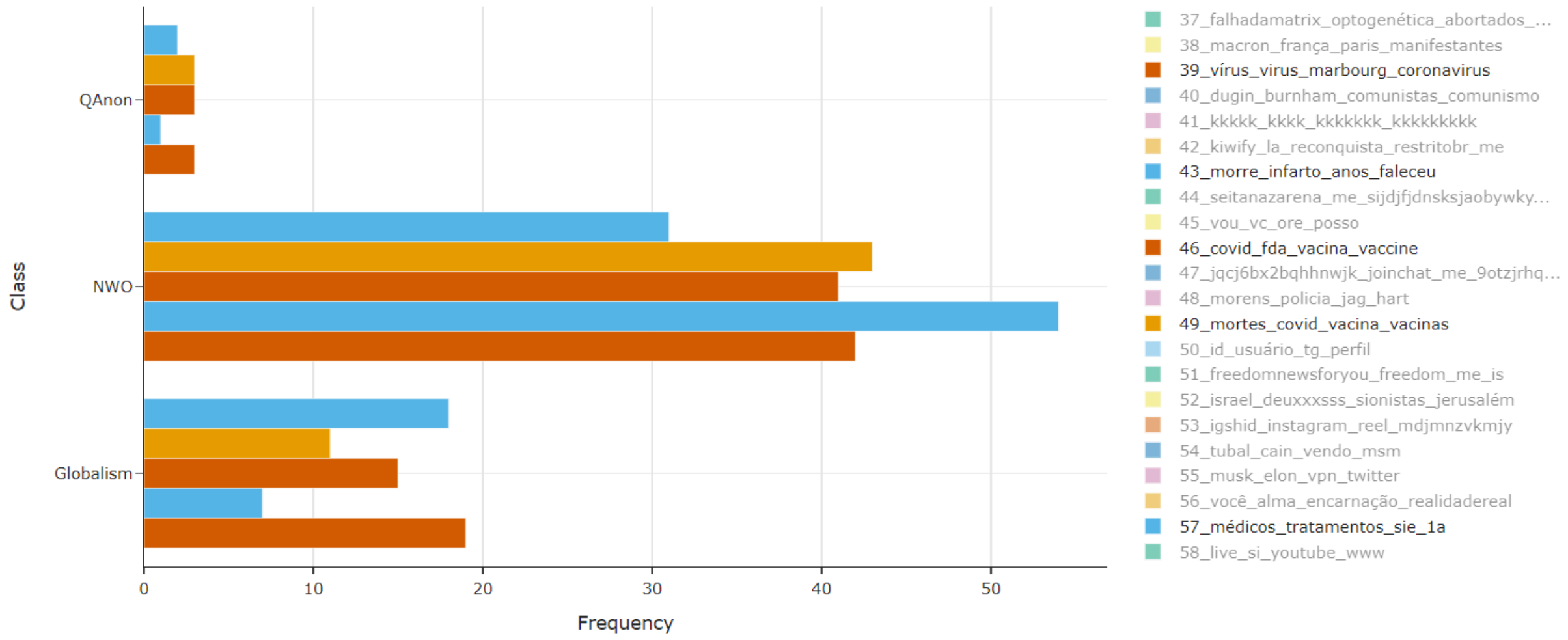

Fonte: Elaboração própria (2024).

Na Figura 16, são discutidas temáticas de negação às vacinas, com foco em narrativas de supostas mortes súbitas causadas por vacinas, dentro das comunidades de QAnon, NWO e Globalismo. Tópicos como "covid", "vacina", e "mortes" são amplamente presentes, sugerindo que essas comunidades utilizam essas narrativas para reforçar a desconfiança em relação às vacinas e ao sistema de saúde global. A prevalência desse discurso, especialmente associado à NWO, reflete uma tentativa de descredibilizar campanhas de vacinação, apresentando-as como parte de um plano para reduzir a população mundial e supostamente exercer controle sobre a humanidade.

## 4. Reflexões e trabalhos futuros

Para responder a pergunta de pesquisa "**como são caracterizadas e articuladas as comunidades de teorias da conspiração brasileiras sobre temáticas de nova ordem mundial, globalismo e QAnon no Telegram?**", este estudo adotou técnicas espelhadas em uma série de sete publicações que buscam caracterizar e descrever o fenômeno das teorias da conspiração no Telegram, adotando o Brasil como estudo de caso. Após meses de investigação, foi possível extrair um total de 217 comunidades de teorias da conspiração brasileiras no Telegram sobre temáticas de nova ordem mundial, globalismo e QAnon , estas somando 5.545.369 de conteúdos publicados entre junho de 2016 (primeiras publicações) até agosto de 2024 (realização deste estudo), com 718.246 usuários somados dentre as comunidades.

Foram adotadas quatro abordagens principais: **(i)** Rede, que envolveu a criação de um algoritmo para mapear as conexões entre as comunidades por meio de convites circulados entre grupos e canais; **(ii)** Séries temporais, que utilizou bibliotecas como “Pandas” (McKinney, 2010) e “Plotly” (Plotly Technologies Inc., 2015) para analisar a evolução das publicações e engajamentos ao longo do tempo; **(iii)** Análise de conteúdo, sendo aplicadas técnicas de análise textual para identificar padrões e frequências de palavras nas comunidades ao longo dos semestres; e **(iv)** Sobreposição de agenda temática, que utilizou o modelo BERTopic (Grootendorst, 2020) para agrupar e interpretar grandes volumes de textos, gerando tópicos coerentes a partir das publicações analisadas. A seguir, as principais reflexões são detalhadas, sendo seguidas por sugestões para trabalhos futuros.

### 4.1. Principais reflexões

**A Nova Ordem Mundial e o Globalismo como catalisadores centrais de teorias conspiratórias no Telegram brasileiro:** O estudo revela que as comunidades de Nova Ordem Mundial (NOM) e Globalismo desempenham papeis centrais na disseminação de teorias conspiratórias no Telegram brasileiro. Com 2.411.003 publicações e 329.304 usuários ativos, as discussões sobre NOM são amplamente difundidas, criando um ambiente de desconfiança global. O Globalismo também se destaca, com 768.176 publicações e 326.596 usuários, mostrando que essas comunidades não apenas perpetuam narrativas conspiratórias, mas também funcionam como *hubs* que ligam diversas outras teorias;

**A interconexão entre NOM, Globalismo e QAnon fortalece uma rede densa de desinformação:** A análise da rede de comunidades mostra que NOM, Globalismo e QAnon estão interconectados de forma a criar uma malha conspiratória coesa. QAnon, embora menos volumosa (531.678 publicações), atua como um “narrativa-*hub*” que conecta teorias, enquanto NOM e Globalismo amplificam essas conexões, reforçando a visão de um mundo controlado por uma elite global. A centralidade dessas temáticas indica que, ao entrar em uma dessas comunidades, os membros são rapidamente expostos a uma rede ampla de desinformação;

**Crescimento expressivo das menções a NOM e Globalismo durante crises globais:** Durante eventos críticos como a Pandemia da COVID-19 e as eleições presidenciais dos EUA em 2020, as menções a NOM e Globalismo cresceram exponencialmente. NOM, por exemplo, registrou um aumento de 3.150% nas menções entre 2019 e 2021. Esse crescimento é atribuído à amplificação das narrativas conspiratórias em resposta a crises globais, refletindo uma desconfiança crescente nas instituições internacionais e nacionais;

**Portas de entrada: NOM e Globalismo atraem novos membros através de outras teorias conspiratórias:** As comunidades de NOM e Globalismo não apenas centralizam narrativas, mas também atuam como portas de entrada para novos membros que já estão imersos em outras teorias conspiratórias. A NOM recebeu 8.699 links de convites de Conspirações Gerais e 6.260 de comunidades antivacinas, mostrando que essas comunidades funcionam como epicentros que atraem e retêm seguidores de outras narrativas, fortalecendo a coesão e o alcance dessas teorias;

**Globalismo como princípio organizador de teorias conspiratórias diversas:** O estudo evidencia que o Globalismo funciona como um princípio organizador que conecta diferentes teorias conspiratórias, como Anti-*Woke*, Revisionismo, e até Mudanças Climáticas. Com 32.640 links recebidos de Conspirações Gerais, o Globalismo se posiciona como uma narrativa integradora, sugerindo que todas as teorias fazem parte de um plano maior, aumentando a interconectividade e a resiliência das comunidades conspiratórias;

**QAnon como uma metanarrativa flexível que absorve outras teorias:** A narrativa de QAnon, apesar de menos volumosa em comparação com NOM e Globalismo, desempenha um papel crucial ao funcionar como uma metanarrativa que absorve e conecta outras teorias conspiratórias. Com 6.308 links de Conspirações Gerais e 5.033 de NOM, QAnon se posiciona como um ponto de convergência para várias crenças, reforçando a visão conspiratória de seus seguidores e dificultando a entrada de informações contraditórias;

**A NOM como um paradigma interpretativo central no universo conspiratório:** A NOM se destaca como um paradigma central que unifica diversas narrativas conspiratórias contra uma suposta elite global. Com um volume expressivo de 3.488.686 publicações, a NOM opera como um aglutinador de desinformação, onde teorias como antivacinas, apocalipse e ocultismo encontram coerência e legitimidade, reforçando uma visão de mundo onde todos os eventos globais são manipulados por uma elite secreta;

**A instrumentalização de temas religiosos nas narrativas conspiratórias:** Termos religiosos como "Deus" e "bíblia" são recorrentes nas discussões sobre NOM e QAnon, indicando uma forte interseção entre religião e conspirações. Essas narrativas utilizam a fé para justificar teorias sobre controle global e manipulação das massas, sugerindo que a luta contra a NOM e o Globalismo é uma batalha espiritual, o que fortalece a coesão ideológica das comunidades e a resistência às informações científicas;

**A rejeição às mudanças climáticas como parte de uma agenda globalista:** As teorias sobre mudanças climáticas são amplamente discutidas nas comunidades de NOM e Globalismo, onde são vistas como parte de uma conspiração global para justificar o controle sobre a população. Com tópicos como "carbono" e "climáticas" frequentemente mencionados, essas comunidades rejeitam a ciência convencional e apresentam as políticas ambientais como uma ameaça às liberdades individuais, contribuindo para a polarização e desinformação;

**A perpetuação da desinformação sobre vacinas e sua ligação com a NOM:** As comunidades de NOM são particularmente ativas na disseminação de desinformação sobre vacinas, utilizando narrativas sobre supostas mortes súbitas causadas por vacinas para reforçar a desconfiança. Essas teorias são frequentemente associadas à ideia de um plano de controle populacional, sugerindo que as campanhas de vacinação fazem parte de uma agenda de manipulação, o que perpetua a resistência às vacinas e amplifica o impacto da desinformação.

### 4.2. Trabalhos futuros

Com base nos principais achados deste estudo, várias direções podem ser sugeridas para futuras pesquisas. A primeira delas seria explorar mais a fundo as conexões entre NOM, Globalismo e QAnon, especialmente em como essas comunidades utilizam crises globais para validar suas narrativas. Investigar como esses eventos são reinterpretados e amplificados por essas comunidades pode oferecer insights valiosos para desenvolver estratégias de combate à desinformação em períodos de crise.

Além disso, a instrumentalização da religião nas comunidades de NOM merece uma investigação mais aprofundada. Entender como a fé é manipulada para justificar e perpetuar teorias conspiratórias pode ajudar a identificar abordagens mais eficazes para desconstruir essas narrativas. Pesquisas que explorem a interseção entre religião e desinformação poderiam fornecer bases para campanhas de correção factual que respeitem as crenças religiosas, mas que ofereçam uma alternativa à narrativa conspiratória. Outro ponto relevante é a análise das conexões entre teorias anti-LGBT e a agenda globalista dentro dessas comunidades. Estudos futuros poderiam se concentrar em mapear como essas narrativas se reforçam mutuamente e como elas são utilizadas para mobilizar apoio contra direitos civis. Compreender essas dinâmicas é crucial para desenvolver intervenções que busquem proteger os direitos humanos enquanto desmantelam as narrativas conspiratórias associadas.

A persistência das teorias conspiratórias no discurso público também aponta para a necessidade de investigar os mecanismos que permitem que essas crenças sejam reintroduzidas e ganhem tração novamente. Estudos poderiam se concentrar em como essas narrativas são mantidas e adaptadas ao longo do tempo, especialmente em face de intervenções externas. Compreender esses ciclos de resiliência pode ser a chave para desenvolver estratégias de longo prazo para combater a desinformação.

Por fim, a criação de uma rede interconectada de desinformação entre NOM, Globalismo e QAnon sugere que há um ciclo de retroalimentação de crenças que precisa ser interrompido. Futuros estudos poderiam investigar como essas redes se formam, como elas se mantêm coesas e quais são os pontos fracos que poderiam ser explorados para desestabilizar essas narrativas. Desenvolver métodos para identificar e neutralizar esses *hubs* de desinformação antes que eles ganhem tração é essencial para reduzir o impacto dessas teorias.

## 5. Referências

## 6. Biografia do autor

**Ergon Cugler de Moraes Silva** possui mestrado em Administração Pública e Governo (FGV), MBA pós-graduação em Ciência de Dados e Análise (USP) e bacharelado em Gestão de Políticas Públicas (USP). Ele está associado ao Núcleo de Estudos da Burocracia (NEB FGV), colabora com o Observatório Interdisciplinar de Políticas Públicas (OIPP USP), com o Grupo de Estudos em Tecnologia e Inovações na Gestão Pública (GETIP USP), com o Monitor de Debate Político no Meio Digital (Monitor USP) e com o Grupo de Trabalho sobre Estratégia, Dados e Soberania do Grupo de Estudo e Pesquisa sobre Segurança Internacional do Instituto de Relações Internacionais da Universidade de Brasília (GEPSI UnB). É também pesquisador no Instituto Brasileiro de Informação em Ciência e Tecnologia (IBICT), onde trabalha para o Governo Federal em estratégias contra a desinformação. Brasília, Distrito Federal, Brasil. Site: https://ergoncugler.com/.